\documentclass[12pt]{article}
\usepackage{amsmath}
\usepackage{graphicx}
\usepackage{enumerate}
\usepackage{natbib}
\usepackage{url} 

\newcommand{\blind}{1}

\RequirePackage{amsthm,amsfonts,amssymb}
\RequirePackage[colorlinks,citecolor=blue,urlcolor=blue, linkcolor=blue]{hyperref}
\RequirePackage{graphicx}
\usepackage{makecell}
\usepackage{multirow}
\usepackage{nicematrix}
\usepackage{xcolor}
\usepackage{enumitem}
\usepackage{lscape}
\usepackage{dsfont}
\usepackage{booktabs}

\usepackage{xcolor} 
\usepackage{tablefootnote}
\usepackage{booktabs,tabularx}
\usepackage{afterpage}

\usepackage[perpage]{footmisc}

\usepackage{subfig}

\usepackage{soul}
\definecolor{amaranth}{rgb}{0.0, 0.0, 1.28}
\colorlet{green}{pink!29}
\colorlet{yellow}{lime!12}
\colorlet{cyan}{cyan!10}

\newcommand{\hlc}[2][yellow]{{%
    \colorlet{foo}{#1}%
    \sethlcolor{foo}\hl{#2}}%
}

\usepackage[linesnumbered,ruled,vlined]{algorithm2e}

\SetKwComment{Comment}{\color{green!80!black!190}// }{}

\SetKwProg{Function}{function}{}{}
\SetKw{Continue}{continue}

\DontPrintSemicolon
\SetAlFnt{\small}
\SetAlgorithmName{Pseudo-code}{algorithmautorefname}

\usepackage{tikz}
\usetikzlibrary{fit,calc}
\newcommand{\boxit}[2]{
    \tikz[remember picture,overlay] \node (A) {};\ignorespaces
    \tikz[remember picture,overlay]{\node[draw, draw=gray!80, line width=0.1pt, yshift=3pt,fill=#1,
    fit={($(A)+(0,0.30\baselineskip)$)($(A)+(.9\linewidth,-{#2}\baselineskip - 0.25\baselineskip)$)}] {};}\ignorespaces
}
\newcommand{\boxitt}[2]{
    \tikz[remember picture,overlay] \node (A) {};\ignorespaces
    \tikz[remember picture,overlay]{\node[draw, draw=gray!80, line width=0.1pt, yshift=3pt,fill=#1,
    fit={($(A)+(0,0.20\baselineskip)$)($(A)+(.858\linewidth,-{#2}\baselineskip - 0.25\baselineskip)$)}] {};}\ignorespaces
}

\usepackage{algpseudocode}
\usepackage{booktabs}  
\usepackage{threeparttable}    
\usepackage{dcolumn}    
\let\estinput=\input 
	
\newcommand{\estauto}[3]{
		\vspace{.75ex}{
			\begin{tabular}{l*{#2}{#3}}
			\toprule
			\estinput{#1}
			\\ \bottomrule          
			\addlinespace[.75ex]
			\end{tabular}
			}
		}

\usepackage{setspace}
\theoremstyle{plain}

\theoremstyle{remark}

\usepackage{tikz,xcolor,hyperref}
\definecolor{lime}{HTML}{A6CE39}
\DeclareRobustCommand{\orcidicon}{%
    \begin{tikzpicture}
    \draw[lime, fill=lime] (0,0) 
    circle [radius=0.16] 
    node[white] {{\fontfamily{qag}\selectfont \tiny ID}};
    \draw[white, fill=white] (-0.0625,0.095) 
    circle [radius=0.007];
    \end{tikzpicture}
    \hspace{-2mm}
}

\newcommand{\orcidAR}{\href{https://orcid.org/0000-0002-3647-7340}{\orcidicon}}
\newcommand{\orcidCM}{\href{https://orcid.org/0000-0002-9208-3194}{\orcidicon}}
\newcommand{\orcidFI}{\href{https://orcid.org/0000-0003-0165-1983}{\orcidicon}}
\newcommand{\orcidAMP}{\href{https://orcid.org/0000-0002-8253-3630}{\orcidicon}}

\begin{document}

\def\spacingset#1{\renewcommand{\baselinestretch}%
{#1}\small\normalsize} \spacingset{1}


\if1\blind
{
  \title{\bf Clustering Hierarchies via a Semi-Parametric Generalized Linear Mixed Model: a statistical significance-based approach}
  \author{Alessandra Ragni$^{1,*}$\orcidAR, Chiara Masci$^{1}$\orcidCM, Francesca Ieva$^{1,2}$\orcidFI\\ and Anna Maria Paganoni$^{1}$\orcidAMP\\
    \small $^1$MOX Lab, Department of Mathematics, Politecnico di Milano, Milan, 20133, Italy\\
    \small $^2$Human Technopole, Health Data Science Center, Milan, 20157, Italy\\
    \small $^*$alessandra.ragni@polimi.it}
    \date{}
  \maketitle
} \fi

\if0\blind
{
  \bigskip
  \bigskip
  \bigskip
  \begin{center}
    {\LARGE\bf Clustering Hierarchies via a Semi-Parametric}
    \LARGE \bf Generalized Linear Mixed Model:
    \LARGE \bf a statistical significance-based approach
\end{center}
  \medskip
} \fi

\bigskip
\begin{abstract}
We introduce a novel statistical significance-based approach for clustering hierarchical data using semi-parametric linear mixed-effects models designed for responses with laws in the exponential family (e.g., Poisson and Bernoulli). 
Within the family of semi-parametric mixed-effects models, a latent clustering structure of the highest-level units can be identified by assuming the random effects to follow a discrete distribution with an unknown number of support points. 
We achieve this by computing $\alpha$-level confidence regions of the estimated support point and identifying statistically different clusters.
At each iteration of a tailored Expectation Maximization algorithm, the two closest estimated support points for which the confidence regions overlap collapse. 
Unlike the related state-of-the-art methods that rely on arbitrary thresholds to determine the merging of close discrete masses, the proposed approach relies on conventional statistical confidence levels, thereby avoiding the use of discretionary tuning parameters.
To demonstrate the effectiveness of our approach, we apply it to data from the Programme for International Student Assessment (PISA - OECD) to cluster countries based on the rate of innumeracy levels in schools. 
Additionally, a simulation study and comparison with classical parametric and  state-of-the-art models are provided and discussed.

\end{abstract}

\vfill

\newpage
\spacingset{1.2} 

\section{Introduction}
\label{sec:intro}

Databases featuring a hierarchical structure contain observations nested within higher-level groups in a tree-like fashion, resulting in interdependence between observations. This type of data, known as multilevel or hierarchical data, is frequently found in repeated measurements and longitudinal studies with grouping factors. Examples include student data in schools or patient data in healthcare centers, which may contain multiple levels of hierarchy, such as students within classrooms within schools.
The hierarchical nature of data requires the use of specialized models like mixed effect models (\cite{pinheiro2000linear}), which account for both random and fixed effects, modeling variability at both group and individual levels.

Classic mixed-effects models assume the random effects to follow a gaussian distribution, but, 
over the past few years, novel semi-parametric mixed-effects models in which the random effects are assumed to follow a discrete distribution have been proposed in literature for continuous (\cite{masci2019semiparametric, masci2021evaluating}) and multinomial responses (\cite{masci2022semiparametric}). The theoretical foundations of this modeling are based on the works proposed in \cite{bock1981marginal, lindsay1983geometry, lindsay1983geometryII}, while the parameters estimation procedure is based on iterative Expectation-Maximization (EM) algorithms are inspired by \cite{aitkin1999general} and \cite{azzimonti2013nonlinear}.
The advantage of this approach relies on the fact that, under the discrete distribution assumption, the random effects collapse within an \textit{a priori} unknown number of support points, identifying a latent clustering structure of the \textit{hierarchy} (groups), e.g., schools or hospitals, where the statistical units are students or patients, respectively. 
Through such an approach, each random effect would represent a cluster of groups, instead of a group itself, leading to several advantages.
Clustering the groups is a valuable dimensionality reduction tool, especially when the cardinality of the groups is huge.
For example, external institutions may want to apply a \textit{limited} number of targeted intervention policies for the performance improvements of given providers, such as schools or hospitals. By fitting a parametric mixed effects model for observations nested within providers (i.e., the groups), the random effects will provide a ranked list of the providers, visualized in a caterpillar plot, that, although widely used in performance monitoring, is associated with some important conceptual issues \citep{mohammed2008context}. 

A semi-parametric mixed effects model, on the other hand, will generate a ranking of clusters of the providers, gaining in interpretability. The identification of clusters and their sizes is also a useful tool for outliers detection, where outliers are intended as very small clusters with respect to others.
Moreover, this approach offers greater flexibility than the parametric version as it does not require the assumption of normal distribution of random effects.
On the other side, the main drawback of this approach regards the collapsing criterion. As the number of iterations of a tailored
developed Expectation Maximization algorithm grows, the support masses of the discrete distribution are made collapsed and a new optimal discrete distribution is identified.
The final number of identified clusters is not selected directly but depends on a threshold that determines the merging of discrete masses with smaller Euclidean distances along the iterations of the algorithm.
However, choosing the threshold is a drawback when this method is applied to real-world data, especially without prior knowledge of the number of clusters to be identified or the difference to be observed across clusters. Tuning the threshold requires multiple runs of the algorithm, making it computationally expensive.

Within this framework, we propose an innovative method to perform the clustering of groups standing on the conventional statistical significance levels, avoiding the use of a discretional tuning parameter for cluster distance. Our proposed approach involves the computation of confidence regions centered on the two closest support points estimated using Maximum Likelihood Estimators (MLEs) and their asymptotic properties. At each iteration of the algorithm, the confidence regions are constructed after maximizing the likelihood. If the regions overlap, the two discrete masses are merged into one. The advantage of this criterion is that it identifies the latent structure solely through a statistical significance-based approach by selecting a level of confidence $\alpha$, rather than an arbitrary and subjective threshold.
Moreover, this approach leads to even more interpretable clusters, as their differences are statistically significant.

More specifically, we address a Semi-Parametric Generalized Linear Mixed-effects Model (SPGLMM) for responses with law in the exponential family.
We recall that Generalized Linear Mixed Models (GLMMs) (\cite{breslow1993approximate}) extend upon Generalized Linear Models (GLMs) (\cite{nelder1972generalized}, \cite{mccullagh1983generalized}) by incorporating 
random effects into the linear predictor in addition to the fixed effects.
Pointedly, we utilize the statistical significance-based approach for Poisson and Bernoulli responses, but the model is readily adaptable to other responses in the exponential family.

To show an example of the proposed model utility, we provide an application with data extracted from the Programme for International Student Assessment (PISA, \cite{oecd2019pisa}) to cluster countries standing on their innumeracy levels, i.e., the levels of mathematical illiteracy, as coined in \cite{evered1990innumeracy}.
The OECD's PISA measures 15-year-olds' knowledge and skills in reading, mathematics, and science to handle real-life challenges. 
Our focus is on mathematical performance, which evaluates students' ability to apply math in various contexts. 
The global indicators for the United Nations Sustainable Development Goals identify a minimum Level of Proficiency - computed on the obtained scores - that all children should acquire by the end of secondary education: students below this level are considered \textit{low-achieving students}.
We aim at investigating the effect the countries involved in the OECD's PISA 2018 survey have on the rate of low-achieving students in mathematics. To do so, we fit a GLMM with non-parametric random effects which provides a random effect for each cluster of countries and auto-tunes the number of clusters according to a chosen level of confidence.

A simulation study is proposed to assess the performance of the SPGLMM in comparison to other existing methods. To the best of our knowledge, there are no models in literature designed to perform clustering of the hierarchies in mixed models with generalized responses.
To evaluate the validity, solidity and benefits of the SPGLMM based on statistical significance, we compare it with an SPGLMM that uses a discretionary threshold and with two parametric GLMMs.

The novelty of the paper is twofold: the development of semi-parametric mixed-effects models for generalized responses and, most importantly, the definition of a new methodological approach that allows identifying the clustering structure at the grouping level building the procedure on the solely statistical significance. This second point is the key point of the proposal and can be employed for any type of response variable.  

The paper is organized as follows: in Section \ref{sec2} we address the SPGLMM for a generalized response and we present the tailored EM algorithm for the estimation of the parameters based on a statistical-significance approach; in Section \ref{sec:case_study} we apply the SPGLMM algorithm to OECD's PISA survey data for clustering the countries standing on their school's innumeracy rates (i.e. assuming a Poisson distributed response);
in Section \ref{sec3} we present the simulation study in which we test the SPGLMM performances within different settings and compare them with the ones obtained by other \textit{state-of-the-art} methods; in Section \ref{sec:concluding_remarks} we draw our conclusions and discuss some future perspectives. 
Further details concerning the methodology, results, proofs and a parallel discussion on a Bernoulli distributed response, can be found in the Supplementary Materials.
Models implementation and results analysis are performed both through the statistical software R (\cite{Rproject}) and Python 3 (\cite{10.5555/1593511}). The code is available upon request.

\section{Methodology} 
\label{sec2}
In this section, we will cover the basics of a GLMM and its extension to non-parametric random effects (Section \ref{sec:21}), describe the EM algorithm for parameters estimation (Section \ref{sec:22}) and present the key method for reducing random effects support (Section \ref{sec:24}).

\subsection{GLMMs with nonparametric random effects}
\label{sec:21}

Our methodology focuses on the case of hierarchical data with nested observations and a single level of grouping, with $N$ groups indexed by $i=1,...,N$, each containing $n_i$ observations indexed by $j=1,...,n_i$, with $\sum_{i=1}^N n_i = J$.  
The vector of responses within the $i^{\text{th}}$ group, $\mathbf{y}_i$, contains (conditionally) independent observations $y_{ij}$ for $j=1,...,n_i$.
The conditional distribution of $\mathbf{y}_i$ given the random effects $\mathbf{b}_i$ belongs to the exponential family with probability mass (or density) function $p(\mathbf{y}_i|\boldsymbol{\beta}, \mathbf{b}_i)$, being $\boldsymbol{\beta}$ the vector of fixed coefficients.
GLMMs are defined such that the expectation of $\mathbf{y}_i$ conditioned on $\mathbf{b}_i$ in the $i^{\text{th}}$ group is related to the linear predictor $\mathbf{X}_i \boldsymbol{\beta} + \mathbf{Z}_i \mathbf{b}_i$ via the monotonic and differentiable \textit{link function} $g(\cdot)$:
\begin{equation}
    g(\, \mathbb{E}[\mathbf{y}_i|\mathbf{b}_i] \,) = g(\boldsymbol{\mu}_i) = \boldsymbol{\eta}_i = \mathbf{X}_i \boldsymbol{\beta} +  \mathbf{Z}_i \mathbf{b}_i \quad \text{for} \quad i=1,...,N
    \label{GLMM1}
\end{equation} 
where $\mathbf{X}_i$ and $\mathbf{Z}_i$ are, respectively, the $n_i \times P$ and $n_i \times Q$ matrices of fixed and random covariates\footnote{In many cases, the matrix $\mathbf{Z}_i$ is created by selecting a subset of appropriate columns of the matrix $\mathbf{X}_i$, i.e. the corresponding fixed and random effects are \textit{coupled} (\cite{galecki2013linear}).} in the $i^{\text{th}}$ group;
$\boldsymbol{\beta}$ is the $P$-dimensional vector of fixed coefficients and $\mathbf{b}_i$ the $Q$-dimensional\footnote{In the theoretical discussion, we address the general case of $Q \in \mathbb{N}\setminus \{ 0 \}$; however, in the simulation study in Section \ref{sec3}, we restrict ourselves to the case $Q\leq2$, for which the model is composed by either a random intercept only, a random slope only (cases $Q=1$) or both ($Q=2$).} vector of random coefficients relative to the $i^{\text{th}}$ group. 
In the parametric framework, the random coefficients  are assumed to be normally distributed, i.e., $\mathbf{b}_i \sim \mathcal{N}_{Q}(\mathbf{0}, \mathbf{\Sigma}_{\mathbf{b}_i})$, $i=1,...,N$.
For the estimation of the model parameters within the frequentist approach, likelihood-based approaches are used. 

Following the approach presented in 
\cite{masci2019semiparametric} and \cite{masci2022semiparametric}, we move to a nonparametric framework, assuming the random effects $\mathbf{b}_1,...,\mathbf{b}_N$ to follow a discrete distribution $\mathcal{P}$ composed by an \textit{a priori} unknown number of support points.
The proposed algorithm starts by assuming a number of discrete masses equal to the number of groups, $N$. It then iteratively reduces the number by combining groups into $M<N$ \textit{clusters}. This process allows for identification of a latent structure in which groups within the same cluster exhibit a certain degree of similarity. The number of discrete masses, $M$, is determined by the algorithm in conjunction with the estimation of other model parameters.
In this nonparametric framework, we define the \textit{latent} variables $\mathbf{c}_1,...,\mathbf{c}_M$ as the set of random coefficients where each $\mathbf{c}_m\in\mathbb{R}^{Q}$ $m=1,...,M$ corresponds to the random coefficient of the $m^{\text{th}}$ cluster. 
These latent variables are related to each previously defined random effect $\mathbf{b}_i$ through the relationship $p(\mathbf{b}_i = \mathbf{c}_m ) = \omega_{m}$ for $m=1,...,M$
where $\omega_1,...,\omega_M$ is a set of weights such that $\sum_{m=1}^M \omega_{m} = 1$ and $\omega_{m} \geq 0 $.
The $i^{\text{th}}$ group  
is assigned with probability $\omega_{m}$ to a cluster $m$ 
with parameter values $\mathbf{c}_m$ allowing the identification of a latent structure among the groups.
Consequently, starting from the GLMM formulation in Eq. (\ref{GLMM1}), we make the dependence on $m$ (and $j$) explicit and we get our SPGLMM formulation:
\begin{equation}
\label{model_form}
    g(\, \mathbb{E}[y_{ij}|\mathbf{c}_m] \,) = \eta_{ijm} = \mathbf{x}'_{ij} \boldsymbol{\beta} +  \mathbf{z}'_{ij} \mathbf{c}_{m} \quad \text{for} \quad i=1,...,N, \; j=1,...,n_i, \; m=1,...,M
\end{equation} 
where $\mathbf{x}_{ij}$ is he $P$-dimensional vector and  $\mathbf{z}_{ij}$ the $Q$-dimensional vector of covariates relative to the $({i,j})^{\text{th}}$ observation.

The marginal likelihood $\mathcal{L}(\boldsymbol{\beta}, \mathbf{b}_1,...,\mathbf{b}_N | \mathbf{y}) =  \prod_{i=1}^N \prod_{j=1}^{n_i} \; p(y_{ij}|\boldsymbol{\beta},\mathbf{b}_i) $, where $p(y_{ij}|\boldsymbol{\beta},\mathbf{b}_i)$ denotes the conditional probability mass (or density) function of $y_{ij}$ given random and fixed effects.
Including the latent variables  with the corresponding contribution of the weights of the mixture (\cite{aitkin1999general}), the loglikelihood $\mathrm{ln}\,\mathcal{L}$ can be expressed as
\begin{equation}
\label{L_c}
    \mathrm{ln}\,\mathcal{L}(\boldsymbol{\beta}, \mathbf{c}_1,...,\mathbf{c}_M | \mathbf{y}) =  \sum_{m=1}^M \omega_m \sum_{i=1}^N \sum_{j=1}^{n_i} \mathrm{ln}\;p(y_{ij}|\boldsymbol{\beta},\mathbf{c}_m).
\end{equation}
By maximizing the quantity in Eq. (\ref{L_c}) we jointly estimate the values of $\boldsymbol{\beta}$ , $(\mathbf{c}_1,..., \mathbf{c}_M)$ and $(\omega_1,..., \omega_M)$. We develop a tailored EM algorithm (\cite{dempster1977maximum}), 
that we discuss in Subsection \ref{sec:22}.
In Appendix \ref{app0}, we express the loglikelihood in Eq. (\ref{L_c}) for the two special cases of Bernoulli and Poisson distributions.

\subsection{EM algorithm for SPGLMM}
\label{sec:22}

Inspired by \cite{aitkin1999general} and \cite{azzimonti2013nonlinear}, we implement an EM algorithm
to obtain the pointwise estimates
$\hat{\boldsymbol{\beta}}$, $(\hat{\mathbf{c}}_1,...,\hat{\mathbf{c}}_M)$ and $(\hat{\omega}_1,...,\hat{\omega}_M)$ of the unknown parameters through the evaluation and the maximization of the (log) likelihood.
Specifically, the EM algorithm is an iterative procedure that alternates between two steps: the \textit{expectation step}, in which the conditional expectation of the log-likelihood ($\mathrm{ln}\,\mathcal{L}$) is computed with respect to the random effects, given the parameters and observations obtained from the previous iteration; and the \textit{maximization step}, in which the conditional expectation of $\mathrm{ln}\,\mathcal{L}$ is numerically maximized. The algorithm terminates when either convergence is achieved or a maximum number of iterations is reached.
The parameters updates are given by:
\begin{equation}
    \hat{\omega}_m^{(up)} = \frac{\sum_{i=1}^N \hat{W}_{im}}{N} \quad \text{for} \quad m=1,...,M
    \label{w_up}
\end{equation}
where
\begin{eqnarray}
\label{W_im}
 \hat{W}_{im}  & = & \frac{\hat{\omega}_m \; p(\mathbf{y}_i|\hat{\boldsymbol{\beta}},\hat{\mathbf{c}}_m)}{\sum_{k=1}^M \hat{\omega}_k \; p(\mathbf{y}_i|\hat{\boldsymbol{\beta}}, \hat{\mathbf{c}}_k)} =   \frac{p(\mathbf{b}_i = \hat{\mathbf{c}}_m) \; p(\mathbf{y}_i|\hat{\boldsymbol{\beta}}, \hat{\mathbf{c}}_m)}{ p(\mathbf{y}_i|\hat{\boldsymbol{\beta}})}  =  \frac{p(\mathbf{y}_i, \mathbf{b}_i =
 \hat{\mathbf{c}}_m |\hat{\boldsymbol{\beta}} )}{ p(\mathbf{y}_i|\hat{\boldsymbol{\beta}})} \\  
                     & = &  p(\mathbf{b}_i=\hat{\mathbf{c}}_m | \mathbf{y}_i, \hat{\boldsymbol{\beta}}) \quad \text{for} \quad m=1,...,M, \; i=1,...N \nonumber 
\end{eqnarray}
and 
\begin{equation}
\label{maximization}
     (\hat{\boldsymbol{\beta}}^{(up)}, \hat{\mathbf{c}}_1^{(up)},..., \hat{\mathbf{c}}_M^{(up)}) = \underset{\boldsymbol{\beta}, \mathbf{c}_m}{\mathrm{arg\;max}} \sum_{m=1}^M \sum_{i=1}^N \hat{W}_{im} \;\;\mathrm{ ln}\; p(\mathbf{y}_i|\boldsymbol{\beta}, \mathbf{c}_m) \; .
\end{equation}
The proof of the increasing likelihood property and the derivation of the updates in Eqs. (\ref{w_up}) and (\ref{maximization}) can be found in Section S1 of Supplementary Materials.
The weight $\hat{\omega}_m^{(up)}$ in Eq. (\ref{w_up})  corresponds to the sample mean over the $N$ groups of all the weights relative to the $m^{\text{th}}$ cluster. $\hat{W}_{im}$ represents the probability that group $i$ belongs to cluster $m$, conditionally on observations $\mathbf{y}_i$ and fixed coefficients $ \hat{\boldsymbol{\beta}}$.
The maximization in Eq. (\ref{maximization}) involves two different steps, performed iteratively: in the first step, we compute $\hat{\mathbf{c}}_m^{(up)}$ maximizing with respect to the support points of the random coefficients $\mathbf{c} = (\mathbf{c}_1,...,\mathbf{c}_M)$, setting $\hat{\boldsymbol{\beta}}$ equal to the values computed at the previous iteration, namely
\begin{equation}
\label{update_cm}
        \hat{\mathbf{c}}_m^{(up)} = \underset{\mathbf{c}}{\mathrm{arg\;max}} \sum_{i=1}^N \hat{W}_{im} \;\;\mathrm{ ln}\; p(\mathbf{y}_i|\hat{\boldsymbol{\beta}}, \mathbf{c}) \quad \text{for} \quad m=1,..., M \; .
\end{equation}
In the second step, we fix the support points of the random coefficients computed in the previous step and we compute the ${\mathrm{arg\;max}}$ of Eq. (\ref{maximization}) with respect to $\boldsymbol{\beta}$, namely:
\begin{equation}
\label{update_beta}
    \hat{\boldsymbol{\beta}}^{(up)} = \underset{\boldsymbol{\beta}}{\mathrm{arg\;max}} \sum_{m=1}^M \sum_{i=1}^N  \hat{W}_{im} \;\;\mathrm{ ln}\; p(\mathbf{y}_i|\boldsymbol{\beta}, \hat{\mathbf{c}}_m) \; .
\end{equation}
In order to compute the point estimate $\hat{\mathbf{b}}_{i}$ of the coefficients  $\mathbf{b}_i$ of the random effects for each group $i=1,...,N$, we maximize over $m$ the conditional probability $p(\mathbf{b}_i=\hat{\mathbf{c}}_m | \mathbf{y}_i, \hat{\boldsymbol{\beta}})$. For Eq. (\ref{W_im}), the estimation of $\hat{\mathbf{b}}_{i}$ is given by the maximization of 
$\hat{W}_{im}$ over $m$, namely:
\begin{equation}
\label{l_tilde}
    \hat{\mathbf{b}}_{i} = \hat{\mathbf{c}}_{\tilde{l}_i} \quad \text{where} \quad \tilde{l}_i =  \underset{m}{\mathrm{arg\;max}} \; \hat{W}_{im}  \quad \text{for} \quad i=1,..., N \; .
\end{equation} 
All details concerning parameters initialization procedure are addressed in Section S2.1 of Supplementary Materials.

\subsection{Support points reduction criterion}
\label{sec:24}

In each of the $k$ iterations of the algorithm, we aim to identify the latent structure composed of $M < N$ clusters by 
reducing the support of the random effects discrete distribution by making points \textit{very close} to each other collapse.
The notion of \textit{very close} needs to be defined.
In the state-of-the-art papers dealing with nonparametric random effects, at each iteration until convergence, the discrete masses with Euclidean distance lower than a chosen threshold (denoted by $t$\footnote{In the following, we will refer to this method as \textit{t-criterion}. Such criterion is deepened 
and discussed in Section \ref{sec3}.}), are made collapsed.
In this work, on the other hand, we suggest identifying the latent cluster structure by only means of the \textit{conventional} confidence levels, gaining in interpretability within the classical framework of the inferential statistics and untying from the choice of a \textit{discretionary} threshold.
More specifically, we propose (i) to compute the confidence regions (intervals) of level 1-$\alpha$ centered in each of the two closest - in terms of Euclidean distance - estimated support points, 
exploiting the properties of the MLEs (Section \ref{sec:241}) and (ii) to collapse the two discrete masses to a unique point, if the two confidence regions (intervals) overlap (Section \ref{sec:242}).




\subsubsection{The computation of the confidence regions (intervals) for a MLE} 
\label{sec:241}
Let $\hat{\boldsymbol{\theta}}$ be a MLE and $\boldsymbol{\theta}_0$ the true value.
The Hessian matrix $H$ of the loglikelihood function 
is defined as $ [H(\boldsymbol{\theta})]_{ij} = [\nabla ^2 \; \mathrm{ln}\;\mathcal{L}(\boldsymbol{\theta})]_{ij} = \dfrac{\partial^2 \; \mathrm{ln}\;\mathcal{L}(\boldsymbol{\theta})}{\partial \theta_i \; \partial \theta_j} $.
The Fisher Information Matrix $\mathcal{I}(\hat{\boldsymbol{\theta}})$ is defined as $ \mathcal{I}(\hat{\boldsymbol{\theta}}) = -\mathbb{E}[H(\boldsymbol{\theta})|\hat{\boldsymbol{\theta}}] $
and the variance-covariance matrix (\cite{king1998unifying}, \cite{long2006regression}) is $\text{var}(\hat{\boldsymbol{\theta}}) = \mathcal{I}^{-1}(\hat{\boldsymbol{\theta}})$.
Given the asymptotic efficiency property of the MLEs (\cite{casella2021statistical}), 
we know that MLEs are asymptotically normal, i.e., $\sqrt{J}(\hat{\boldsymbol{\theta}}_J-\boldsymbol{\theta}_0) \xrightarrow[J \rightarrow \infty]{d} N(\mathbf{0}, \mathcal{I}^{-1}(\boldsymbol{\theta}_0))$, where $\xrightarrow[]{d}$ denotes the convergence in distribution. 

Let $\hat{\boldsymbol{\theta}}^{(k)}$ now be the MLE computed at each iteration $k$ of our iterative algorithm.
We deduce that, when $\hat{\theta}^{(k)}$ is 1-dimensional, the asymptotic confidence region is an interval $CI$ of level $1-\alpha$ for $\hat\theta^{(k)}$ given by
    $CI_{1-\alpha}(\hat\theta^{(k)}) = \left[\hat\theta^{(k)} \pm z_{1-\frac{\alpha}{2}} \; \frac{1}{\sqrt{\text{var}(\hat\theta^{(k)})}}\right] $.
Instead, when $\hat{\boldsymbol\theta}^{(k)}$ is $Q$-dimensional with $Q>1$ and the symmetric $\text{var}(\hat{\boldsymbol \theta}^{(k)})$ is positive definite\footnote{If not, the same formula holds by replacing $[\text{var}(\hat{\boldsymbol \theta}^{(k)})]^{-1}$ with the generalized inverse and by substituting the degrees of freedom of the $\chi^2_{1-\alpha}$ from $Q$ to the rank of $\text{var}(\hat{\boldsymbol \theta}^{(k)})$.}, we get a confidence region with an ellipsoidal shape defined by $CR_{1-\alpha}(\hat{\boldsymbol{\theta}}^{(k)}) = \{\boldsymbol{\theta} \; : \; (\boldsymbol{\theta}-\hat{\boldsymbol{\theta}}^{(k)})'\; [\text{var}(\hat{\boldsymbol \theta}^{(k)})]^{-1} \; (\boldsymbol{\theta}-\hat{\boldsymbol{\theta}}^{(k)}) \leq \chi_{1-\alpha}^2(Q)\}$ (\cite{johnson2002applied}).

\subsubsection{\texorpdfstring{$\alpha$}{TEXT}-criterion}
\label{sec:242}
At each iteration $k$ of the SPGLMM algorithm, the MLE $\hat{\mathbf{c}}_m^{(k)}$ for $m=1,...,M$ is estimated as shown in Eq. (\ref{update_cm}).
The elements $D_{l,m}^{(k)}$ of the matrix $\mathbf{D}^{(k)}$, composed by the Euclidean distances between the two MLEs $\hat{\mathbf{c}}_l^{(k)}$ and $\hat{\mathbf{c}}_m^{(k)} \; \forall l,m=1,...,M$, are computed as follows
   $ D_{l,m}^{(k)} = \sqrt{\sum_{h=1}^Q(\hat{c}_{lh}^{(k)}-\hat{c}_{mh}^{(k)})^2} \quad \forall l,m=1,...,M $.
The two mass points $\hat{\mathbf{c}}_l^{(k)}$ and $\hat{\mathbf{c}}_m^{(k)}$ with minimum Euclidean distance are selected and the confidence regions (intervals) of level $1-\alpha$ centered in those mass points are computed as explained in Section \ref{sec:241}.
Thus, we check whether $CR_{1-\alpha}(\hat{\mathbf{c}}_l^{(k)})$ overlaps $CR_{1-\alpha}(\hat{\mathbf{c}}_m^{(k)})$ through the following \textit{overlapping condition}, addressed separately for the unidimensional and multidimensional cases.
In the unidimensional case, the two confidence intervals do overlap if the following inequality is satisfied:
$\text{max}\{ \text{min}\{CI_{1-\alpha}(\hat{c}_l^{(k)})\} ,  \text{min}\{CI_{1-\alpha}(\hat{c}_m^{(k)})\} \} < 
\text{min}\{  \text{max}\{CI_{1-\alpha}(\hat{c}_l^{(k)})\} , \text{max}\{CI_{1-\alpha}(\hat{c}_m^{(k)})\} \} $.
In the $Q$-dimensional case, when $Q=2$, we determine if one ellipse is entirely contained within the other. This can occur when among the two closest MLEs $\hat{\mathbf{c}}_l^{(k)}$ and $\hat{\mathbf{c}}_m^{(k)}$, one exhibits higher values for the Information Matrix and thus has a larger confidence region.
To assess this in our code, we use a \textit{sufficient} condition that checks if the
Euclidean distance between the centers of the two ellipses is smaller than the difference between the semi-minor axis length $(\chi_{1-\alpha}^2(2) \cdot \lambda_{min})^{1/2}$ of the larger ellipse and the semi-major axis length $(\chi_{1-\alpha}^2(2) \cdot \lambda_{max})^{1/2} $
of the smaller ellipse. 
If the sufficient condition is not met, indicating that one ellipse is not entirely inside the other, we use the Fast Ellipsoid Intersection Test\footnote{\url{https://github.com/NickAlger/nalger_helper_functions/blob/master/tutorial_notebooks/ellipsoid_intersection_test_tutorial.ipynb}} by performing a unidimensional minimization. A detailed description of the Test is addressed in Section S3 of Supplementary Materials. Such a method works for all the $Q$-dimensional cases in which $Q>1$. 

If $CR_{1-\alpha}(\hat{\mathbf{c}}_l^{(k)})$ overlaps $CR_{1-\alpha}(\hat{\mathbf{c}}_m^{(k)})$, the two points $\hat{\mathbf{c}}_l^{(k)}$ and $\hat{\mathbf{c}}_k^{(k)}$ collapse to a unique point, result of the \textit{weighted}\footnote{In the earlier proposed literature, 
Eq. (\ref{update_masses}) was a classical (non-weighted) mean for the $t$-criterion.
In our methodology, we propose a weighted mean for the $alpha$-criterion. 
This enables us to progressively approach the desired mass point as we iterate towards convergence.
} mean among $\hat{\mathbf{c}}_l^{(k)}$ and $\hat{\mathbf{c}}_m^{(k)}$: 
\begin{equation}
    \label{update_masses}
\hat{\mathbf{c}}_{l,m}^{(k)} = \frac{\hat{\omega}_l^{(k)} \hat{\mathbf{c}}_l^{(k)} + \hat{\omega}_m^{(k)} \hat{\mathbf{c}}_m^{(k)}}{\hat{\omega}_l^{(k)} + \hat{\omega}_m^{(k)}}  
\end{equation}  
and the weight is updated with the sum of the weights of the two points:
\begin{equation}
    \label{update_weights}
    \hat{\omega}_{l,m}^{(k)} = \hat{\omega}_l^{(k)} + \hat{\omega}_m^{(k)} \; .
\end{equation}

\begin{figure}
\centering
\caption{Support masses collapse procedure through $\alpha$-\textit{criterion} for $Q=1$.}
\includegraphics[width=16cm]{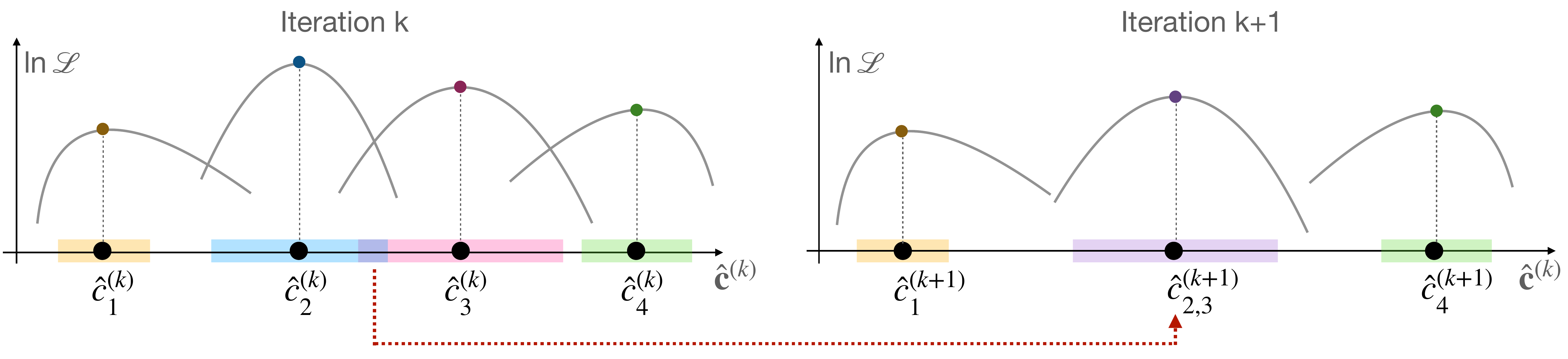} \\
\footnotesize
Notes: In the left-side chart, 
the confidence intervals centered in four different support points (black dots), 
estimated at a given iteration $k$, are displayed 
on the horizontal axis. The 
confidence intervals centered in $\hat{c}_2^{(k)}$ and $\hat{c}_3^{(k)}$
meet the \textit{overlapping condition}, thus $\hat{c}_2^{(k)}$ and $\hat{c}_3^{(k)}$ are merged.
The right-side chart shows the scenario in iteration $k+1$, where the support points have been reduced to three and the new confidence intervals have been recomputed.
\label{fig:MLE_1dim}
\end{figure}

\begin{figure}
\centering
\caption{Support masses collapse procedure through $\alpha$-\textit{criterion} for $Q=2$.}
\includegraphics[width=16cm]{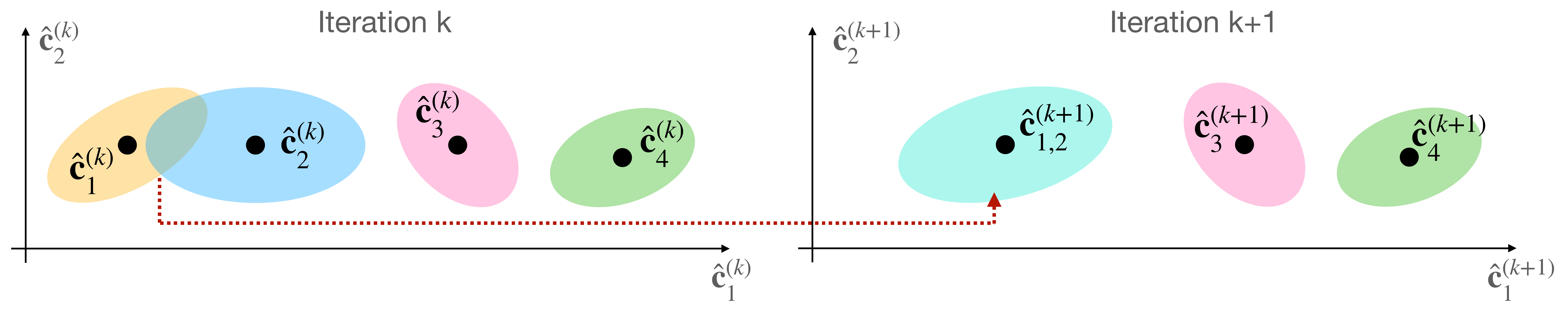} \\
\footnotesize
Notes: In the left-side chart, 
the confidence regions centered in four different support points (black dots), 
estimated at a given iteration $k$, are displayed 
in the plan. The ellipses centered in $\hat{\mathbf{c}}_1^{(k)}$ and $\hat{\mathbf{c}}_2^{(k)}$ do intersect, hence $\hat{\mathbf{c}}_1^{(k)}$ and $\hat{\mathbf{c}}_2^{(k)}$ are collapsed.
The right-side chart shows the situation in iteration $k+1$, where the support points have been reduced to three and the new confidence regions have been recomputed.
\label{fig:ellipses}
\end{figure}

In this way, at each iteration $k$ we compute $\hat{\mathbf{c}}_{l,m}^{(k)}$ and $\hat{\omega}_{l,m}^{(k)}$, which will be used in iteration $k+1$ as new mass and weight.
Instead, if the two confidence regions centered on the two closest mass points do not overlap, the \textit{overlapping condition} is then checked for all other pairs of mass points (ordered by increasing Euclidean distance) until either two confidence regions overlap or all pairs have been checked.
A graphical representation of the masses collapse procedure for $\alpha$-\textit{criterion} is reported in Figures \ref{fig:MLE_1dim} and \ref{fig:ellipses} for $Q=1$ and $Q=2$, respectively.

We report the pseudo-code for the SPGLMM with $\alpha$\textit{-criterion} in Algorithm \ref{pseudo_code}.
For easier comparison with the state-of-the-art collapsing criterion, also the \textit{t-criterion} is reported and the main differences between the two criteria are highlighted.

\begin{algorithm}
\caption{SPGLMM}
\label{pseudo_code}
\footnotesize{
  \Function{SPGLMM(Initial estimates for $(\hat{\mathbf{c}}_1^{(0)},..., \hat{\mathbf{c}}_{M=N}^{(0)})$, $(\hat{\omega}_1^{(0)},..., \hat{\omega}_{M=N}^{(0)})$ and $\hat{\boldsymbol \beta}^{(0)}$; Tolerance parameters \hlc[green]{$t$ (only for $t$\textit{-criterion})}, \hlc[yellow]{$\alpha$ (only for $\alpha$\textit{-criterion})}, 
$\text{\texttt{K}}$, $\text{\texttt{K1}}$, \hlc[yellow]{$\text{\texttt{K2}}$ (only for $\alpha$\textit{-criterion})}, $\text{\texttt{itmax}}$, $\text{\texttt{tR}}$, $\text{\texttt{tF}}$
)}{
    
    $k \gets 1$; $conv1 \gets 0$; $conv2 \gets 0$ \; 
\While{($conv1$ is 0 or $conv2$ is 0) and $k<\text{\texttt{K}}$}{ 
  \boxit{green}{7.2}
  $\mathbf{D}^{(k)} \gets$ compute distance matrix $\mathbf{D}$ as in Section \ref{sec:242} \Comment{\textbf{only for $t$\textit{-criterion}}}\\
  \While {$\sum_{l,m}(D_{l,m}^{(k)}<t) \neq M^2$ and $\sum_{l,m}(D_{l,m}^{(k)}<t) > M$}{
  Select $\hat{\mathbf{c}}_l^{(k)}$ and $\hat{\mathbf{c}}_m^{(k)}$ for which $D_{l,m}^{(k)}$ is minimum\;
  Collapse $\hat{\mathbf{c}}_l^{(k)}$ and $\hat{\mathbf{c}}_m^{(k)}$ to unique mass point and update as in Eq. (\ref{update_masses})\;
  Update weights as in Eq. (\ref{update_weights})\;
  $\mathbf{D}^{(k)} \gets$ Compute distance matrix $\mathbf{D}$ as in Section \ref{sec:242}\;
  $M \gets M-1$\
  }
  
  $\hat{\mathbf{W}}^{(k)} \gets$ compute $\hat{\mathbf{W}}$ as in Eq. (\ref{W_im})\;
  
 $M$, $(\hat{\mathbf{c}}_1^{(k)},.., \hat{\mathbf{c}}_M^{(k)})$, $ (\hat{\omega}_1,^{(k)}.., \hat{\omega}_M^{(k)})$, $conv2$, $conv1 \gets $
 \textbf{\texttt{Check weights 
 }} \Comment{See Suppl. Mat. S2.2}\;

 $\hat{\mathbf{W}}^{(k)} \gets$ update $\hat{\mathbf{W}}$ as in Eq. (\ref{W_im})\;
 $it \gets 1$ ; $\hat{\mathbf{c}}^{(it-1)} \gets \hat{\mathbf{c}}^{(k)}$;
 $\hat{\boldsymbol{\beta}}^{(it-1)} \gets \hat{\boldsymbol{\beta}}^{(k)}$\;
 $\hat{\mathbf{c}}^{(it)} \gets$ Update the M support points according to Eq. (\ref{update_cm}), keeping $\hat{\boldsymbol\beta}$ fixed\; 
 $\hat{\boldsymbol \beta}^{(it)} \gets$ Update $\hat{\boldsymbol\beta}$ according to Eq. (\ref{update_beta}), keeping $\hat{\mathbf{c}}$ fixed\;
 \While{$\sum(|\hat{\boldsymbol \beta}^{(it)}-\hat{\boldsymbol \beta}^{(it-1)}|>\text{\texttt{tF}})$ is 0 and $\sum(|\hat{\mathbf{c}}^{(it)}-\hat{\mathbf{c}}^{(it-1)}|>\text{\texttt{tR}})$ is 0 and $it<\text{\texttt{itmax}}$}{
    $it \gets it+1$\;
    $\hat{\mathbf{c}}^{(it)} \gets$ Update the M support points according to Eq. (\ref{update_cm}), keeping $\hat{\boldsymbol\beta}$ fixed\; 
 $\hat{\boldsymbol \beta}^{(it)} \gets$ Update $\hat{\boldsymbol\beta}$ according to Eq. (\ref{update_beta}), keeping $\hat{\mathbf{c}}$ fixed\;
 }
  $\hat{\mathbf{c}}^{(k)} \gets \hat{\mathbf{c}}^{(it)}$;
  $\hat{\boldsymbol \beta}^{(k)} \gets \hat{\boldsymbol \beta}^{(it)}$\;
  
\boxit{yellow}{13.3} 
\If (\Comment{\textbf{only for $\alpha$\textit{-criterion}}}){$k>\text{\texttt{K2}}$}{
$\text{not\_merged} \gets 1$\;
$\mathbf{D}^{(k)} \gets$ compute distance matrix $\mathbf{D}$ as in Section \ref{sec:242}\;
\While{\text{not\_merged} is 1 and sum(\text{NaN} values in $\mathbf{D}$)<dim($\mathbf{D}$)}{

Select $\hat{\mathbf{c}}_l^{(k)}$ and $\hat{\mathbf{c}}_m^{(k)}$ for which $D_{l,m}^{(k)}$ is minimum\;
\If{\textit{overlapping condition} is satisfied
}{
Collapse $\hat{\mathbf{c}}_l^{(k)}$ and $\hat{\mathbf{c}}_m^{(k)}$ to unique mass point and update as in Eq. (\ref{update_masses})\;
Update weights as in Eq. (\ref{update_weights})\;
$M \gets M-1$\;
$\text{not\_merged} \gets 0$\;
    }
   \Else{
    Set $D_{l,m}^{(k)}=\text{NaN}$\;
    }
    }
  }

  \If{$\sum(|\hat{\boldsymbol \beta}^{(k)}-\hat{\boldsymbol \beta}^{(k-1)}|>\text{\texttt{tF}})$ is 0 and $\sum(|\hat{\mathbf{c}}^{(k)}-\hat{\mathbf{c}}^{(k-1)}|>\text{\texttt{tR}})$ is 0}{
    \boxitt{yellow}{0.2} 
    \If(\Comment{\textbf{only for $\alpha$\textit{-criterion}}}){sum(\text{NaN} values in $\mathbf{D}$)==dim($\mathbf{D}$)}{
      $conv1 \gets 1$\;
    }
 }
 $k \gets k+1$\;
}
    \Return{Final estimates of $(\hat{\mathbf{c}}_1^{(k)},..., \hat{\mathbf{c}}_M^{(k)})$, $(\hat{\omega}_1^{(k)},..., \hat{\omega}_M^{(k)}) $ and $\hat{\boldsymbol \beta}^{(k)}$; $\hat{\mathbf{W}}^{(k)}$}
    }
}
\end{algorithm}

Final checks concerning the support reduction and details regarding convergence criteria are addressed in Sections S2.2 and S2.3 of Supplementary materials.

\section{Case study: application to innumeracy rates}
\label{sec:case_study}

In this section, we apply the SPGLMM to data extracted from PISA survey of 2018, available online at \href{https://www.oecd.org/pisa/data/2018database/}{https://www.oecd.org/pisa/data/2018database/}.
Students' scores in mathematics tests are divided into Levels of Proficiency, as described in Chapter 6 of \cite{oecd2019pisa}, 
with Level 2 being the minimum required by global indicators for the United Nations Sustainable Development Goals to be acquired by the end of secondary education.
Level 2 proficiency only provides a basic understanding of math for simple real-life situations and does not prepare students for decision-making requiring mathematical literacy. For this reason, students below such a level of proficiency are considered as \textit{low-achieving students}.
We develop a model to predict the percentage number of low-achieving students in each school and country, taking into account school characteristics like size and socio-economic status.
Additionally, we aim to identify groups of countries that have a similar impact on the rate of low-achievers by using a SPGLMM with Poisson response.

The survey provides, among others, data both at the \textit{student} and at the \textit{school level}. Table \ref{Descriptive_variables_initial} reports the selected variables extracted from OECD PISA dataset, together with their description. 
Low-achieving students are identified by \texttt{PV1MATH} scores below 482.38, i.e., the ones with Proficiency levels strictly less than Level 3 (\cite{oecd2019pisa}).

In the following sections, we will address the data preprocessing (Section \ref{subsec:data_preproc}), the model formulation (Section \ref{subsec:model_form}) and the results obtained by fitting the model with Poisson response (Section \ref{poisson_case_study}).
Parallel handling for the Bernoulli response is addressed in Section S4 of Supplementary materials.

\begin{table}
\caption{List, description and summary statistics of variables extracted from OECD PISA dataset.}
\label{Descriptive_variables_initial}
\renewcommand{\arraystretch}{1.5} 
\footnotesize
\begin{tabularx}{\textwidth}{@{}lXlX@{}}
\toprule
Variable 
& Description
& Type 
& Summary statistics\\
\midrule
\texttt{ESCS}    &Index of economic, social and cultural status [\textit{student level}] & Continuous & mean = -0.29, sd = 1.11, median = -0.18, [min; max] = [-8.17; 4.21]. 14379 NaNs (2.35\%)\\
\texttt{PV1MATH} &Score$^{**}$ in mathematics [\textit{student level}] &Continuous & mean = 461.88, sd = 104.49, median = 461.39, [min; max] = [24.74; 888.06]\\
\texttt{SCHSIZE} &School size (sum) [\textit{school level}]&Continuous &mean = 839.65, sd = 869.61, median = 624, [min; max] = [1; 13400]. 3582 NaNs (16.35\%)\\
\texttt{CNTSCHID}&International school id [\textit{school level}, \textit{student level}] &Categorical & 21903-levels factor\\
\texttt{CNT}     &Country code 3-character [\textit{school level}, \textit{student level}] &Categorical & 82-levels factor\\
\bottomrule
\end{tabularx}
$^{**}$More precisely, we considered Plausible Value 1 (\cite{oecd2019pisa}); Plausible Values are a selection of likely proficiencies for students' attained scores, i.e., multiple imputations of the unobservable latent achievement for each student. 
\end{table}

\subsection{Data preprocessing}
\label{subsec:data_preproc}
After having discarded missing values, continuous variables at the \textit{student level} are aggregated at the \textit{school level}: specifically, for each school we consider (i) \texttt{avg\_ESCS\_std}, the average students' \texttt{ESCS}, subsequent to a standardization (mean 0 and standard deviation 1) within the country of the school, for keeping into account differences between countries and (ii) 
\texttt{Y\_MATH}, the rounded percentage of students with a proficiency level strictly less than Level 3 (low-achieving students).
The analysis is restricted to schools with a minimum of 10 students to ensure more accurate results. A dataset at the school level containing information on 12620 schools (the variable \texttt{CNTSCHID} becomes a 12620-levels factor) nested within 50 countries (\texttt{CNT} becomes a 50-levels factor) is created.
Moreover, the two predictors \texttt{avg\_ESCS\_std} and \texttt{SCHSIZE} are further standardized.



\subsection{Model formulation}
\label{subsec:model_form}
We consider a two-level SPGLMM, as in Eq. (\ref{model_form}), and we employ the $\alpha$-\textit{criterion}. For each country $i$, with $i=1,...N$, and each school $j$, with $j = 1,..., n_i$, given that $N=50$ is the total number of countries and $J = 12620$ the total number of schools, the model is
\begin{equation}
\label{model_form_case_study}
g(\, \mathbb{E}[y_{ij}|c_{m}] \,) = \eta_{ijm} = \mathbf{x}'_{ij} \boldsymbol{\beta} 
+ c_{m} \quad \text{for} \quad i=1,...,N, \; j=1,...,n_i, \; m=1,...,M_\alpha
\end{equation}
where $M_\alpha$ is the total number of clusters the model identifies and depends on the level of confidence $\alpha$ chosen for the $\alpha$-\textit{criterion}; 
$\mathbf{x}'_{ij}$ is the two-dimensional vector of fixed effects covariates at the school level that contains \texttt{SCHSIZE}$_{ij}$ and \texttt{avg\_ESCS\_std}$_{ij}$;
$\boldsymbol{\beta} = [\beta_1, \beta_2]'$ is the two-dimensional vector of fixed effects coefficients;
$c_{m}$ is the random intercept relative to the $m^{\text{th}}$ cluster.
The response $\mathbf{y}_{i}$ is given by \texttt{Y\_MATH}$_{i}$ and is assumed to be Poisson distributed. To validate this assumption, a Chi-Square goodness of fit test was conducted and the distribution was visually inspected using Q-Q plots (\cite{wilk1968probability}) for Poisson distribution.
The link function is assumed to be the canonical $g(\cdot ) = \text{ln}(\cdot)$ (see Appendix \ref{app0Poi}).

We run the SPGLMM algorithm with $\text{\texttt{K}}=60$, $\text{\texttt{K1}}=20$, $\text{\texttt{K2}}=5$, $\text{\texttt{itmax}} = 20$, $\text{\texttt{tR}} = \text{\texttt{tF}} = 10^{-5}$ 
and, in turn, $\alpha = 0.01, 0.05 \text{ and } 0.10$.
The algorithm starts with $M=N=50$ support points and the support weights are uniformly initialized on the $M$ support points, as explained in Section S2.1 of Supplementary Materials.
To better assess the validity and the robustness of the obtained results, we report them together with the estimates of the parametric GLMM, fitted through \textit{glmer} R function from package \textit{lme4} (\cite{lme4}, \cite{Rproject}), which assumes the following formulation:
\begin{equation}
\label{GLMM_casestudy}
g(\, \mathbb{E}[y_{ij}|b_{i}] \,) = \eta_{ij} = \mathbf{x}'_{ij} \boldsymbol{\beta} 
+ b_{i} \quad \text{for} \quad i=1,...,N, \; j=1,...,n_i
\end{equation}
All the terms are the same as in Eq. (\ref{model_form_case_study}),
except for $b_{i}$, 
which is the random intercept relative to the $i^{\text{th}}$ country. Also in this case, the fixed intercept is not included in the model.

\subsection{Results}
\label{poisson_case_study}

SPGLMM outputs for Poisson response with $\alpha=0.01, 0.05, 0.10$ are addressed in Table \ref{Poisson_casestudy_random_intercept_slope}.
We get $\hat{M}_{0.01} = 13$, $\hat{M}_{0.05} = 16$ and $\hat{M}_{0.10} = 18$. As expected, at higher values of $\alpha$ correspond higher values of $M$. Indeed, the higher is $\alpha$, the smaller the confidence intervals and less likely to overlap (see Section \ref{sec3} for a further discussion).
The nomenclature of $\hat{\mathbf{c}}$ in the leftmost column of Table \ref{Poisson_casestudy_random_intercept_slope} is in harmony with the estimates obtained for $\alpha=0.10$ (i.e. the highest $\alpha$ for which the algorithm is run), where the random intercepts $\hat{c}_1,..., \hat{c}_{18}$ are reported on $18$ different rows.
For $\alpha$ equal to $0.01$ and $0.05$, the algorithm identifies fewer clusters and the estimated random intercept is reported in between two rows, to indicate that two distinct clusters were merged into one.
The gray or white backgrounds indicate whether discrepancies between the outputs with different $\alpha$ occur\footnote{For instance, with $\alpha=0.01$ and $\alpha=0.05$, in correspondence of $\hat{c}_2$ and $\hat{c}_3$ (reported on white background), only one value of random intercept is identified. 
We will denote this random intercept with $\hat{c}_{2+3}$, meaning for simplicity the random intercept associated with cluster $2+3$.
This coefficient turns out to be a weighted mean between $\hat{c}_{2}$ and $\hat{c}_{3}$, as expected from Eq. (\ref{update_masses}). 
Remarkable is the case in correspondence of $\hat{c}_{12}$, $\hat{c}_{13}$ and $\hat{c}_{14}$ (all reported on white background), where the countries in the cluster $13$, when $\alpha=0.01$, are partially assigned to the cluster $12$ and partially to the cluster $14$.}.
In addition, on the rightmost column of Table \ref{Poisson_casestudy_random_intercept_slope}, we report the results obtained with the parametric GLMM of Eq. (\ref{GLMM_casestudy}) with Poisson response.
In the first 18 rows, we report the means of the random intercepts $b_1,..., b_{50}$ computed by the GLMM within each cluster $m=1, ..., 18$.
We can appreciate that the means of the random intercepts $\hat{b}_i$ in each cluster are slightly lower in absolute value than the SPGLMM estimates. Anyhow, we observe huge coherence between the two models.
The second part of the table is dedicated to the fixed effects $\hat{\boldsymbol{\beta}}$.
Both the two fixed slopes are negative, meaning that the percentage of low-achieving students in mathematics is inversely proportional to the school size and the index of economic, social and cultural status. Specifically, the higher the value of \texttt{SCHSIZE} and \texttt{avg\_ESCS\_std}, the lower the percentage of low-achieving students, though \texttt{avg\_ESCS\_std} has a lower impact than \texttt{SCHSIZE} (the former slope is $\beta_2 = -0.09$ compared to the latter one of  $\beta_1 = -1.36$).
In general, we can conclude that also for the fixed slopes, SPGLMM and GLMM provide coherent results in the estimates, the standard errors and the p-values (estimated through the likelihood-ratio test).

\begin{table*}[!t]
\caption{SPGLMM estimates and comparison with GLMM output.}
\label{Poisson_casestudy_random_intercept_slope}
\resizebox{\textwidth}{!}{%
\begin{NiceTabular}{c c c c c c}
\CodeBefore
  \rowcolors{2}{gray!15}{white}
  \columncolor{white!15}{1,1}
\Body
\toprule
\multicolumn{2}{c}{\textbf{Coeff. estimates}} 
& \multicolumn{3}{c}{\textbf{SPGLMM}} & \multicolumn{1}{c}{\textbf{GLMM}}\\
\cmidrule(lr){3-5} 
& & \multicolumn{1}{c}{$\alpha=0.01$} & \multicolumn{1}{c}{$\alpha=0.05$}
& \multicolumn{1}{c}{$\alpha=0.10$} &  
\\
\midrule
\multirow{18}{*}{$\hat{\mathbf{c}}$ $[\hat{\omega}]$}

& $\hat{c}_{1}$ [0.02]  & -0.668 (0.092) &  -0.668 (0.092) & -0.668 (0.092) & -0.664 \\

& \makecell{$\hat{c}_2$ [0.06] \\ $\hat{c}_3$ [0.02]} & 0.082 (0.020) & 0.082 (0.020) & \makecell{ 0.060 (0.022) \\ 0.183 (0.044)} & \makecell{ 0.050 \\ 0.180 } \\

& \makecell{$\hat{c}_4$ [0.10] \\ $\hat{c}_5$ [0.02]} & 0.350 (0.018) & 0.350 (0.018) & \makecell{ 0.337 (0.020) \\ 0.445 (0.041) } & \makecell{ 0.327 \\ 0.448 } \\

& $\hat{c}_{6}$ [0.06]  & 0.549 (0.021) & 0.549 (0.021)  & 0.559  (0.021) &  0.563\\

& $\hat{c}_{7}$ [0.12]  & 0.656 (0.015) & 0.655 (0.015) &  0.656 (0.016) &  0.662\\

& \makecell{$\hat{c}_8$ [0.04] \\ $\hat{c}_9$ [0.04]} & 0.831 (0.014) & \makecell{ 0.813 (0.015)\\ 0.917 (0.033) } & \makecell{ 0.813 (0.015) \\ 0.917 (0.033) } & \makecell{ 0.821 \\ 0.921} \\

& \makecell{$\hat{c}_{10}$ [0.02] \\ $\hat{c}_{11}$ [0.12]} & 1.122 (0.010) & \makecell{ 1.048 (0.030) \\ 1.132 (0.011) } & \makecell{ 1.048 (0.03) \\ 1.133 (0.011) } & \makecell{ 1.038 \\ 1.125} \\

& \makecell{$\hat{c}_{12}$ [0.10] \\ $\hat{c}_{13}$ [0.04] \\ $\hat{c}_{14}$ [0.06]} & \makecell{ 1.229 (0.011) \\ 1.361 (0.012) } & \makecell{ 1.217 (0.012) \\ 1.296 (0.022) \\ 1.369 (0.012)} & \makecell{ 1.217 (0.012) \\ 1.296 (0.022)  \\ 1.369 (0.012) } & \makecell{ 1.216 \\ 1.297 \\ 1.379} \\

& $\hat{c}_{15}$ [0.04]  & 1.482 (0.020) &  1.483 (0.020) & 1.483 (0.02) &  1.481 \\

& $\hat{c}_{16}$ [0.08]  & 1.654 (0.023) &  1.654 (0.023)) & 1.655 (0.023) & 1.657 \\

& $\hat{c}_{17}$ [0.04]  & 1.920 (0.017) &  1.920  (0.017) & 1.920 (0.017) &  1.919\\

& $\hat{c}_{18}$ [0.02]  & 2.268 (0.041) & 1.268 (0.041)  & 2.268 (0.041) & 2.265 \\

\cmidrule(lr){1-6}
\multirow{2}{*}{$\hat{\boldsymbol{\beta}}$} & $\hat{\beta}_1$        
&  -1.361 (0.007) ***  & -1.361 (0.007) *** & -1.361 (0.007) ***  & -1.359 (0.012) *** \\
& $\hat{\beta}_2$  
&  -0.094 (0.004) ***  & -0.094 (0.004) *** & -0.094 (0.004) ***  & -0.095 (0.004) *** \\
\bottomrule
\end{NiceTabular}
}
\begin{tablenotes}
      \footnotesize
      \item Notes: The estimated random intercepts $\hat{\mathbf{c}}$ are presented in increasing order, together with their respective weights $\hat{\omega}$ in brackets, as well as the fixed effects $\hat{\boldsymbol{\beta}}$ for both SPGLMM (with $\alpha=0.01, \, 0.05, \, 0.10)$ and GLMM (for each row of $\hat{c}_m$, the average of the $\hat{b}_i$s in each cluster $m$ is reported). 
      In parenthesis, the standard error is computed by square rooting the inverse of the Fisher Information Matrix.
      For $\hat{\boldsymbol{\beta}}$, the p-value is estimated by means of likelihood-ratio test (* p-value $< 0.1$; ** p-value $< 0.01$; *** p-value $< 0.001$). 
    \end{tablenotes}
\end{table*}

\begin{figure}[!t]
\centering
\caption{Caterpillar plots reporting the comparison between the 50 random intercepts estimated by GLMM and the clusters obtained by SPGLMM, for $\alpha=0.01, 0.05, 0.10$.}
    \centering \includegraphics[width=16cm]
    {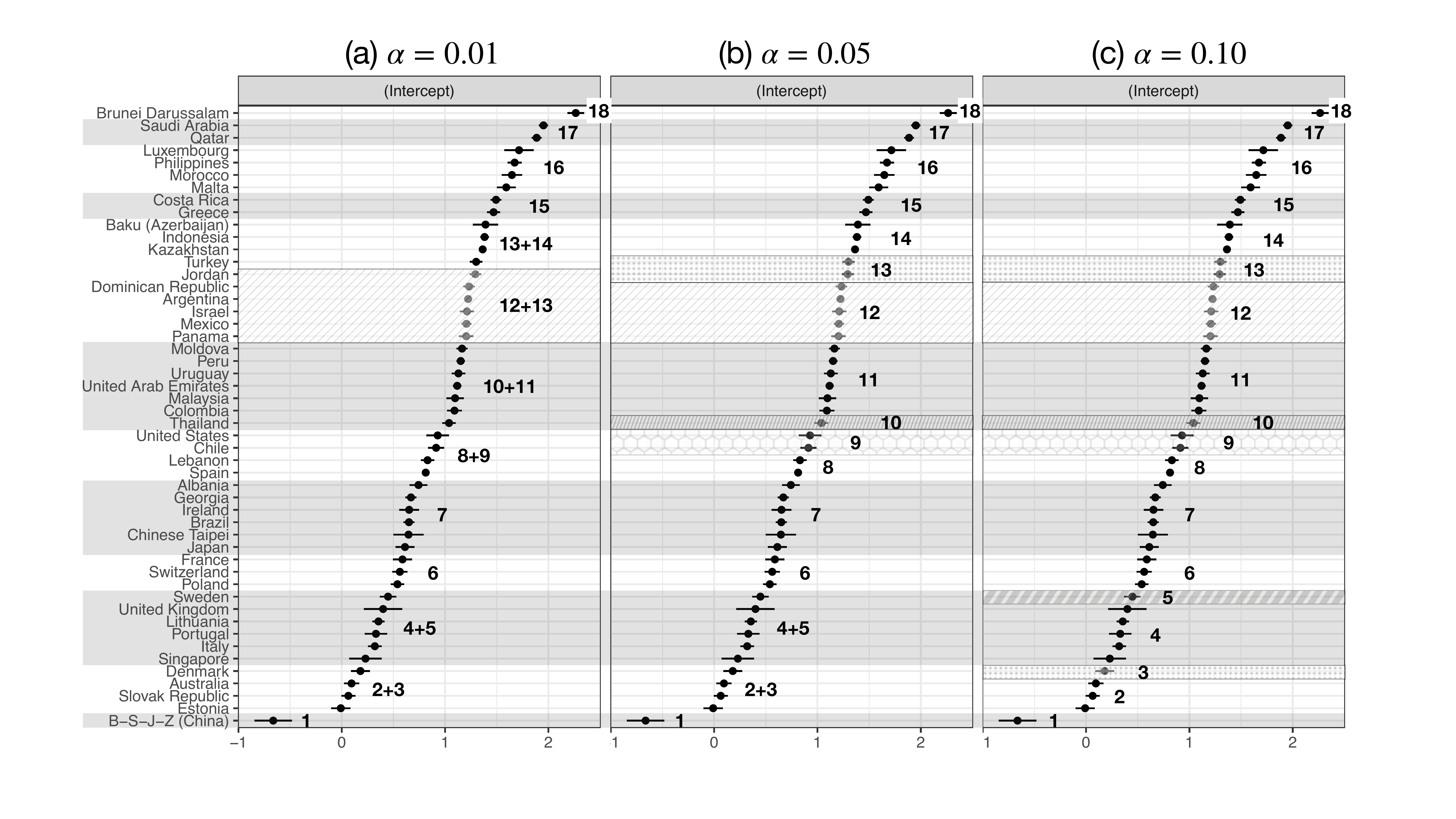}\\
\footnotesize
Notes: To ease the comparison with Table \ref{Poisson_casestudy_random_intercept_slope},
colours used to highlight clusters of countries in panel (a) are equal to the ones used in the table. Panels (b) and (c) adopt different textures in order to better highlight the differences in the detection of the more numerous clusters.
Next to the random intercepts, we report the number of the cluster, following the nomenclature of Table \ref{Poisson_casestudy_random_intercept_slope} (i.e., the plain enumeration $1,..., M_{0.10}$).
\label{fig:Poi_random_intercept}
\end{figure}

\begin{figure}
\centering
\caption{Choropleth map of the clusters of countries identified by the random intercepts in SPGLMM with $\alpha=0.01$.}
\vspace{-0.5cm}
\includegraphics[width=14cm]{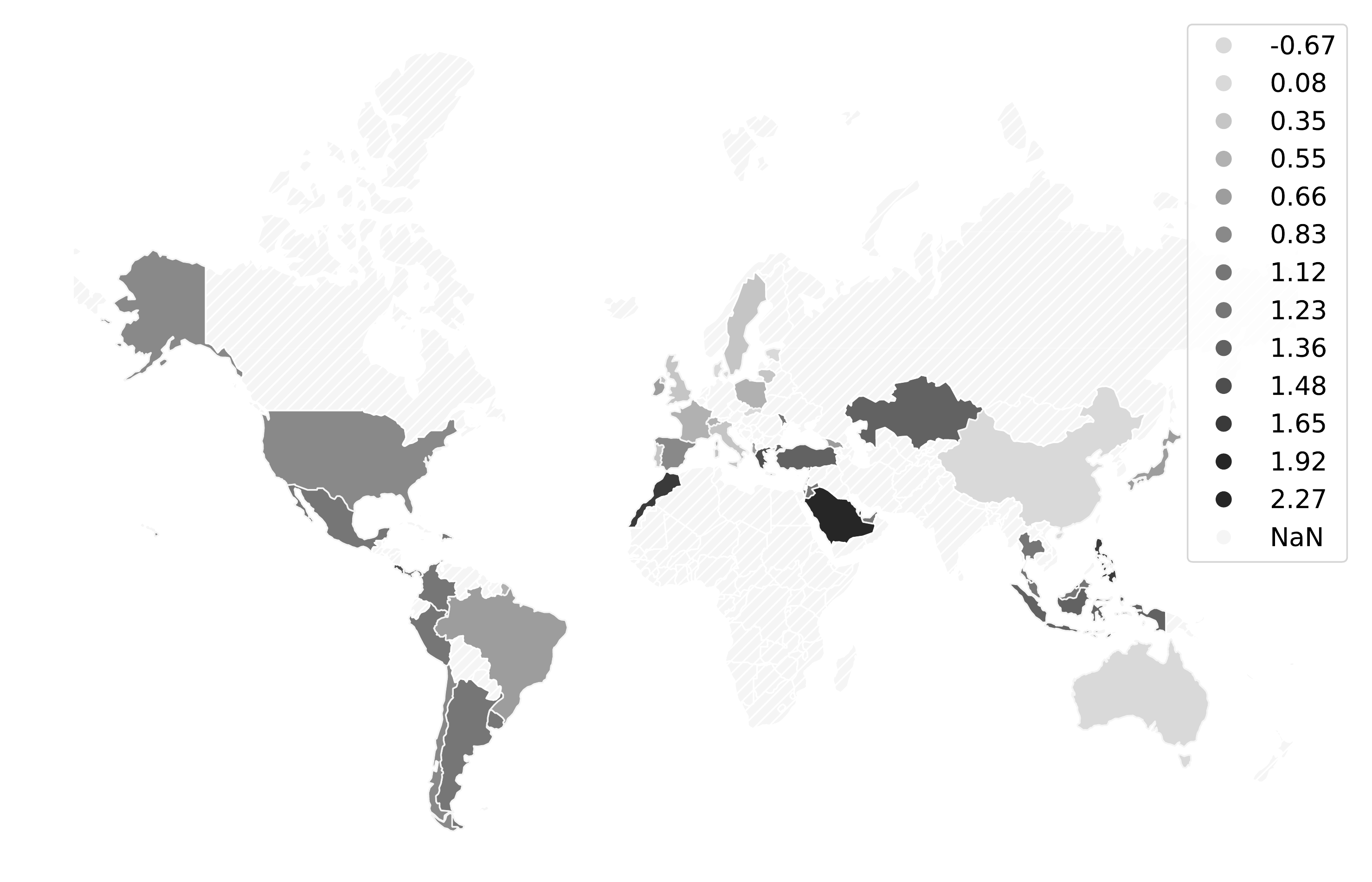} \\
\footnotesize
Notes: Countries represented with the same color belong to the same cluster. The lighter, the lower the random intercept.
Light grey-striped countries are the ones for which the survey was not performed, or which were presenting missing values.%
    \label{fig:Poi_world_16}
\end{figure}

In the three panels in Figure \ref{fig:Poi_random_intercept}, we display the caterpillar plots for the random intercepts (together with their confidence intervals) of the 50 countries obtained through parametric GLMM with Poisson response. 
On each panel, we highlight the identified clusters of countries, both for $\alpha=0.01$ in panel (a), $\alpha=0.05$ in panel (b) and $\alpha=0.10$ in panel (c).  
We remind that each country $i$ is assigned to the cluster $m$ by maximizing the posterior conditional weight $\hat{\omega}_{im}$, as shown in Eq. (\ref{l_tilde}).  
Results can be interpreted as follows: the lower the estimated random intercept for a cluster (i.e., the bottom countries in the caterpillar plots), the lower the percentage of low-achieving students in mathematics in the schools of the countries of that cluster, and vice-versa.
For better visualization of the clusters of countries identified by the SPGLMM, we highlight with the same shade of gray on the map in Figure \ref{fig:Poi_world_16} the countries identified by the same random intercept (i.e. the countries in the same cluster), for $\alpha=0.01$. We notice that B-S-J-Z (China) and Australia, net of the other features, decrease the percentage of low-achieving students in mathematics. After them, the European countries slightly increase it, while the Americas and other middle-east countries have a wider impact.

\noindent
For the Goodness of Fit (GoF) evaluation, we consider the following metrics for integer responses: the MSE of responses ($\frac{1}{J} \sum_{i,j} (y_{ij} - \hat{y}_{ij})^2$), the MSE of log responses ($\frac{1}{J} \sum_{i,j} (log(y_{ij}+1) - log(\hat{y}_{ij}+1))^2$)
and the Chi-Squared Error $\frac{1}{J} \sum_{i,j} \frac{(y_{ij} - \hat{y}_{ij})^2 }{\hat{y}_{ij} + 1}$ (\cite{mccullagh1989generalized}). 
The $+1$ at the denominator and inside the logarithm is added because $y_{ij}$ and $\hat{y}_{ij}$ could possibly assume value $0$. 
The same data used for training the models are also utilized for computing the predictions and the aforementioned indexes, to ensure a fair comparison between the predictive abilities of SPGLMM and GLMM.
In Table \ref{tab:GOF_POI_case_study} we report on three different rows the three metrics obtained comparing  $\mathbf{y} = \texttt{Y\_MATH} $ and the predicted $\hat{\mathbf{y}}$ (retrieved by rounding $\hat{\boldsymbol{\mu}}$ to the closest integer) for each of the three SPGLMM and GLMM.
The two methods reveal similar predictive performances. The SPGLMM with Poisson response does not worsen the predictions with respect to the parametric GLMM, whereas, it further provides in output a clustering of the hierarchies (i.e., the countries in our case study), revealing the inner structure the model assumes.
This result is further discussed in Section \ref{sec3}, analyzing results obtained through a simulation study.

\begin{table}[!t]
\caption{GoF metrics estimates 
for the case study, fitted via SPGLMM and GLMM.}
\label{tab:GOF_POI_case_study}
\centering
\begin{tabular}{@{}lcccc@{}}
\toprule
& \multicolumn{3}{c}{\textbf{SPGLMM}} & \textbf{GLMM} \\
\cline{2-4}
& 
\multicolumn{1}{c}{$\alpha = 0.01$} &
\multicolumn{1}{c}{$\alpha = 0.05$} &
\multicolumn{1}{c}{$\alpha = 0.10$} & \\
\midrule
\textit{MSE of responses}     & 27.461  & 27.405 & 27.398 & 27.406 \\
\textit{MSE of log responses} & 0.284  & 0.283 & 0.282 & 0.282  \\
\textit{Chi-Squared Error}    & 2.376  & 2.363 & 2.361 & 2.369  \\
\bottomrule
\end{tabular}
\end{table}

\section{Discussion and comparison with existing methods}
\label{sec3}

In this section, we discuss the accuracy and reliability of the SPGLMM with $\alpha$-\textit{criterion} proposed in Section \ref{sec2}, proving its well-performance against other state-of-the-art methods.
More specifically, we propose a simulation study for generating sets of data with an \textit{a priori} built latent grouping structure on which to test our SPGLMM with $\alpha$-\textit{criterion} under different settings and easily compare its results with the ones obtained by other models, i.e. SPGLMMs with \textit{t}-\textit{criterion} and parametric GLMMs.
In fact, as briefly introduced in Section \ref{sec:24}, all the papers in literature dealing with discrete random effects make the collapsing step of the algorithm relying on a priori chosen threshold $t$ (from here, the denomination \textit{t}-\textit{criterion}). 
At each iteration until convergence, the discrete masses with Euclidean distance lower than $t$ are made collapsed.
Through such methodology, the number of obtained clusters $M$ is not chosen \textit{a priori}, but it intrinsically depends on the choice of $t$.
The higher is $t$, the lower the number of clusters and the less homogeneous the groups within each cluster: $t$ should be chosen depending on the required homogeneity level within clusters.
The selection of the threshold $t$ represents the main drawback when these methods are applied to real-world data, 
especially when no prior information on the heterogeneity among clusters is available: results could be very sensitive to $t$. 
For dealing with such a choice, we conduct in Section S6 of Supplementary Materials a sensitivity analysis and we propose a criterion for driving the choice of the threshold $t$ based on the individuation of an elbow in the plot of the average entropy of the conditional weights matrix; nevertheless, such a criterion has the drawback to be computationally expensive, especially when dealing with massive amounts of data (e.g., the numerosity of the case study addressed in Section \ref{sec:case_study}).

In our simulation study, which we perform both for Poisson response (addressed in Section \ref{simPoi}) and Bernoulli response (addressed in Section S5 of Supplementary Materials), we run the SPGLMM with $\alpha$-\textit{criterion} for different confidence levels $\alpha$ (i.e., $\alpha=0.01, 0.05$ and $0.10$)
and the \textit{t}-\textit{criterion} for different values of $t$ 
and we compare their performances. 
The SPGLMM with $t$-\textit{criterion} is, to the best of our knowledge, the only method in literature able to fit a GLMM with discrete random effects (i.e., able to cluster \lq\lq the hierarchies"). As a matter of fact, similarly to what we have done for the case study, we challenge our method with two different versions of a parametric GLMM, taking into account that parametric GLMM proposes a different interpretation of the random effects, estimating a single coefficient for each group rather then clustering them.


In the next two sections, we address the set-up of the simulation study (Section \ref{sec:sim_study_set_up}) and more specifically the Poisson response (Section \ref{simPoi}).

\subsection{The simulation study set-up}
\label{sec:sim_study_set_up}
For our simulation study, we consider $N=10$ groups of data\footnote{This choice is driven by 
a trade-off between
the need of having enough groups for making SPGLMM detect non-trivial clusters
and the constraint of not introducing too many groups in the generative models, which would lead to more complexity and less interpretability to the results of our simulation.} and,
since  SPGLMM can handle different number of observations within groups, we sample the number of observations $n_i$ from a uniform distribution $n_i \sim \mathcal{U}(70,100)$ for $i = 1,..., N$.
We simulate the data by inducing the presence of 3 clusters.
We set $\text{\texttt{K}}=60$, $\text{\texttt{K1}}=20$, $\text{\texttt{itmax}} = 20$, $\text{\texttt{tR}} = \text{\texttt{tF}} = 10^{-5}$ 
and $\text{\texttt{K2}}=5$.

\subsection{Poisson response case}
\label{simPoi}
For the case of a Poisson response distribution, we simulate a model presenting only a random intercept\footnote{This choice is due to the fact that in a GLMM with Poisson response, the exponential as inverse canonical link function (i.e., $\boldsymbol{\mu_{i}} = \text{exp}(\boldsymbol{\eta_{i}})$), makes the identification of the model coefficients computationally difficult. Simulation study for the Binary response includes also the case of only random slope and both random intercept and slope.}
with either one or two fixed slopes.
The second fixed slope will be indicated in parenthesis in the following equations.
The linear predictor $\boldsymbol{\eta}_{i} = \boldsymbol{\beta}_1 \mathbf{x}_{1i} \; + ( \boldsymbol{\beta}_2 \mathbf{x}_{2i}) +  c_{1i} \mathds{1}_{n_{i}} $ is defined by the following DGP:
    \begin{equation}
    \label{Poi_int_2}
    \boldsymbol{\eta}_i = \begin{cases} 
    0.3 \mathbf{x}_{1i} \; + ( 0.9 \mathbf{x}_{2i}) + 2.5\;\mathds{1}_{n_{i}} & \mbox{if } i = 1,2, \\ 
    0.3 \mathbf{x}_{1i} \; + ( 0.9 \mathbf{x}_{2i}) + 1 \; \mathds{1}_{n_{i}} & \mbox{if } i = 3,4,5,6,7, \\
    0.3 \mathbf{x}_{1i} \; + ( 0.9 \mathbf{x}_{2i}) -1 \; \mathds{1}_{n_{i}} & \mbox{if } i = 8,9,10
          \end{cases}
\end{equation}
Variables $\mathbf{x}_{1i}$ and $\mathbf{x}_{2i}$ are normally distributed with mean equal to 0 and standard deviation equal to 1.
The choice of the coefficients values is arbitrary. In this case, they are chosen in order to simulate different situations in which we obtain a different skewness with respect to the zero, but also avoiding generating too high numbers, which could cause numerical issues in the computation of the poisson density.
After the computation of $\boldsymbol{\eta}_{i}$ according to DGP in Eq. (\ref{Poi_int_2}), we retrieve $\mu_{ij}=\text{exp}(\eta_{ij})$ 
and we compute $y_{ij} \sim \text{Poi}(\mu_{ij})$ for $i=1,...,N$ and $j=1,...,n_i$. Namely, $y_{ij}$ is extracted from a Poisson distribution with mean equal to the retrieved $\mu_{ij}$.
Afterwards, we apply SPGLMM with both \textit{t-} and $\alpha$\textit{-criterion}s, performing 500 runs for the setting with one fixed slope shown in Eq. (\ref{Poi_int_2}), for different values of $t$ and $\alpha$. The values of $t$ are chosen simmetrically around the value that maximizes the times in which the true number of clusters is identified, in order to best analyze and visualize the model behaviour.

\begin{figure}
\centering
\caption{Barplot for the frequency a certain number of clusters is identified over 500 runs, across different values of $alpha$ and $t$, for DGP for Poisson response with one fixed slope.}
    \subfloat[\centering $\alpha$-\textit{criterion}]{
    {\includegraphics[width=7.7cm]{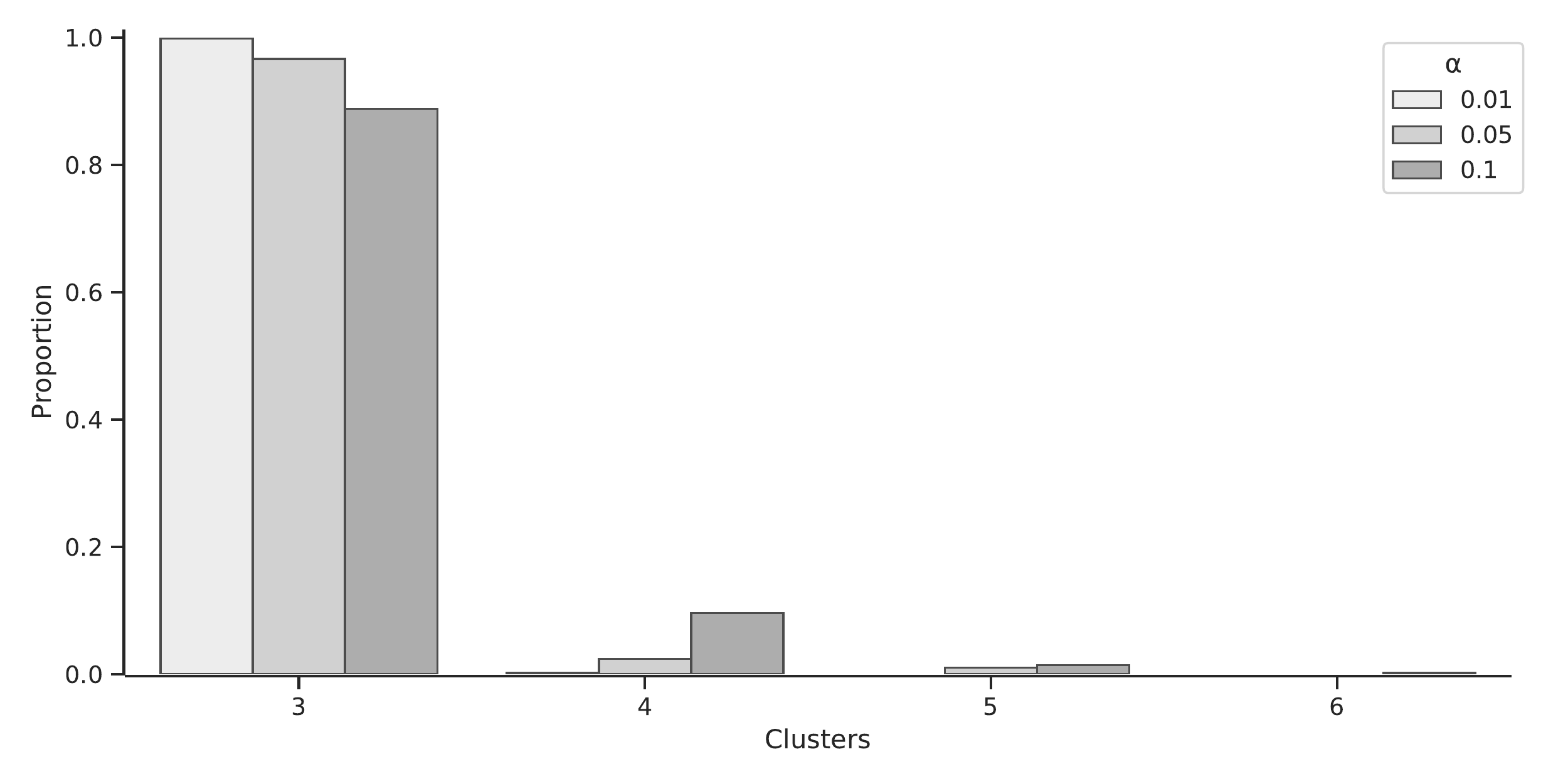} }
    } 
    \subfloat[\centering \textit{t-criterion}]{
    {\includegraphics[width=7.7cm]{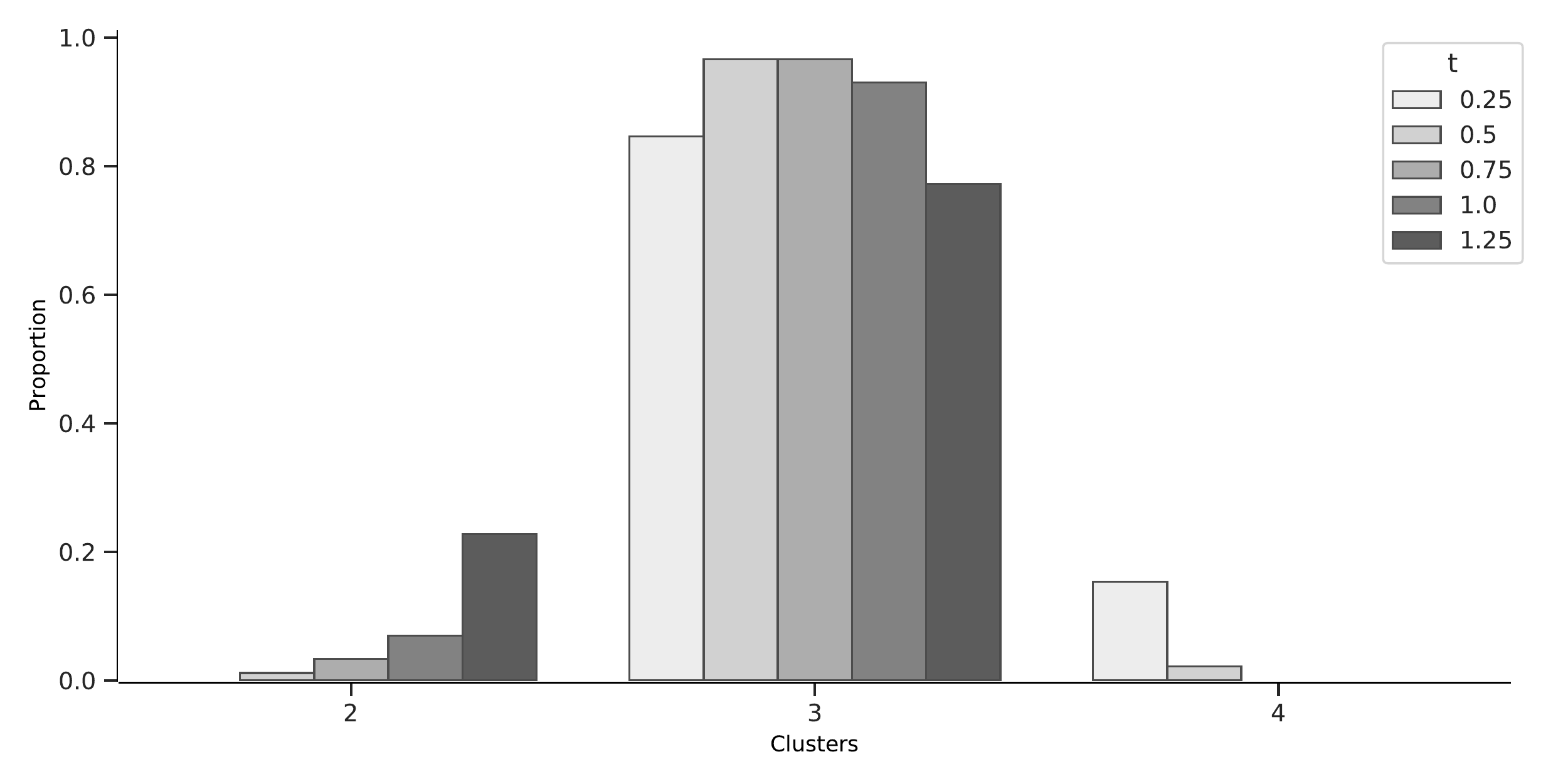} }
    }
    \\
    \footnotesize
    Notes: To ease the comparison, the frequencies on y-axis are reported on the same scale. The number of identified clusters is reported on the x-axis. In panel (a), results across $alpha=0.01, \, 0.05, \, 0.1$ are represented with different shadows of grey. Similarly, in panel (b) results across $t=0.25,\, 0.5, \, 0.75, \, 1, \, 1.25$ are addressed.
\label{fig:proportions_Poi}
\end{figure}

\begin{figure}
\centering
\caption{Boxplots for the random intercept $\mathbf{c}_1$ distribution for the DGP for Poisson response, with one fixed slope.}
    \subfloat[\centering $\mathbf{c}_1$ (2 clusters) - $\alpha$-\textit{criterion}]{
    {\includegraphics[width=7.7cm]{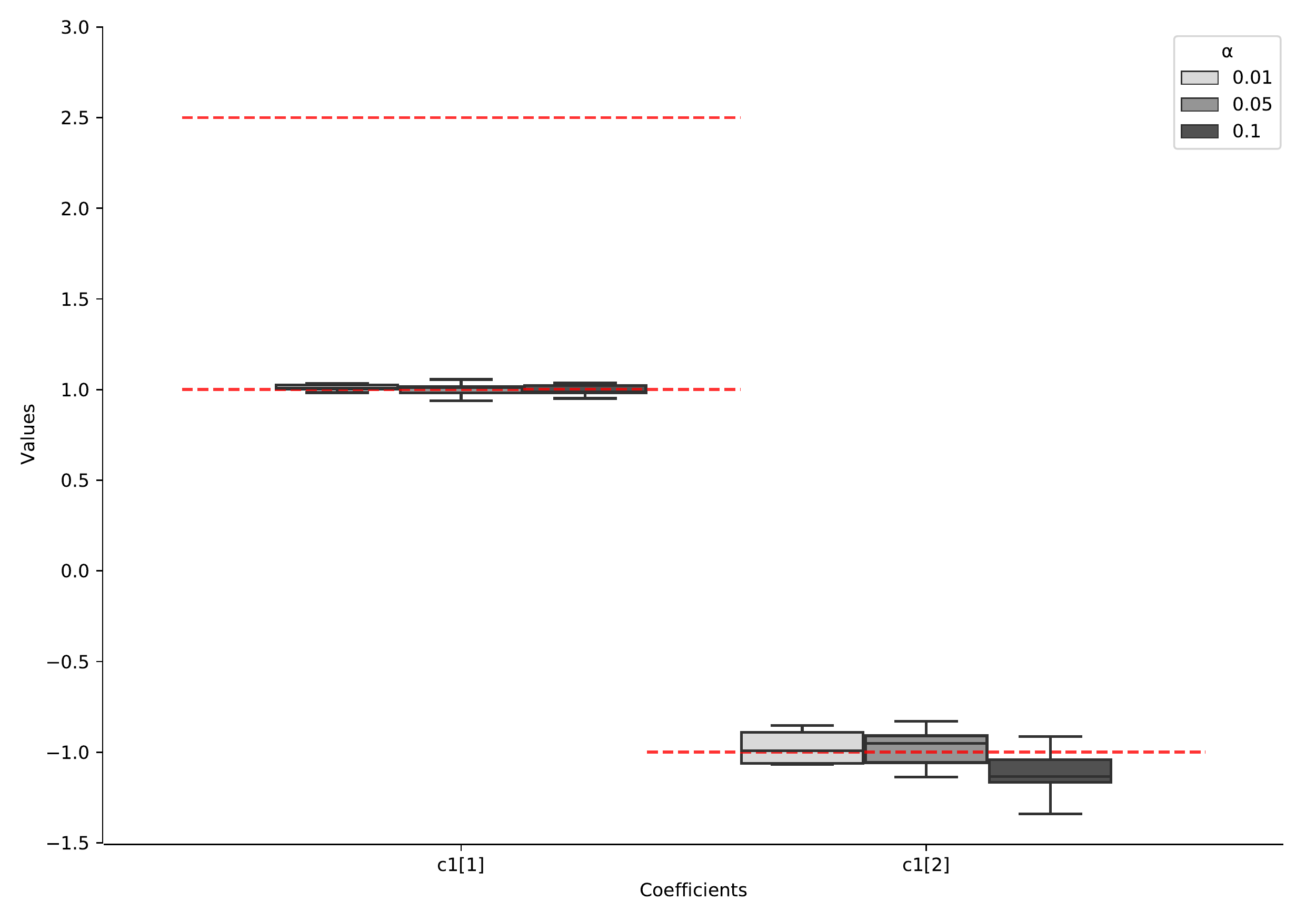} }
    } 
    \subfloat[\centering $\mathbf{c}_1$ (2 clusters) - \textit{t}-\textit{criterion}]{
    {\includegraphics[width=7.7cm]{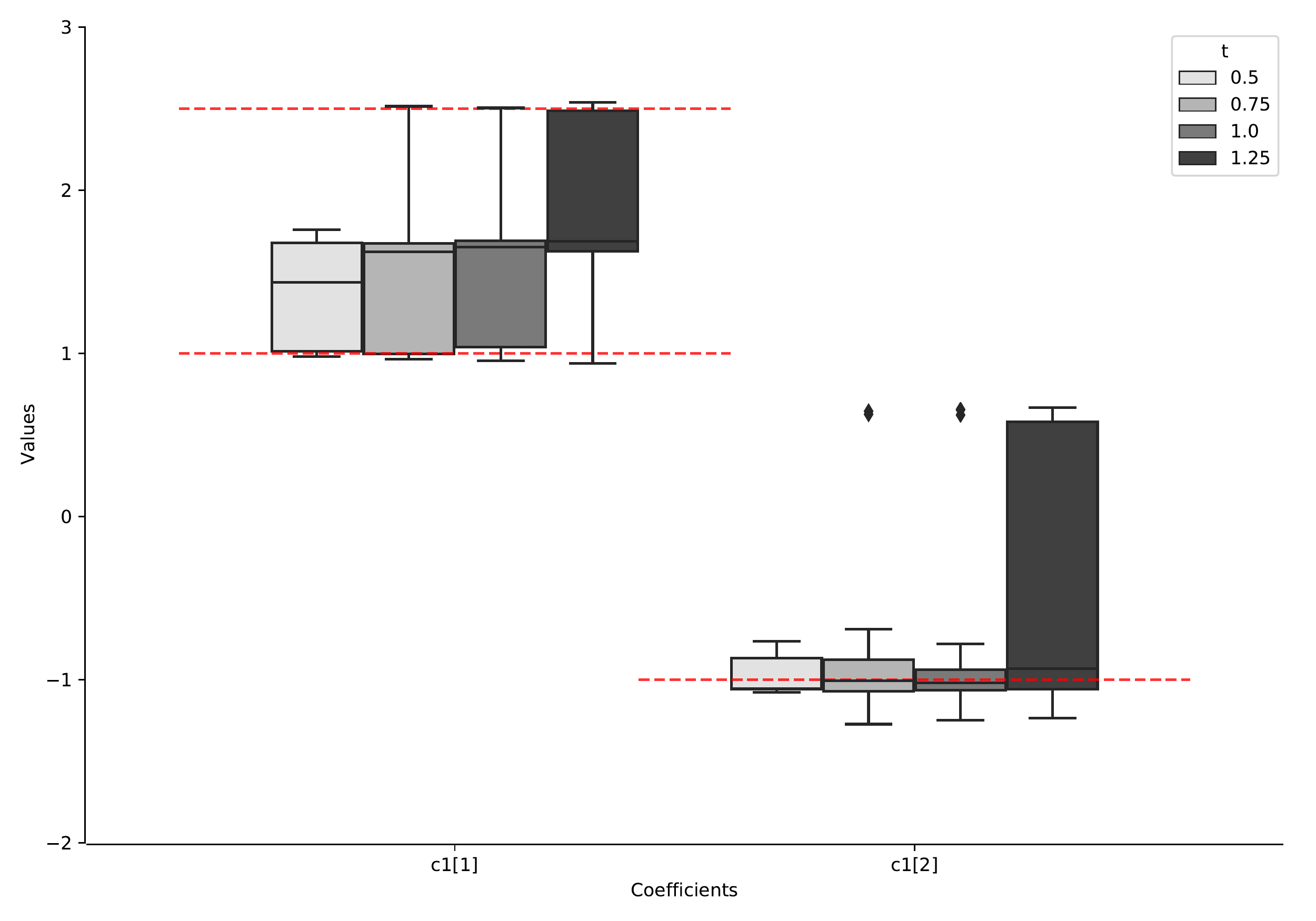} }
    }
    \\
    \subfloat[\centering $\mathbf{c}_1$ (3 clusters) - $\alpha$-\textit{criterion}]{
    {\includegraphics[width=7.7cm]{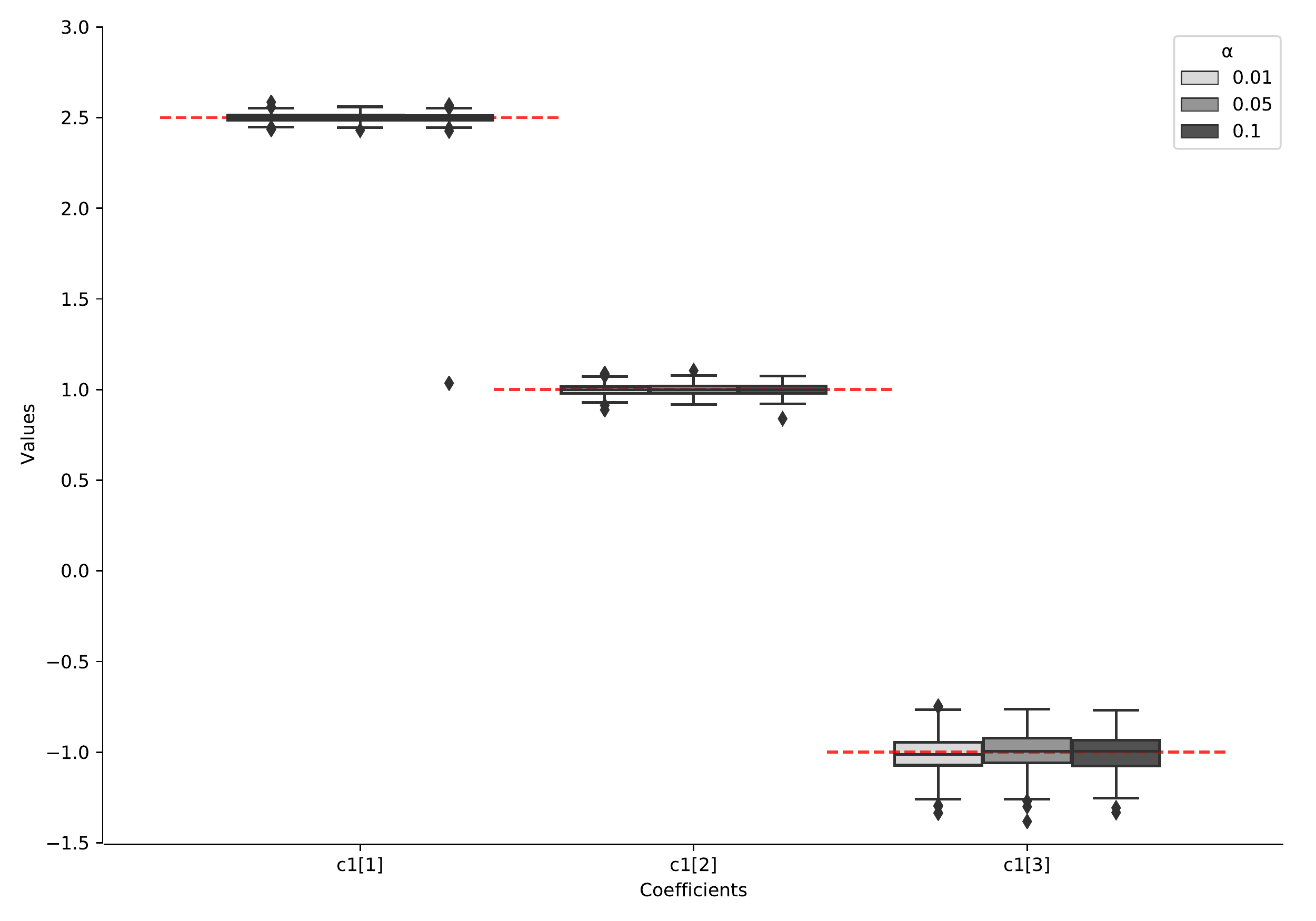} }
    } 
    \subfloat[\centering $\mathbf{c}_1$ (3 clusters) - \textit{t}-\textit{criterion}]{
    {\includegraphics[width=7.7cm]{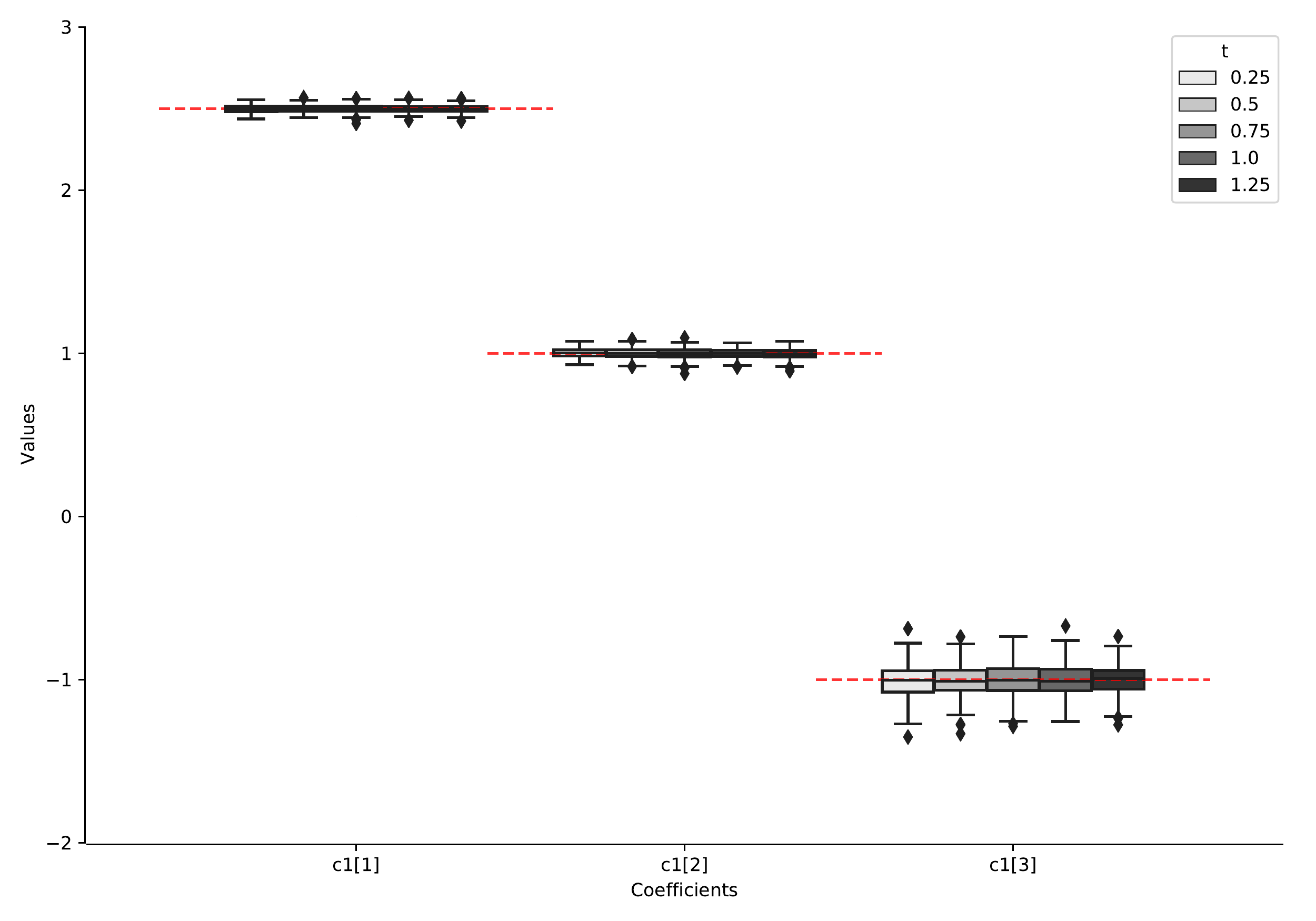} }
    }
    \\
    \subfloat[\centering $\mathbf{c}_1$ (4 clusters) - $\alpha$-\textit{criterion}]{
    {\includegraphics[width=7.7cm]{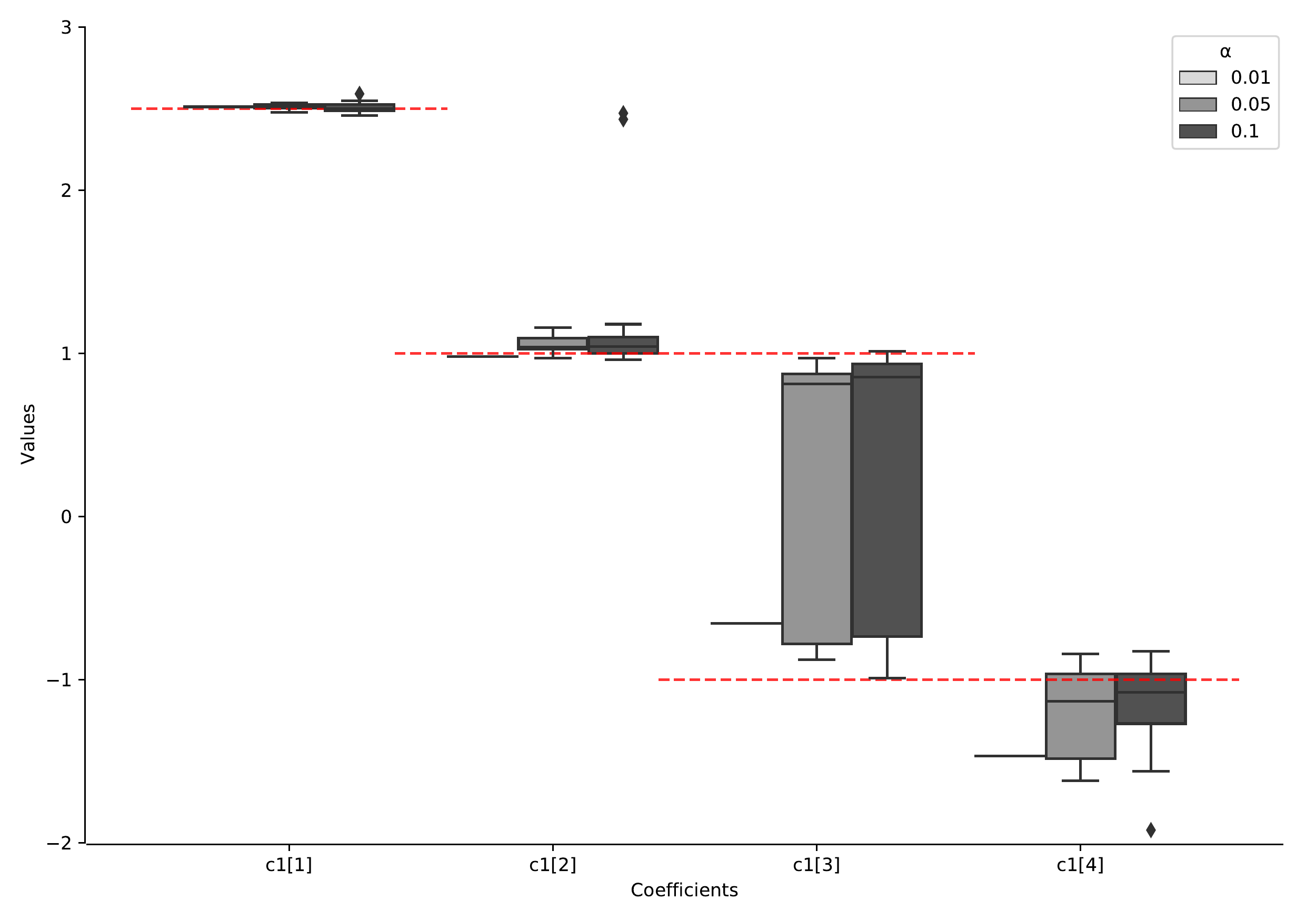} }
    } 
    \subfloat[\centering $\mathbf{c}_1$ (4 clusters) - \textit{t}-\textit{criterion}]{
    {\includegraphics[width=7.7cm]{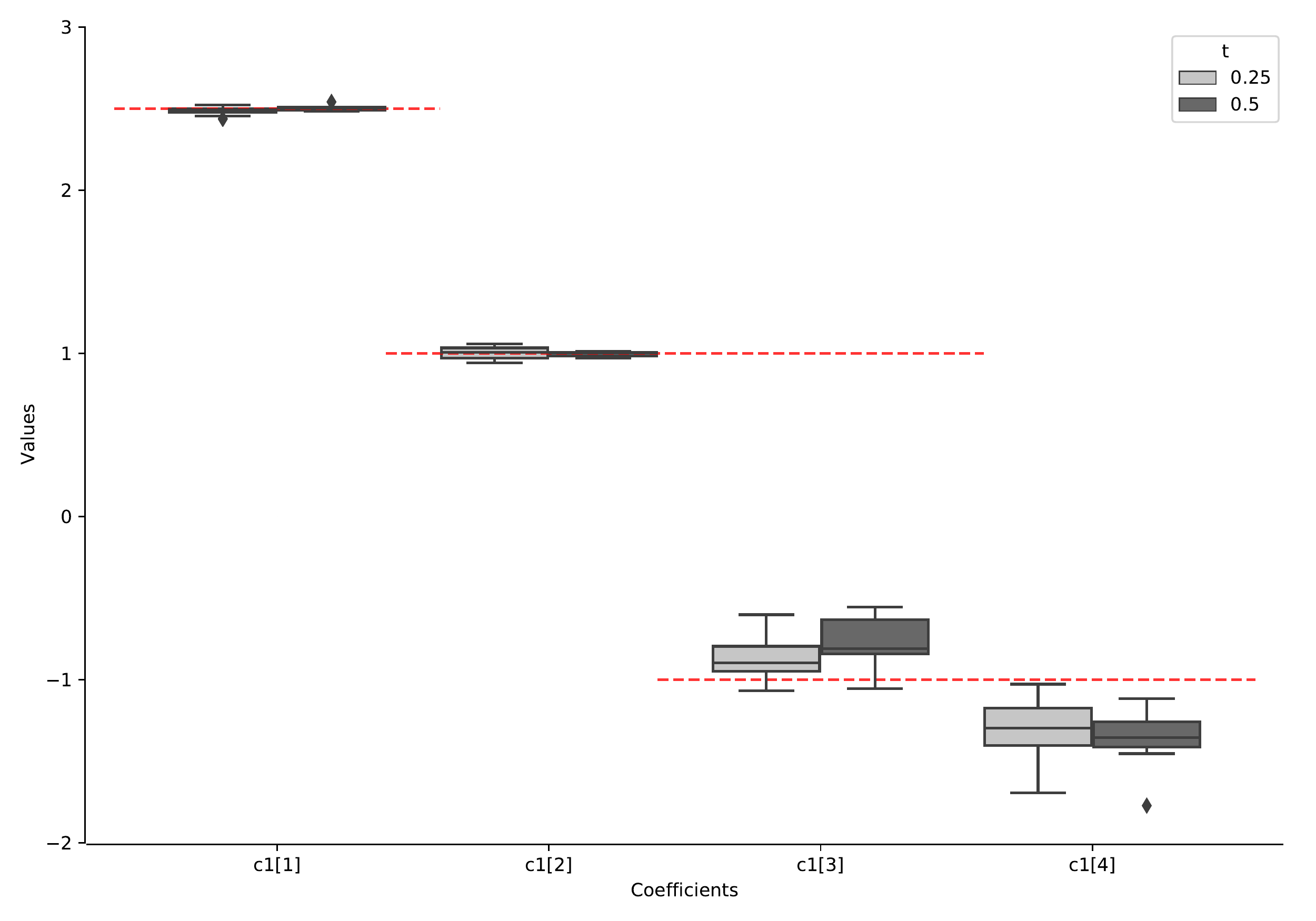} }
    }
    \\
    \footnotesize
Notes: For each of the 500 runs with a chosen threshold, we represent boxplots for the values of the components of the random intercept $\mathbf{c}_1$ (y-axis), separately, according to the number of identified clusters (panels (a) and (b) for 2 clusters, panels (c) and (d) for 3 clusters, panels (e) and (f) for 4 clusters).
In the left panels, we report the results of the SPGLMM run via $\alpha$-\textit{criterion}, while in the right panels the results obtained via \textit{t}-\textit{criterion}. The horizontal dotted lines indicate the simulated coefficients.
\label{fig:plots_c1_Poi}
\end{figure}

\begin{figure}
\centering
\caption{Boxplots for the fixed slope $\beta_1$ distribution for the DGP for Poisson response, with one fixed slope.}
    \subfloat[\centering $\beta_1$ - $\alpha$-\textit{criterion}]{
    {\includegraphics[width=7.7cm]{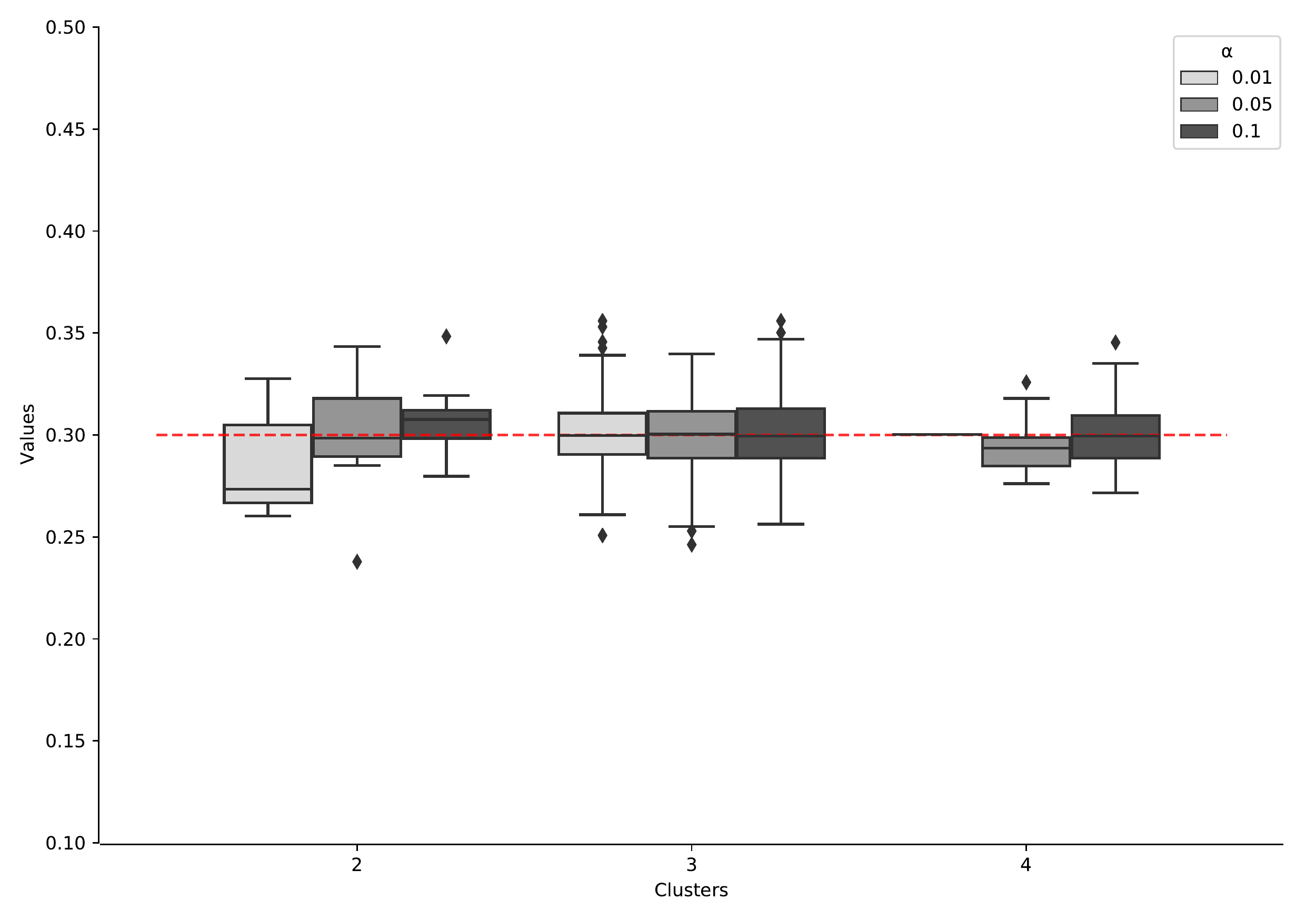} }
    } 
    \subfloat[\centering $\beta_1$ - \textit{t}-\textit{criterion}]{
    {\includegraphics[width=7.7cm]{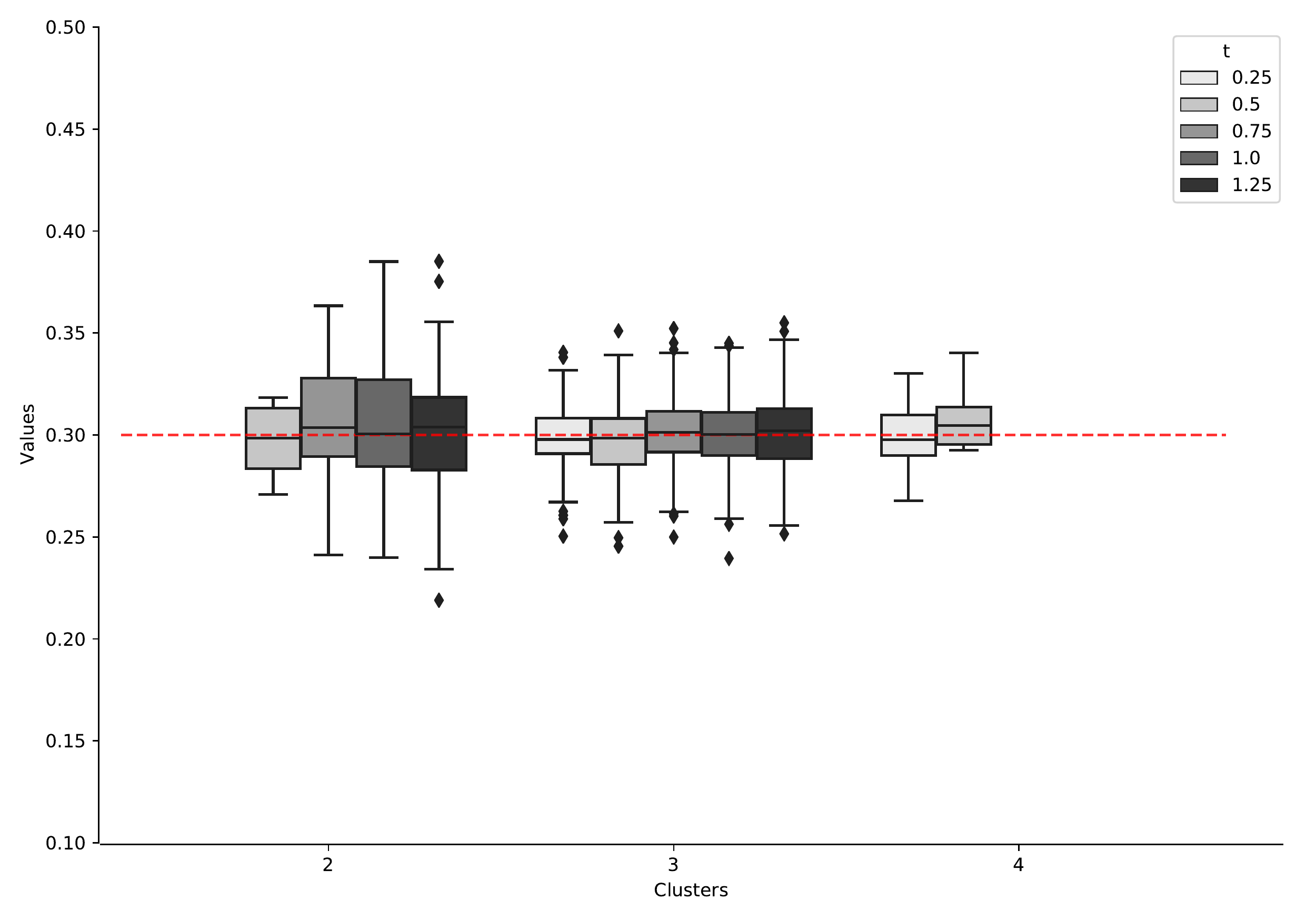} }
    }
    \\
    \footnotesize
Notes: For each of the 500 runs with a chosen threshold, we represent boxplots of the value for the fixed slope $\beta_1$ (y-axis) according to the number of identified clusters (indicated on the x-axis).
In the left panel, we report the results of the SPGLMM via $\alpha$-\textit{criterion}, while in the right panel the results are obtained via \textit{t}-\textit{criterion}. The horizontal dotted lines indicate the simulated coefficients.
\label{fig:plots_beta1_Poi}
\end{figure}

The obtained results with only one fixed slope are collected in Figures \ref{fig:proportions_Poi}, \ref{fig:plots_c1_Poi} and \ref{fig:plots_beta1_Poi}.
By looking at Figure \ref{fig:proportions_Poi}, we appreciate that the algorithm correctly identifies the true number of clusters (i.e., three) in the majority of the runs.
If \textit{t-criterion} is adopted, for all the DGPs, the higher the threshold, the higher the proportion of the cases in which the algorithm identifies fewer clusters than the true number; vice versa, the lower the threshold, the higher the proportion of the cases in which the algorithm identifies more clusters than the true number.
In fact, decreasing the value of $t$, the support points of the random effects coefficients distribution with distances lower than $t$ collapse to a unique point, and the SPGLMM algorithm is more sensitive to the variability among the groups, identifying a higher number of clusters, and vice versa.
It follows that, for each DGP, there exists an \textit{optimal} threshold, for which the proportion of the correctly identified true number of clusters is maximized.
When the $\alpha$-\textit{criterion} is adopted, the proportion of such number is always very high, higher than the optimal case of the \textit{t-criterion}; moreover, the higher is $\alpha$, the higher the proportion of times in which the algorithm identifies more clusters than the true number. This happens because, as anticipated in Section \ref{sec:case_study}, the higher is $\alpha$, the smaller the confidence region (interval) of level $1-\alpha$ is; this induces the confidence regions (intervals) relative to different support points to overlap less and the collapse effect to have a lower impact.
In Figure \ref{fig:plots_c1_Poi}, we clearly see that when the SPGLMM identifies 3 clusters, the estimated coefficients are very close to the simulated ones and their variability is low.
When the algorithm identifies a higher number of clusters with respect to the true one, it generally splits a cluster in two; vice versa, when SPGLMM identifies a lower number of clusters, it merges two clusters into one. 
The estimates for the fixed effects coefficients represented in Figure \ref{fig:plots_beta1_Poi} result to be only marginally affected by the identified number of clusters, being their estimates quite robust with respect to the random effects. 

For completeness, the results obtained for DGP with both one and two fixed slopes are reported respectively in Tables S7.1 and S7.2 in Section S7 of Supplementary Materials.

\begin{figure}[!t]
\centering
\caption{Boxplots representing the distribution of the fixed slope $\beta_1$ and random intercepts for DGP for Poisson response, for 100 iterations of the DGP.
}
    \subfloat[\centering GLMM (10 groups)]{{\includegraphics[width=15.5cm]{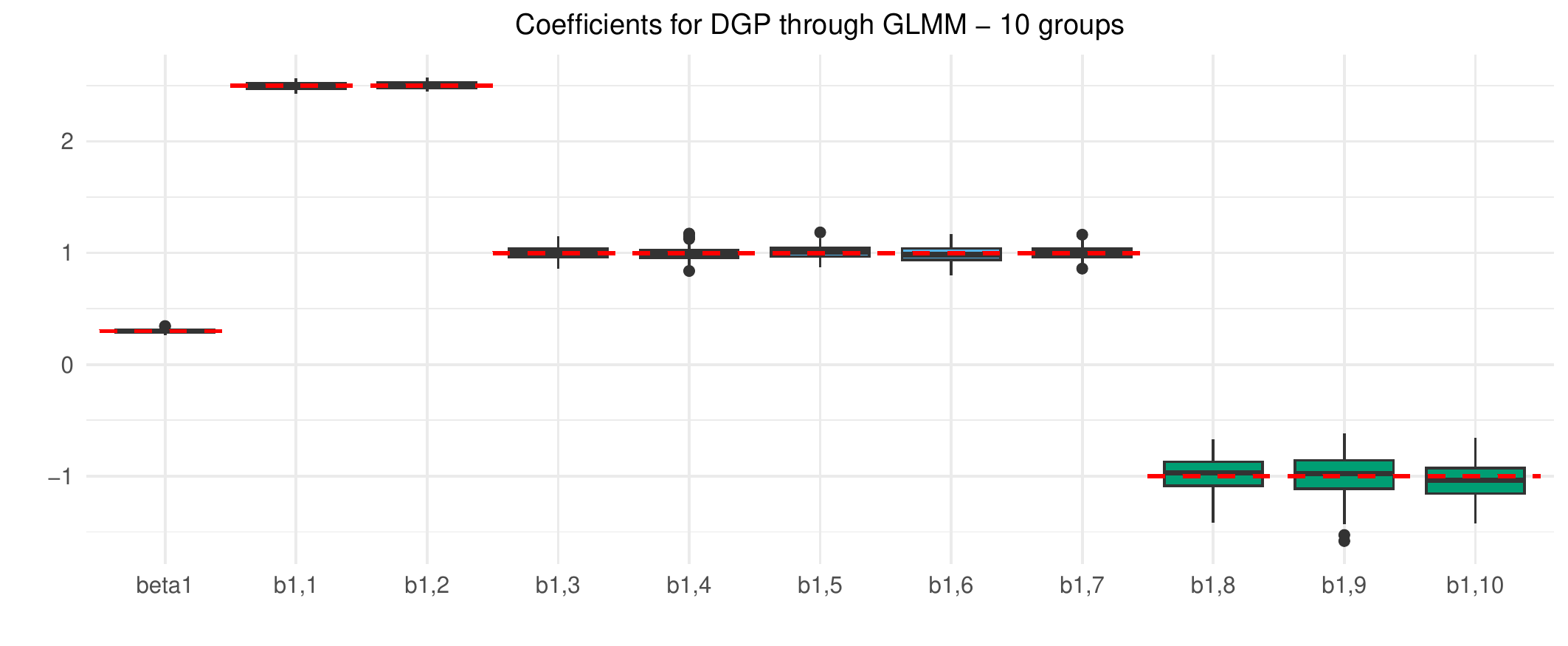} }} \\
    \subfloat[\centering GLMM (3 clusters)]{{\includegraphics[width=7.5cm]{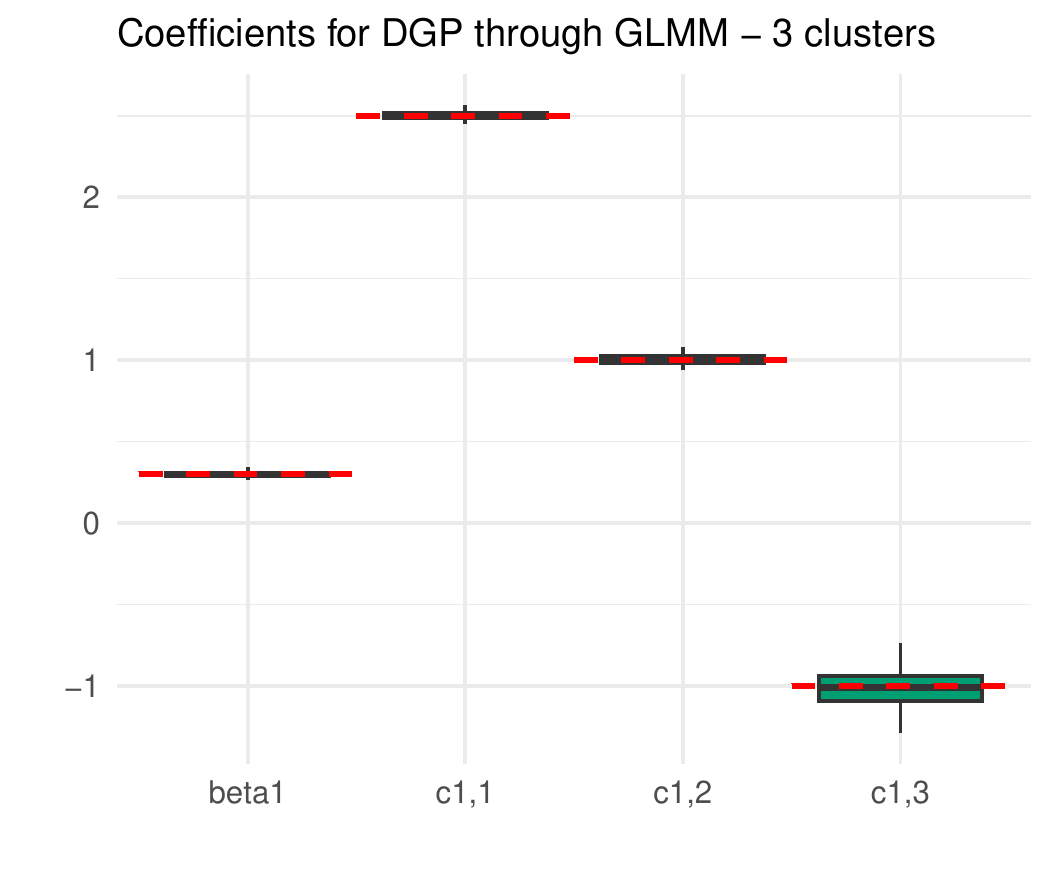} }}
    \;
    \subfloat[\centering SPGLMM (3 clusters)]{{\includegraphics[width=7.5cm]{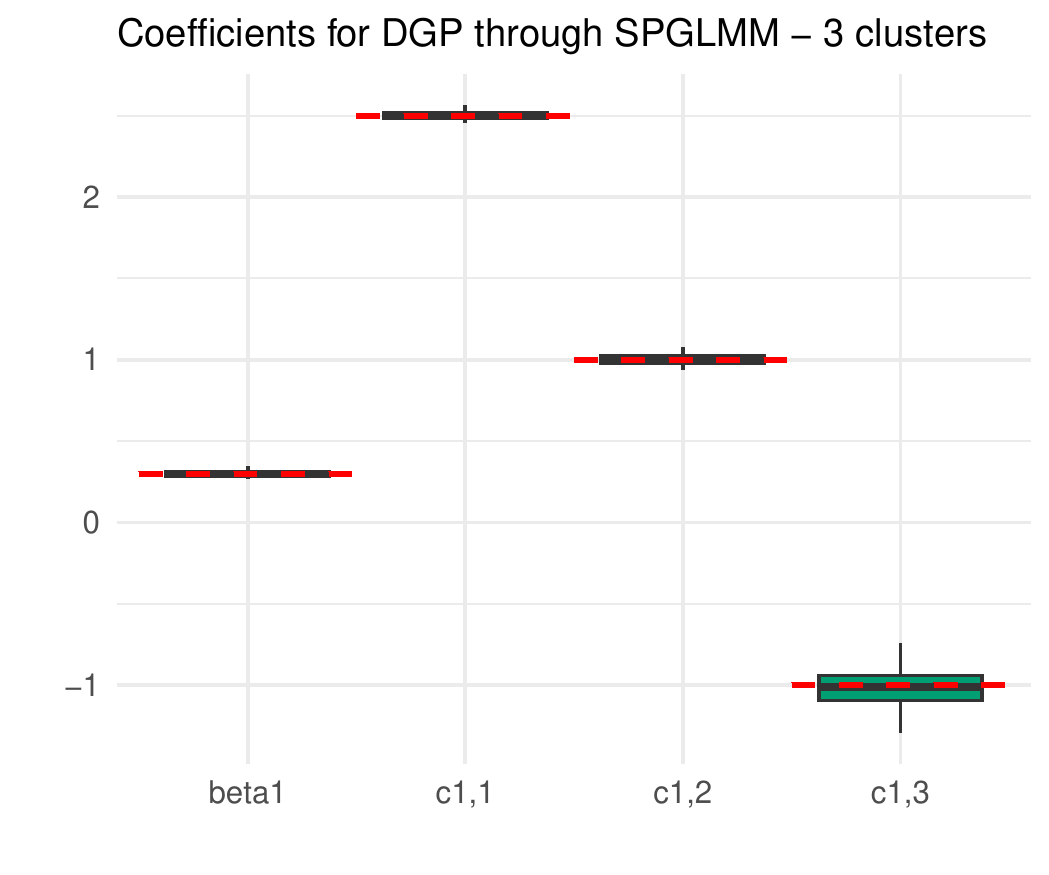} }} \\
    \footnotesize
Notes: Random intercepts are denoted by $b_{1i}$ for $i=1,...,10$ in panel (a) and $c_{1m}$ for $m=1,2,3$ in panels (b) and (c). The horizontal dotted lines indicate the simulated coefficients.
\label{fig:10glmer_3GLMM_3SPGLMM_coef_POI}
\end{figure}

To assess the goodness of fit of SPGLMM, we compare our results to the ones obtained by a parametric GLMM fitted through the function \textit{glmer} 
in R package \textit{lme4} (\cite{lme4}, \cite{Rproject}).
We simulate 100 different DGPs as in Eq. (\ref{Poi_int_2}) and we fit, in turn, two different GLMMs, 
the first one considering a random intercept for each group $i=1,...,10$ and the second one with a random intercept for each cluster $m=1,2,3$. 
Moreover, we fit the SPGLMM with $\alpha$-\textit{criterion} with $\alpha=0.05$, knowing that with such a value the algorithm identifies 3 clusters in the $94.8\%$ of the times, as highlighted in Table S7.1 in Section S7 of Supplementary Materials.
For each of the three models (i.e., GLMM with 10 random intercepts, GLMM with 3 random intercepts and SPGLMM), we represent through boxplots the distribution of the obtained coefficients across the 100 DGPs, emphasizing the simulated values through dotted lines, respectively in Figure \ref{fig:10glmer_3GLMM_3SPGLMM_coef_POI}, panels (a), (b) and (c).
We report in Table \ref{table:GoF_poisson} the summary statistics of the GoF metrics,
which are computed comparing  $y_{ij}$ of the DGP and the predicted $\hat{y}_{ij}$, retrieved by rounding $\hat{\mu}_{ij}$ to the closest integer.
As in Section \ref{sec:case_study}, we consider the MSE of responses ($\frac{1}{J} \sum_{i,j} (y_{ij} - \hat{y}_{ij})^2$), the MSE of log responses ($\frac{1}{J} \sum_{i,j} (log(y_{ij}+1) - log(\hat{y}_{ij}+1))^2 $) and Chi-Squared Error ($\frac{1}{J} \sum_{i,j} \frac{(y_{ij} - \hat{y}_{ij})^2 }{\hat{y}_{ij} + 1}$). 

Results in Table \ref{table:GoF_poisson} show that the SPGLMM has similar performance to GLMM ones in which three clusters are provided to the parametric model. This confirms that our semiparametric model is able to recognize the three clusters as efficiently as when we give this information in input to the model.
The case in which we run a GLMM with 10 groups performs slightly better, as expected, since the models have more flexibility to adapt to the differences between the groups.
Concerning the computation of the coefficients, in the boxplots in Figure \ref{fig:10glmer_3GLMM_3SPGLMM_coef_POI} we can appreciate that the true values are correctly identified in all the cases, with a slightly higher presence of outliers when a GLMM with 10 groups is fitted (panel (a)).

\begin{table}
\caption{Summary statistics of the GoF metrics estimates for DGP for Poisson response, with GLMM (10 groups and 3 clusters) and SPGLMM (3 clusters, $\alpha$-\textit{criterion}, $\alpha = 0.05$).}
\label{table:GoF_poisson}
\centering
\small
\begin{tabular}{@{}lcrcrrr@{}}
\toprule
&& & &\multicolumn{3}{c}{Quantile} \\
\cmidrule(lr){5-7}
Model &Metric  &
\multicolumn{1}{c}{Mean} &
Std. dev.&
\multicolumn{1}{c}{25\%} &
\multicolumn{1}{c}{50\%}&
\multicolumn{1}{c@{}}{75\%} \\
\midrule

\multirow{3}{*}{\textbf{GLMM}, 10 groups}  & \textit{MSE of responses}     & 4.0995   & 0.3502 & 3.8153 &4.0987  & 4.3132 \\
                    & \textit{MSE of log responses} & 0.2024   & 0.01269 & 0.1944 &0.2023  & 0.2099 \\
                    & \textit{Chi-Squared Error}    & 0.6873   & 0.0388 & 0.6620 &0.6764  & 0.7184 \\
 \addlinespace                          
\multirow{3}{*}{\textbf{GLMM}, 3 clusters} & \textit{MSE of responses}     & 4.1377 & 0.3519 & 3.8847 & 4.1328 & 4.3753 \\
                    & \textit{MSE of log responses} & 0.2044 & 0.0129 & 0.1962 & 0.2042 & 0.2131 \\
                    & \textit{Chi-Squared Error}    & 0.6962 & 0.0385 & 0.6704 & 0.6871 & 0.7239 \\
  \addlinespace                        
\multirow{3}{*}{\textbf{SPGLMM}, 3 clusters}& \textit{MSE of responses}  & 4.1382 & 0.3519 & 3.8775 & 4.1334 & 4.3764\\
                    & \textit{MSE of log responses} &  0.2044 & 0.0129 &  0.1963 & 0.2043 & 0.2127\\
                    & \textit{Chi-Squared Error}    & 0.6965 & 0.0384 & 0.6709 & 0.6870 & 0.7248\\
\bottomrule
\end{tabular}
\end{table}

%

\section{Conclusions}
\label{sec:concluding_remarks}

In this work, a novel statistical significance-based approach for implementing semi-parametric generalized linear mixed models with discrete random effects is proposed.
When dealing with hierarchical data, the so-called $\alpha$-\textit{criterion} induces a clustering of the groups in which observations are nested, having no priors on the number of clusters to be identified.
This approach results to be suitable and extremely useful when dealing with a high number of groups and reducing the dimensionality by identifying a latent structure at the group level is of interest. 
By setting a  confidence level $\alpha$, only clusters that are statistically different are identified within the iterations of a tailored EM algorithm.

This work enters into the literature about mixed-effects models with discrete random effects in which the state-of-the-art collapsing criterion of clusters was based on a discretional threshold - the \textit{t-criterion}. This approach is suitable if the user knows a priori the distance to be observed between clusters or if a huge computational effort is put into running the algorithm under different thresholds in order to identify either the 
sought number of clusters or an elbow on the average entropy. Even in these cases, the \textit{t-criterion} approach does not provide any insight about the statistical significance of the difference between the identified clusters. Besides these advantages, the simulation study in Section \ref{sec3} shows that the $\alpha$-\textit{criterion} performs better than the \textit{t-criterion} in most of the cases, also when the choice of the threshold is driven.

A further contribution of this work regards the generalization of mixed-effects models with discrete random effects to responses with law in the exponential family, in particular, binary and Poisson. This allows to enlarge the families of response variables that this type of models can handle, enriching the potential areas of applications.
When tested on real data, the proposed methodology achieves a good prediction performance holding on the comparison with the parametric version of the model, still performing the extra task of clustering the groups.

Some limitations of the work and, consequently, possible future directions regard the parameters initialization procedure and the definition of a clear objective function. Being the maximization step of the EM algorithm based on numerical approximations, the range of the parameters in which numerical methods look for the maximum plays a determinant role. Fitting a GLM within each group 
requires balanced responses within each group. Further research should be dedicated to the identification of a more straightforward and flexible initialization procedure. Regarding the objective function, we prove the increasing property of the likelihood when $M$ is fixed, but we do not define an objective function against which to assess whether the identified latent structure is the optimum one. In this perspective, the definition of an objective function that takes into account the likelihood and the model complexity would serve the purpose.

The analysis of illiteracy rates constitutes a simple and interpretable case study, but the method is applicable to different contexts. The identification of clusters of groups might be useful in the healthcare context to discover latent structures of patients with different levels of vulnerability standing on their repeated measurements or in the efficiency analysis context to evaluate different types of providers standing on the performance of their consumers or, again, in any situation in which dealing with clusters of groups is preferable that dealing with groups themselves. 
 

\begin{appendix}

\numberwithin{equation}{section}
\numberwithin{table}{section}
\numberwithin{figure}{section}

\section{Loglikelihood for generalized responses}
\label{app0}

We here express the loglikelihood in Eq. (\ref{L_c})
for the two special cases of Bernoulli and Poisson distributions, exploiting their canonical link function.

\subsection{Bernoulli case}
\label{app0Ber}
Bernoulli distribution models binary response variables.
Given the random effects $\mathbf{b}_i$, the binary responses $\mathbf{y}_1,...,\mathbf{y}_N$ are conditionally independent such that $\mathbf{y}_i | \mathbf{b}_i \sim \mathrm{Be}(\boldsymbol{\mu}_i)$
for $i=1,...,N$, where
 $  \text{logit}(\boldsymbol{\mu}_i) = \text{ln} \Big( \frac{\boldsymbol{\mu}_i}{1-\boldsymbol{\mu}_i} \Big)= \boldsymbol{\eta}_i = \mathbf{X}_i \boldsymbol{\beta} + \mathbf{Z}_i \boldsymbol{b}_i $.
Then $
    p(y_{ij}|\boldsymbol{\beta}, \mathbf{b}_i) = \frac{\mathrm{exp} ( y_{ij} \; \eta_{ij})}{ 1 + \mathrm{exp} (\eta_{ij} ) }
$
and the marginal loglikelihood is
$    \mathrm{ln}\,\mathcal{L}(\boldsymbol{\beta}, \mathbf{c}_1,...,\mathbf{c}_M | \mathbf{y}) =  \sum_{m=1}^M \omega_m \sum_{i=1}^N \sum_{j=1}^{n_i} y_{ij} \; \eta_{ijm} - \mathrm{ln} ( \; 1 + \mathrm{exp} ( \eta_{ijm} ) \; ) $.

\subsection{Poisson case}
\label{app0Poi}
Poisson distribution models counts as outcomes.
Given the random effects $\mathbf{b}_i$, the counts $\mathbf{y}_1,...,\mathbf{y}_N$ are conditionally independent such that $\mathbf{y}_i | \mathbf{b}_i \sim \mathrm{Poi}(\boldsymbol{\mu}_i)$
for $i=1,...,N$, where
    $\text{ln}(\boldsymbol{\mu}_i) = \boldsymbol{\eta}_i = \mathbf{X}_i \boldsymbol{\beta} + \mathbf{Z}_i \boldsymbol{b}_i $.
Then
    $p(y_{ij}|\boldsymbol{\beta}, \mathbf{b}_i) = 
    \frac{ \mathrm{exp}( y_{ij} \; \eta_{ij} ) \; \; \mathrm{exp}( \; - \mathrm{exp} (\eta_{ij} ) \; ) }{y_{ij}!}$
and the marginal loglikelihood is 
    $\mathrm{ln}\,\mathcal{L}(\boldsymbol{\beta}, \mathbf{c}_1,...,\mathbf{c}_M | \mathbf{y}) =  \sum_{m=1}^M \omega_m \sum_{i=1}^N \sum_{j=1}^{n_i} y_{ij} \; \eta_{ijm} - \mathrm{exp} (\eta_{ijm}) - \mathrm{ln}( y_{ij}! ) $.
\end{appendix} 



\pagebreak
\begin{center}
\textbf{\LARGE Supplementary Materials}
\end{center}

\setcounter{section}{0}
\renewcommand{\thesection}{S-\Roman{section}}

\renewcommand\thesection{S\arabic{section}}
\renewcommand\thesubsection{\thesection.\arabic{subsection}}
\numberwithin{equation}{section}
\numberwithin{table}{section}
\numberwithin{figure}{section}

\section{Proof of the increasing likelihood property}\label{appA}

Following the setting presented in \cite{azzimonti2013nonlinear} and \cite{masci2022semiparametric}, we prove the increasing likelihood property of the EM algorithm for SPGLMM, given a fixed number of clusters M.
We want to prove that $\mathcal{L}(\boldsymbol{\beta}^{(up)}, \mathbf{y}) \geq \mathcal{L}(\boldsymbol{\beta}, \mathbf{y}) $
where $\boldsymbol{\beta}^{(up)}$ is the updated fixed effect. From the marginal likelihood definition, we get:
\begin{equation}
\label{eqlnL}
    \mathrm{ln} \bigg[ \frac{\mathcal{L}(\boldsymbol{\beta}^{(up)}| \mathbf{y})} { \mathcal{L}(\boldsymbol{\beta}| \mathbf{y})} \bigg] = \sum_{i=1}^N  \mathrm{ln} \bigg[ \frac{p(\mathbf{y}_i|\boldsymbol{\beta}^{(up)})}{p(\mathbf{y}_i|\boldsymbol{\beta})} \bigg]
\end{equation}
so that
\begin{equation*}
    \mathrm{ln} \bigg[ \frac{p(\mathbf{y}_i|\boldsymbol{\beta}^{(up)})}{p(\mathbf{y}_i|\boldsymbol{\beta})} \bigg] = Q_i(\boldsymbol{\theta}^{(up)}, \boldsymbol{\theta}) - Q_i(\boldsymbol{\theta}, \boldsymbol{\theta})
\end{equation*}
Thanks to the convexity of the logarithm function and some algebraic passages, we obtain:
\begin{eqnarray}
\label{eqlnp}
 \mathrm{ln} \bigg[ \frac{p(\mathbf{y}_i|\boldsymbol{\beta}^{(up)})}{p(\mathbf{y}_i|\boldsymbol{\beta})} \bigg]   & = & 
 \mathrm{ln} \sum_{m=1}^M \frac{\omega_m^{(up)} \; p(\mathbf{y}_i|\boldsymbol{\beta}^{(up)}, \mathbf{c}_m^{(up)}) }{p(\mathbf{y}_i|\boldsymbol{\beta})} \\
 & = & \mathrm{ln} \sum_{m=1}^M 
 \bigg( \frac{\omega_m \; p(\mathbf{y}_i|\boldsymbol{\beta}, \mathbf{c}_m)}{p(\mathbf{y}_i|\boldsymbol{\beta})} \bigg) 
 \bigg( \frac{\omega_m^{(up)} \; p(\mathbf{y}_i|\boldsymbol{\beta}^{(up)}, \mathbf{c}_m^{(up)})}{\omega_m \; p(\mathbf{y}_i|\boldsymbol{\beta}, \mathbf{c}_m)}\bigg)
 \nonumber \\
 & \geq & \sum_{m=1}^M \bigg( \frac{\omega_m \; p(\mathbf{y}_i|\boldsymbol{\beta}, \mathbf{c}_m)}{p(\mathbf{y}_i|\boldsymbol{\beta})} \bigg) \; \mathrm{ln} \bigg( \frac{\omega_m^{(up)} \; p(\mathbf{y}_i|\boldsymbol{\beta}^{(up)}, \mathbf{c}_m^{(up)})}{\omega_m \; p(\mathbf{y}_i|\boldsymbol{\beta}, \mathbf{c}_m)}\bigg) \nonumber\\
 & = & \mathcal{Q}_i(\boldsymbol{\theta}^{(up)}, \boldsymbol{\theta}) - \mathcal{Q}_i(\boldsymbol{\theta}, \boldsymbol{\theta}) \nonumber
\end{eqnarray}
where 
\begin{equation*}
    \mathcal{Q}_i(\boldsymbol{\theta}^{(up)}, \boldsymbol{\theta}) = \sum_{m=1}^M \bigg( \frac{\omega_m \; p(\mathbf{y}_i|\boldsymbol{\beta}, \mathbf{c}_m)}{p(\mathbf{y}_i|\boldsymbol{\beta})} \bigg) \; \mathrm{ln} \big( \omega_m^{(up)} \; p(\mathbf{y}_i|\boldsymbol{\beta}^{(up)}, \mathbf{c}_m^{(up)}) \big)
\end{equation*}
and 
\begin{equation*}
    \mathcal{Q}_i(\boldsymbol{\theta}, \boldsymbol{\theta}) = \sum_{m=1}^M \bigg( \frac{\omega_m \; p(\mathbf{y}_i|\boldsymbol{\beta}, \mathbf{c}_m)}{p(\mathbf{y}_i|\boldsymbol{\beta})} \bigg) \; \mathrm{ln} \big( \omega_m \; p(\mathbf{y}_i|\boldsymbol{\beta}, \mathbf{c}_m) \big) \; .
\end{equation*}
Defining $ \mathcal{Q}(\boldsymbol{\theta}^{(up)}, \boldsymbol{\theta}) = \sum_{i=1}^N \mathcal{Q}_i(\boldsymbol{\theta}^{(up)}, \boldsymbol{\theta})$ and $  \mathcal{Q}(\boldsymbol{\theta}, \boldsymbol{\theta}) = \sum_{i=1}^N \mathcal{Q}_i(\boldsymbol{\theta}, \boldsymbol{\theta})$, 
we get a lower bound for the quantity of interest thanks to Eqs. (\ref{eqlnL}) and (\ref{eqlnp}):
\begin{equation*}
    \mathrm{ln} \bigg[ \frac{p(\mathbf{y}_i|\boldsymbol{\beta}^{(up)})}{p(\mathbf{y}_i|\boldsymbol{\beta})} \bigg] \geq \mathcal{Q}(\boldsymbol{\theta}^{(up)}, \boldsymbol{\theta}) -  \mathcal{Q}(\boldsymbol{\theta}, \boldsymbol{\theta}) \; .
\end{equation*}
We must now show that 
    $ \mathcal{Q}(\boldsymbol{\theta}^{(up)}, \boldsymbol{\theta}) \geq \mathcal{Q}(\boldsymbol{\theta}, \boldsymbol{\theta}) $.
This can be proved by defining $\boldsymbol{\theta}^{(up)}$ as
\begin{equation*}
    \boldsymbol{\theta}^{(up)} = \underset{\tilde{\boldsymbol{\theta}}}{\mathrm{arg\;max}} \;  \mathcal{Q}(\tilde{\boldsymbol{\theta}}, \boldsymbol{\theta}) \quad \forall \; \boldsymbol{\theta} \; \text{  fixed}.
\end{equation*}
    Defining $W_{im}$ as in Eq. (5), we get
\begin{eqnarray*}
 \mathcal{Q}(\tilde{\boldsymbol{\theta}}, \boldsymbol{\theta})
 & = & \sum_{i=1}^N \sum_{m=1}^M 
    \bigg( \frac{\omega_m \; p(\mathbf{y}_i|\boldsymbol{\beta}, \mathbf{c}_m)}{\sum_{k=1}^M \omega_k \; p(\mathbf{y}_i|\boldsymbol{\beta}, \mathbf{c}_k)} \bigg) \; \mathrm{ln} \big( \tilde{\omega}_m \; p(\mathbf{y}_i|\tilde{\boldsymbol{\beta}}, \tilde{\mathbf{c}}_m) \big)  \\
 & = & \sum_{i=1}^N \sum_{m=1}^M W_{im} \; \mathrm{ln} \big( \tilde{\omega}_m \; p(\mathbf{y}_i|\tilde{\boldsymbol{\beta}}, \tilde{\mathbf{c}}_m) \big) \nonumber \\
  & = & \sum_{i=1}^N \sum_{m=1}^M W_{im} \; \mathrm{ln}\; \tilde{\omega}_m + \sum_{i=1}^N \sum_{m=1}^M W_{im} \; p(\mathbf{y}_i|\tilde{\boldsymbol{\beta}}, \tilde{\mathbf{c}}_m)  \nonumber \\
  & = & \mathcal{J}_1 (\tilde{\omega}_1,...,\tilde{\omega}_M)  + \mathcal{J}_2(\tilde{\boldsymbol{\beta}}, \tilde{\mathbf{c}}_1,..., \tilde{\mathbf{c}}_M) \nonumber
\end{eqnarray*}
The functionals $\mathcal{J}_1$ and $\mathcal{J}_2$ can be maximized separately.
Eq. (4) is obtained by maximinzing $\mathcal{J}_1$ in closed form.
More specifically, the functional $\mathcal{J}_1$ could be rewritten as follows:
\begin{eqnarray*}
    \mathcal{J}_1 (\tilde{\omega}_1,...,\tilde{\omega}_M) 
    & = & \sum_{m=1}^{M-1} \sum_{i=1}^N W_{im}  \; \mathrm{ln}\; \tilde{\omega}_m + \sum_{i=1}^N W_{iM} \; \mathrm{ln} \; \tilde{\omega}_M \\
    & = & \sum_{m=1}^{M-1} \sum_{i=1}^N W_{im}  \; \mathrm{ln}\; \tilde{\omega}_m + \sum_{i=1}^N W_{iM} \; \mathrm{ln} \bigg( 1- \sum_{m=1}^{M-1} \tilde{\omega}_m \bigg) \nonumber 
\end{eqnarray*}
and, imposing the gradient equal to zero, we obtain:
\begin{equation*}
    \dfrac{\partial \mathcal{J}_1}{\partial \tilde{\omega}_m} =
    \frac{\sum_{i=1}^N W_{im}}{\tilde{\omega}_m} - \frac{\sum_{i=1}^N W_{iM}}{1- \sum_{m=1}^{M-1} \tilde{\omega}_m}
    = \frac{\sum_{i=1}^N W_{im}}{\tilde{\omega}_m} - \frac{\sum_{i=1}^N W_{iM}}{\tilde{\omega}_M} = 0 \quad \forall \; m = 1,..., M-1
\end{equation*}
that is equivalent to $
    \frac{\sum_{i=1}^N W_{im}}{\tilde{\omega}_m} =  \frac{\sum_{i=1}^N W_{ik}}{\tilde{\omega}_k} \quad \forall \; m,k = 1,..., M $. Summing on $k = 1,..., M$, since $\sum_{m=1}^M W_{im} = 1$, we obtain Eq. (4).

On the other hand, the update in Eq. (6) for $\boldsymbol{\beta}$ and $(\mathbf{c}_1,...,\mathbf{c}_M)$ is obtained maximizing the functional $\mathcal{J}_2$ through numerical approximations.

\section{Methodology: further details}

\subsection{Parameters initialization}
\label{sec:23}

SPGLMM starts by considering $N$ discrete masses and iteratively reduces the number by grouping them into $M<N$ clusters. 
However, like most clustering algorithms, SPGLMM is highly sensitive to the initial placement of the starting points.
To avoid any biases in the clustering procedure, it is important to initialize the $N$ discrete masses in a robust way. 
To achieve this, we fit a GLM within each group $i$, for $i=1,...,N$ and obtain $N$ distinct models.
We then extract the $Q$ parameters distributions, composed of the estimates of the intercept and slopes of the $N$ models.
The $M=N$ starting values of the $h^{\text{th}}$ random coefficient, 
for $h=1,...,Q$,
are obtained by computing the first and third quartiles $q_{0.25,h}$ and $q_{0.75,h}$ from the $N$-dimensional distribution composed by the coefficients in each of the $N$ fitted models corresponding to the $h^{\text{th}}$ random coefficient. Inspired by the boxplot whiskers definition, we construct the interval $r_h = [r_{min,\,h} \;\; ; \;\; r_{max,\,h}] = [q_{0.25,\,h}-1.5 \cdot IQR_h \;\; ; \;\; q_{0.75,\,h}+1.5 \cdot IQR_h]$, where $IQR_h$ stands for the interquartile range $q_{0.75,\,h}-q_{0.25,\,h}\;$. From the interval $r_h$, we then randomly select $M=N$ support points as follows: 
   $ r_{min,\,h} + (r_{max,\,h} - r_{min,\,h}) \cdot \mathcal{U}(0,1)$,
where $\mathcal{U}(0,1)$ stands for the uniform distribution between $0$ and $1$.
At the initial step, the weights $\hat{\omega}_1^{(0)},...,\hat{\omega}_M^{(0)}$ are uniformly distributed on the $M=N$ support points: each of them is initialized at $1/N$.

Following the method proposed in
\cite{masci2019semiparametric}, 
if the number of groups $N$ is extremely large (e.g., $N>100$) the algorithm could be slowed down.
For this reason, in the computation of $\hat{\mathbf{c}}_h^{(0)}$, $h=1,...,Q$,
we can alternatively fix a boundary $\tilde{N}$ and randomly select 
$M=\tilde{N} < N$ (instead of $N$) support points within the $r_h$ interval and uniformly distribute the weights on these $\tilde{N}$ support points.
The fixed coefficients are also initialized by exploiting the $N$ distinct fitted models.
Specifically, the starting value of the $p^{\text{th}}$ fixed coefficient $\hat{\beta}_p^{(0)}$ , for $p=1,...,P$, is assigned to the median of the $N$-dimensional distribution composed by the $p^{\text{th}}$ fixed coefficients. 

\subsection{Beyond the support reduction: a check on the weights}
\label{sec:25}
Beyond the support point reduction, a check on the weights is performed.
A sketch of the pseudo-code related to this check is addressed in Algorithm \ref{pseudo_code_weights}. 
Specifically, at each iteration $k$, if 
the $m^{\text{th}}$ cluster has weight $\hat{\omega}_m^{(k)}$ equal to zero (i.e., the m$^{\text{th}}$ column of $\hat{\mathbf{W}}^{(k)}$ contains all zero elements), the cluster $m^{\text{th}}$ is removed and the total number of clusters is updated accordingly. Moreover, the remaining weights are then renormalized in such a way that they sum up to $1$, as follows:
\begin{equation}
\label{reparametrize_weights}
    \hat{\omega}_m^{new} = \frac{\hat{\omega}_m^{old}}{S_{\hat{\omega}}} \; \forall m=1,...,M_{new} \text{ where } S_{\hat{\omega}} = \sum_{m=1}^{M_{new}} \hat{\omega}_m^{old} \;.
\end{equation}

In addition, 
when convergence is reached (see Section \ref{sec:26} for the convergence conditions), 
we remove the empty clusters, i.e., the support points to which no groups are associated (see $\tilde{l}_i$ in Eq. (9) for the association between groups and clusters).
The remaining weights are then renormalized as in Eq. (\ref{reparametrize_weights}).
If no mass points are deleted, 
the algorithm can terminate; otherwise, at least another iteration is required, in order to make the algorithm update the mass points.



\begin{algorithm}
\caption{Support reduction: final check on weights}
\label{pseudo_code_weights}
\footnotesize{
  \Function{Check weights($conv1$,$M$,$N$,\texttt{K1},$k$, $\hat{\mathbf{W}}^{(k)}$,$(\hat{\mathbf{c}}_1^{(k)},..., \hat{\mathbf{c}}_M^{(k)})$,$ (\hat{\omega}_1^{(k)},..., \hat{\omega}_M^{(k)})$ 
  )}{
  \If(\Comment{we delete clusters with $\hat{\omega}_m^{(k)}=0$}){M is not 1}{
    Delete null columns of $\hat{\mathbf{W}}^{(k)}$ (if any)\;
    $M \gets M-$ number of deleted columns (if any)\;
    Reparametrize weights as in Eq. (9)\;
 }
\If{$conv1$ is 1 or $k\geq \text{\texttt{K1}}$}{
    \For{i=1,...,N} {
    $\tilde{l}_i \gets$ $\tilde{l}_i$ as in Eq. (9)\;
    }
    $flag \gets 1$\;
    \For (\Comment{we delete empty clusters 
    }) 
    {m=1,...,M} 
    {
    \If{
    None of $[\tilde{l}_1,...,\tilde{l}_N]$ is equal to m }{
        Delete the mass point $\hat{\mathbf{c}}_m^{(k)}$ satisfying such condition\;
        $M \gets M-1$\;
        Reparametrize weights as in Eq. (\ref{reparametrize_weights})\;
        $flag \gets 0$\;
        $conv1 \gets 0$\;
    }
    }
    \If{$flag$ is 1}{ 
       $conv2 \gets 1$\;
    }
}
\Return{$M$, $(\hat{\mathbf{c}}_1^{(k)},..., \hat{\mathbf{c}}_M^{(k)})$, $ (\hat{\omega}_1^{(k)},..., \hat{\omega}_M^{(k)})$, $conv2$, $conv1$}
}}
\end{algorithm}

\subsection{Convergence criteria}
\label{sec:26}
At each iteration $k$ of the SPGLMM, the updated number of mass points $M$ is estimated: the EM algorithm computes the updates of both fixed and random effects 
within each $it^{\text{th}}$ \textit{sub}-iteration of $k$,
until either a maximum number of \textit{a priori} fixed iterations \texttt{itmax} is reached (worst case of not convergence)
or all the differences of fixed and random parameters estimates at two consecutive iterations $it$ and $it-1$ are smaller than fixed tolerance values \texttt{tF} and \texttt{tR}, respectively (i.e., if
$|\hat{c}_{mh}^{(it)}-\hat{c}_{mh}^{(it-1)}|<\text{\texttt{tR}} \; \forall m=1,...,M, \; \forall h=1,...,Q$ and $|\hat{\beta}_p^{(it)}-\hat{\beta}_p^{(it-1)}|<\text{\texttt{tF}} \; \forall p=1,...,P$).



The support points reduction based on $\alpha$\textit{-criterion} starts after the first \texttt{K2} iterations 
since before the estimates get stabilized.
When all the differences between the estimates of the parameters at two consecutive iterations $k$ and $k-1$ are smaller than fixed tolerance values (i.e. if $|\hat{c}_{mq}^{(k)}-\hat{c}_{mq}^{(k-1)}|<\text{\texttt{tR}} \; \forall m=1,...,M, \; \forall h=1,...,Q$ and $|\hat{\beta}_p^{(k)}-\hat{\beta}_p^{(k-1)}|<\text{\texttt{tF}} \; \forall p=1,...,P$) and no more confidence regions overlap, the value of the dummy variable \texttt{conv1} switches from 0 to 1.
When convergence is reached (i.e., \texttt{conv1} is 1 or after a given number of iterations \texttt{K1}), the empty clusters, if present, are removed (see Section \ref{sec:25} and Algorithm \ref{pseudo_code_weights}). 
The algorithm stops when both \texttt{conv1} is 1 and no more empty clusters are present, or, in the case of no convergence, when $k\geq \texttt{K}$, where \texttt{K} is an \textit{a priori} fixed threshold.

\section{Fast Ellipsoid Intersection Test}\label{app1}
Consider the scalar function $K: (0,1)\rightarrow \mathbb{R}$
\begin{equation*}
    K(s) := 1- (\hat{\mathbf{c}}_l^{(k)} - \hat{\mathbf{c}}_m^{(k)})^{\text{T}} \bigg( \frac{1}{1-s} \text{var}(\hat{\mathbf{c}}_m^{(k)})^{-1} + \frac{1}{s} \text{var}(\hat{\mathbf{c}}_l^{(k)})^{-1}\bigg) (\hat{\mathbf{c}}_l^{(k)} - \hat{\mathbf{c}}_m^{(k)})
\end{equation*}
where $^{\text{T}}$ stands for the transpose.

The Fast Ellipsoid Intersection Test states that
$CR_{1-\alpha}(\hat{\mathbf{c}}_l^{(k)})$ intersects $CR_{1-\alpha}(\hat{\mathbf{c}}_m^{(k)})$ if and only if $K(s)\geq0 \; \forall s\in(0,1)$, as proven in Proposition 2 of \cite{gilitschenski2012robust}.
Therefore, through fast one-dimensional optimization methods, we can minimize $K(s)$ on $(0,1)$ and check the sign at minimizing point $s^*$. One of the following three cases applies: (i) if $K(s^*)>0$, the ellipsoids intersect; (ii) if $K(s^*)<0$ they do not intersect while (iii) if $K(s^*)=0$ the ellipsoids touch their boundaries.

The computation of $K(s)$ can be simplified exploiting the generalized eigenvalues $\lambda_h$ and generalized eigenvectors $\phi_h$ of the generalized eigenvalue problem (\cite{parlett1998symmetric}), for $h=1,...,Q$, defined by $\text{var}(\hat{\mathbf{c}}_m^{(k)}) \; \phi_h = \lambda_h \; \text{var}(\hat{\mathbf{c}}_l^{(k)}) \; \phi_h$ such that $\phi_h^{\text{T}} \; \text{var}(\hat{\mathbf{c}}_l^{(k)}) \; \phi_h = 1$.
In fact, it is well established that $\boldsymbol\Phi^{\text{T}} \; \text{var}(\hat{\mathbf{c}}_m^{(k)})  \; \boldsymbol\Phi = \boldsymbol\Lambda$ and $\boldsymbol\Phi^{\text{T}}  \; \text{var}(\hat{\mathbf{c}}_l^{(k)}) \; \boldsymbol\Phi = \mathbf{I}$, where $\boldsymbol\Phi$ is the $Q\times Q$ matrix whose $h^{\text{th}}$ column is the vector $\phi_h$, $\boldsymbol\Lambda$ is the $Q\times Q$ diagonal matrix whose $h^{\text{th}}$ diagonal entry is $\lambda_h$ and $\mathbf{I}$ is the $Q\times Q$ identity matrix.
Since $\text{var}(\hat{\mathbf{c}}_m^{(k)})$ and $\text{var}(\hat{\mathbf{c}}_l^{(k)})$ in the basis of generalized eigenvectors are diagonal, through algebraic manipulation, we can rewrite $K(s)$ as follows:
\begin{equation}
\label{conf_reg_criteria_intersection}
    K(s) := 1- \frac{1}{\chi_{Q}^2(1-\alpha)} \sum_{h=1}^Q v_h^2 \frac{s(1-s)}{1+s(\lambda_h-1)}
\end{equation}
where $v_h:= \boldsymbol \Phi^{\text{T}} (\hat{\mathbf{c}}_{mh}^{(k)} - \hat{\mathbf{c}}_{lh}^{(k)})$.

\section{Case study results for the Bernoulli response}
\label{bernoulli_case_study}

For the case of a Bernoulli response, SPGLMM aims to predict the presence of low-achieving students 
identifying  clusters of countries.
Starting from the data pre-processed as in Section 3.1, we create the variable Bernoulli-distributed \texttt{Y\_BIN\_MATH}, which assumes value 1 if \texttt{Y\_MATH} is strictly greater than 2$\%$, 0 otherwise. 2$\%$ is chosen because it is the minimum threshold ensuring that in each country (the groups) there are both 0s and 1s.
We get 6125 schools with class 1 ($>2\%$ of low-achieving students) and 6495 schools with class 0 ($\leq2\%$ of low-achieving students).
By assuming the model formulations described in Section 3.2, we now consider the response $y$ as \texttt{Y\_BIN\_MATH} and the canonical link function as $g(\cdot) = \text{logit}(\cdot)$.

Similarly to the Poisson case, results for Bernoulli response are addressed in Table \ref{Bernoulli_casestudy_random_intercept_slope}. 
We get $\hat{M}_{0.05} = \hat{M}_{0.10} = 10$ and $\hat{M}_{0.01} = 8$. By comparing the SPGLMM estimates with a parametric GLMM as explained in Section 3.2, we appreciate that the means of the random intercepts $b_i$ in each cluster are slightly lower in absolute value than the estimates for SPGLMM. Anyhow, we observe huge coherence between the two models.
In addition, the random intercept obtained for $\alpha=0.01$ in correspondence of the clusters $1$ and $2$ - that we denote with $\hat{c}_{1+2}$ for simplicity -
turns out to be a weighted mean between $\hat{c}_{1}$ and $\hat{c}_{2}$; the same happens for $\hat{c}_{7+8}$ (i.e., the random intercept obtained for $\alpha=0.01$ in correspondence of the clusters $7$ and $8$).
Moreover, given the negative sign of $\beta_1$, we can observe that the school size is inversely proportional to the probability of the presence of more than $2\%$ of low-achieving students: the higher the value of \texttt{SCHSIZE}, the lower the estimated probability. Similarly, the higher the value of \texttt{avg\_ESCS\_std}, the lower the estimated probability (see the negative sign of $\beta_2$), even if \texttt{avg\_ESCS\_std} has a much lower impact than \texttt{SCHSIZE} (the former slope is $\beta_2 = -0.62$ compared to the latter one of  $\beta_1 = -3.26$).
In general, we can conclude that also for the fixed slopes, SPGLMM and GLMM provide coherent results in the estimates, the standard errors and the p-values (estimated through likelihood-ratio test).

\begin{table*}
\caption{SPGLMM estimates for Bernoulli response and comparison with GLMM output.}
\label{Bernoulli_casestudy_random_intercept_slope}
\resizebox{\textwidth}{!}{%
\begin{NiceTabular}{c c c c c c}
\CodeBefore
  \rowcolors{2}{gray!15}{white}
  \columncolor{white!15}{1,1}
\Body
\toprule
\multicolumn{2}{c}{\textbf{Coeff. estimates}} 
& \multicolumn{3}{c}{\textbf{SPGLMM}} & \multicolumn{1}{c}{\textbf{GLMM}}\\
\cmidrule(lr){3-5} 
& & \multicolumn{1}{c}{$\alpha=0.01$} & \multicolumn{1}{c}{$\alpha=0.05$}
& \multicolumn{1}{c}{$\alpha=0.10$} &  
\\
\midrule
\multirow{10}{*}{$\hat{\mathbf{c}}$ $[\hat{\omega}]$} 

& \makecell{$\hat{c}_1$ [0.02] \\ $\hat{c}_2$ [0.10]} & -2.747 (0.078) & \makecell{-3.533 (0.320) \\ -2.679 (0.080)} & \makecell{-3.533 (0.320) \\ -2.680 (0.080)} & \makecell{ -3.415\\ -2.635} \\

& $\hat{c}_{3}$ [0.12]  & -2.091 (0.069) & -2.088 (0.070)              & -2.088 (0.070)  & -2.029 \\
& $\hat{c}_{4}$ [0.08]                 & -1.591 (0.078) & -1.590 (0.078)              & -1.591 (0.078) &-1.562 \\

& $\hat{c}_{5}$ [0.14] & -0.984 (0.054) & -0.984 (0.054)              & -0.984 (0.054 & -0.954  \\
& $\hat{c}_{6}$ [0.04] & -0.536 (0.103) & -0.563 (0.106)              & -0.564 (0.106) & -0.470  \\

& \makecell{$\hat{c}_7$ [0.06] \\ $\hat{c}_8$ [0.18]} & 0.318 (0.059) & \makecell{0.022 (0.105)\\ 0.437 (0.070)}   & \makecell{0.022 (0.105) \\ 0.437 (0.069)} & \makecell{-0.020 \\ 0.448} \\   
& $\hat{c}_9$ [0.14]              & 1.239 (0.075) & 1.245 (0.076)                      & 1.245 (0.076) &1.162 \\
& $\hat{c}_{10}$ [0.12]                 & 2.237 (0.137) & 2.237 (0.137)                       & 2.237 (0.137) &2.057  \\
\cmidrule(lr){1-6}
\multirow{2}{*}{$\hat{\boldsymbol{\beta}}$} & $\hat{\beta}_1$        
&  -3.262 (0.052) ***  & -3.261 (0.052) *** & -3.261 (0.052) ***  & -3.223 (0.073) *** \\
& $\hat{\beta}_2$  
&  -0.618 (0.028) ***  & -0.620 (0.028) *** & -0.620 (0.028) ***  & -0.623 (0.029) *** \\
\bottomrule
\end{NiceTabular}
}

\begin{tablenotes}
      \small
      \item Notes: The estimated random intercepts $\hat{\mathbf{c}}$ are presented in increasing order, together with their respective weights $\hat{\omega}$ in brackets, as well as the fixed effects $\hat{\boldsymbol{\beta}}$ for both SPGLMM (with $\alpha=0.01, \, 0.05, \, 0.10)$ and GLMM (for each row of $\hat{c}_m$, the average of the $\hat{b}_i$s in each cluster $m$ is reported). 
      In parenthesis, the standard error is computed by square rooting the inverse of the Fisher Information Matrix.
      For $\hat{\boldsymbol{\beta}}$, the p-value is estimated by means of likelihood-ratio test (* p-value $< 0.1$; ** p-value $< 0.01$; *** p-value $< 0.001$). 
    \end{tablenotes}
\end{table*}

\begin{figure}[!t]
\centering
\caption{Caterpillar plots representing the comparison between the 50 random intercepts estimated by GLMM and the clusters obtained through SPGLMM  for $\alpha=0.01$ and $0.05, \, 0.10$ with Bernoulli response.}
    \centering \includegraphics[width=16cm]
    {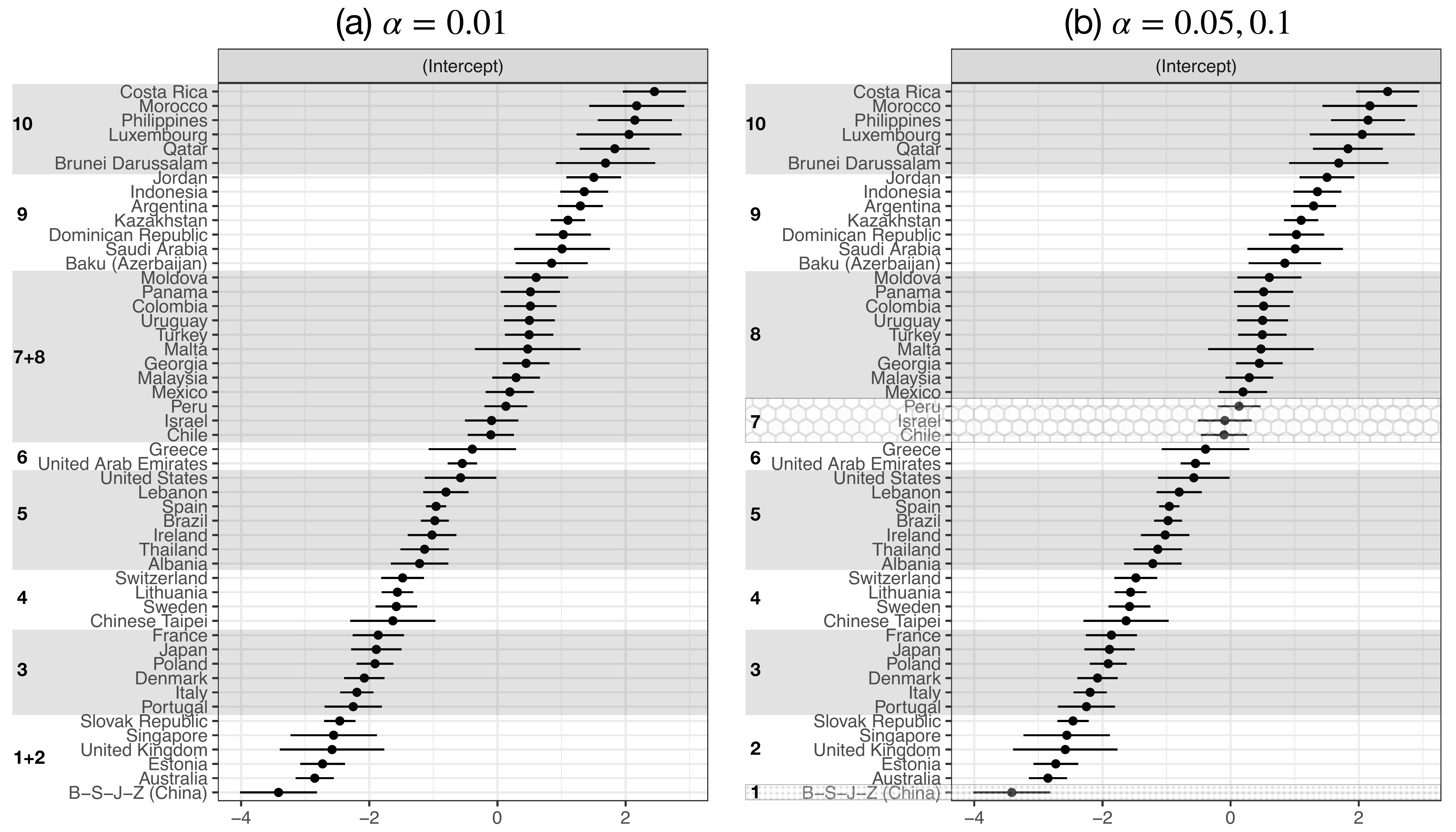}\\
    \footnotesize
Notes: To ease the comparison with Table \ref{Bernoulli_casestudy_random_intercept_slope}, the same table colours are used in panel (a) to highlight the clusters of countries. In panel (b), clusters 1 and 7 are highlighted with different texture.
\label{fig:Ber_random_intercept}
\end{figure}

In Figure \ref{fig:Ber_random_intercept}, both in panels (a) and (b), we display the caterpillar plot for the random intercepts (together with their confidence intervals) of the 50 countries, obtained through parametric GLMM. 
On each of the two panels, we highlight the identified clusters of countries, both for $\alpha=0.01$ in panel (a) and $\alpha=0.05, \, 0.10$ in panel (b).   
Results can be interpreted as follows: the lower the estimated random intercept for a cluster, the less likely (with respect to the average) the presence of at least 2\% of low-achieving students in mathematics in the schools of the countries of that cluster.

\begin{figure}
\centering
\caption{Choropleth map of the clusters of countries identified by the random intercepts in SPGLMM with $\alpha=0.01$ for Bernoulli response.}
\vspace{-0.5cm}
\includegraphics[width=14cm]{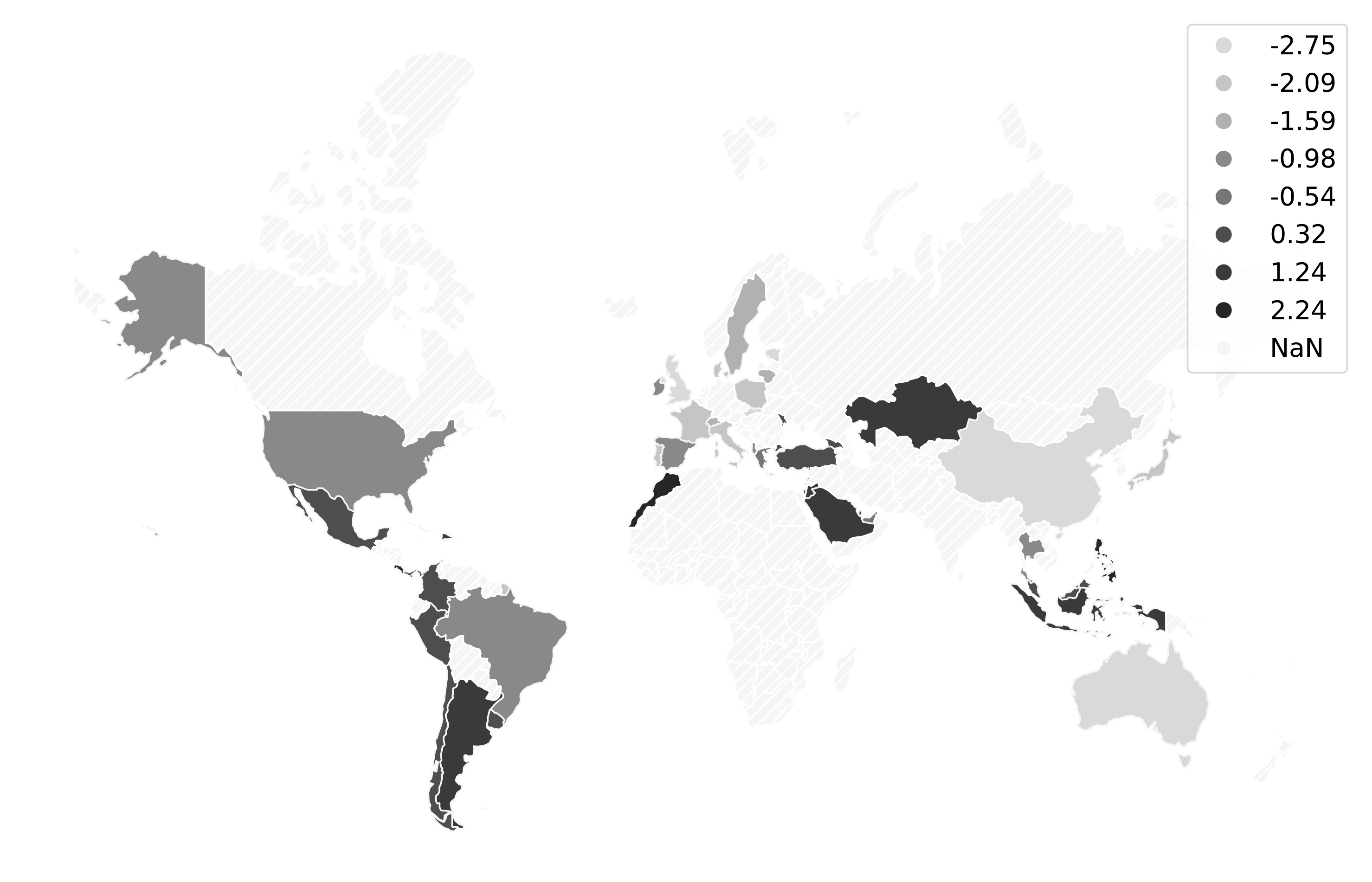} \\
\footnotesize
Notes: Countries represented with the same colour belong to the same cluster. The
lighter, the lower the random intercept.
Light grey-striped countries are the ones for which the survey was not performed, or
which were presenting missing values.
    \label{fig:Ber_world_8}
\end{figure}

For better visualisation of the clusters of countries identified by the SPGLMM, we highlight with the same shade of grey on the map in Figure \ref{fig:Ber_world_8} the countries identified by the same random intercept (i.e. the countries in the same cluster), for $\alpha=0.01$. We notice that B-S-J-Z (China) and Australia, net of the other features, decrease the probability of innumeracy presence (more specifically, decrease the probability of having at least $2\%$ of low-achieving students in mathematics). After them, the European countries decrease the probability with less impact, while in the Americas the probability of innumeracy presence increases.

\begin{figure}
\centering
\caption{ROC curves for the \textbf{Bernoulli response}, for the SPGLMM ($\alpha=0.05$) and GLMM.}
    \subfloat{{\includegraphics[width=7.8cm]{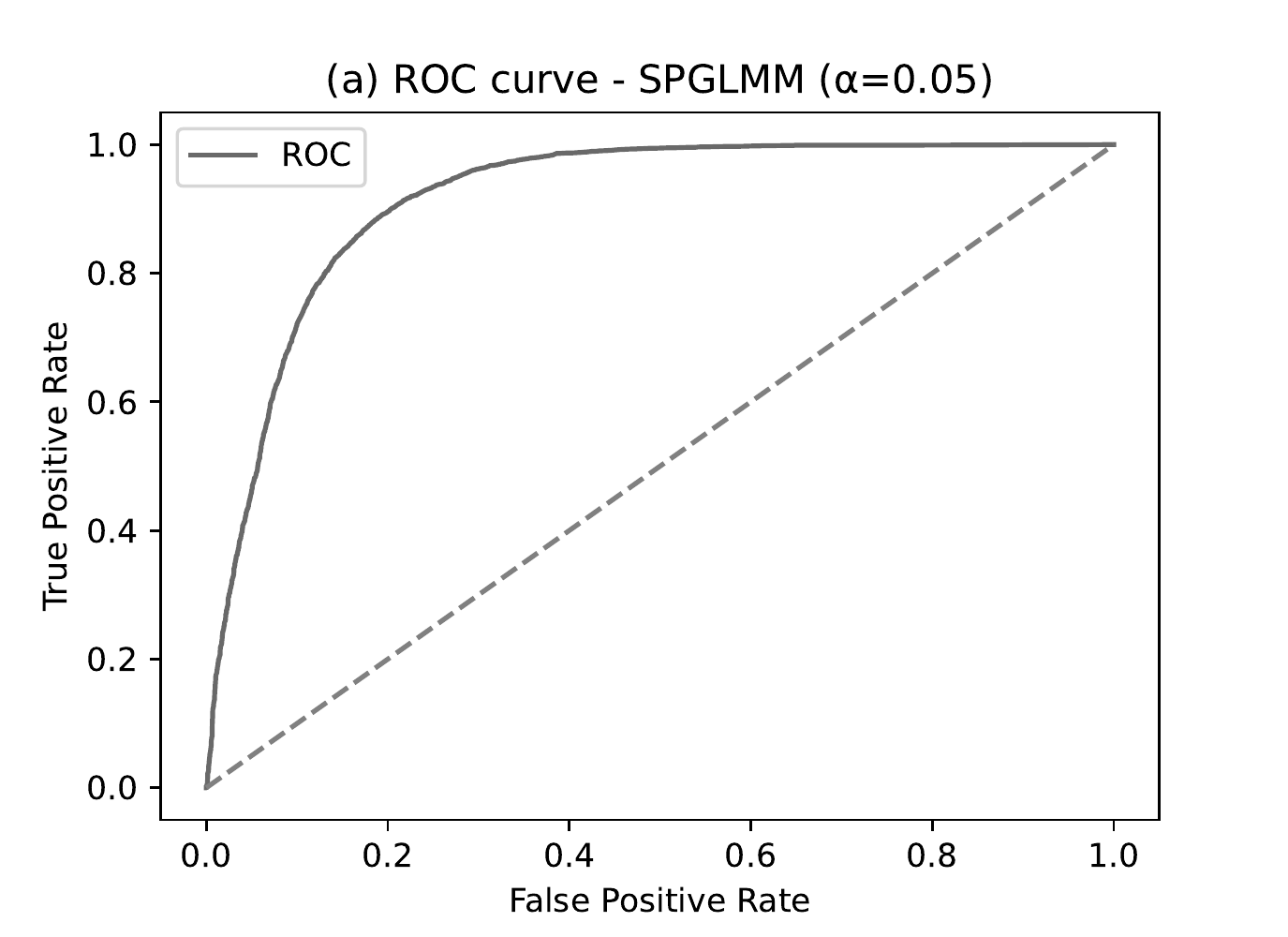} }}
    \,\,\,
    \subfloat{{\includegraphics[width=7.8cm]{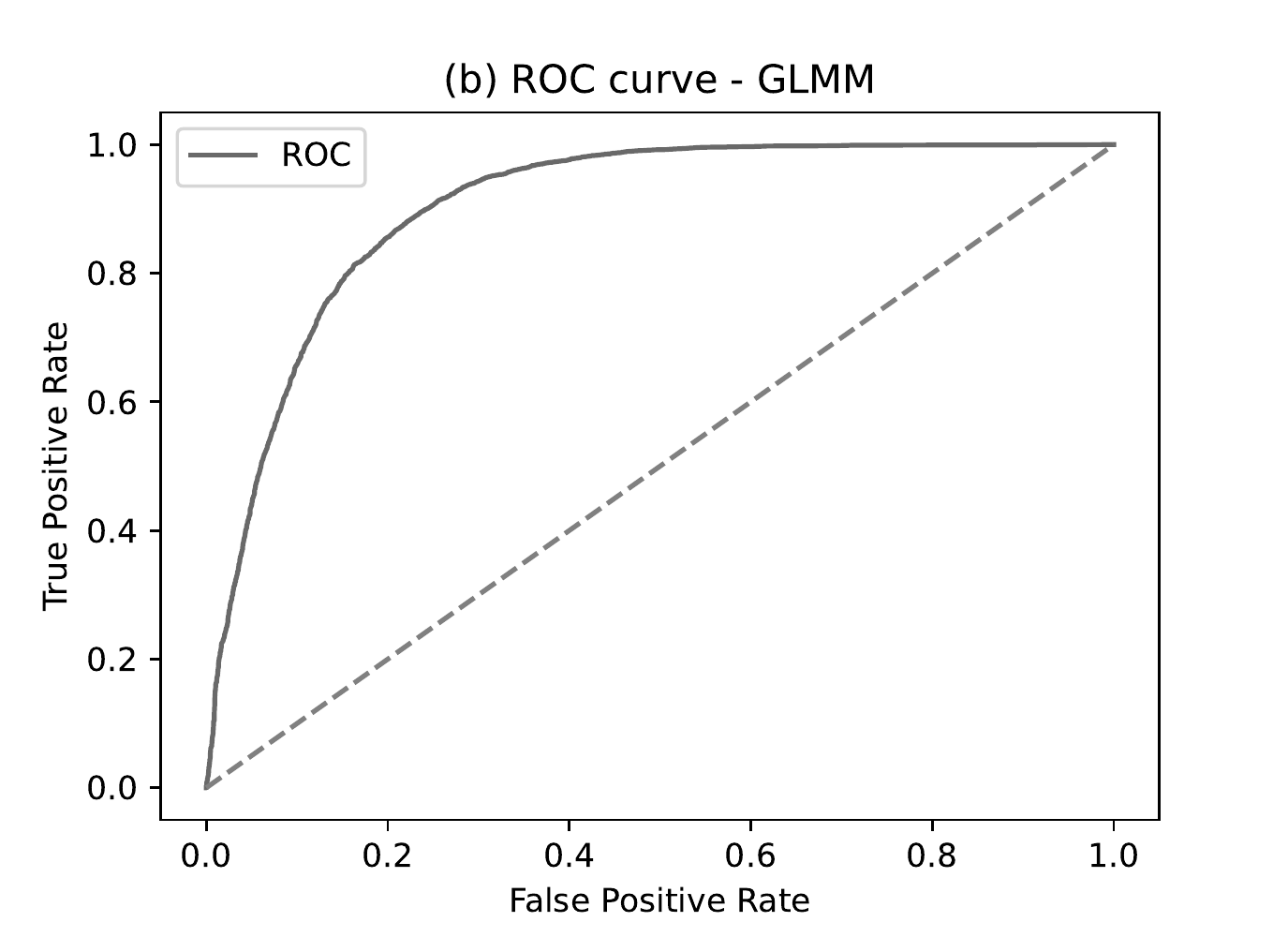} }} \\
\label{fig:ROC_curves}
\end{figure}

For the Goodness of Fit (GoF) evaluation, we use the same data we used to train the models since our main purpose is to compare on equal terms the predictive power of SPGLMM and GLMM. 
In Figure \ref{fig:ROC_curves} the Receiver Operating Characteristic (ROC) curve is displayed both for the SPGLMM (with $\alpha=0.05$) and GLMM.
Moreover, in Table \ref{tab:GOF_BER_case_study} we report for each case of the SPGLMM and for the GLMM, the Area Under Curve (AUC), the optimal identified threshold and the Sensitivity, Specificity and Accuracy computed by assuming the chosen threshold.
The two methods reveal similar predictive performances. Both the ROC curves and the results in the table drive us to the conclusion that the SPGLMM with Bernoulli response does not underperform with respect to the parametric classical GLMM. Furthermore, it also provides as output a clustering of the hierarchies (i.e., the countries in our case study), revealing the inner structure the model assumes.

\begin{table}
\caption{GoF metrics estimates 
for the case study with Bernoulli response, fitted via SPGLMM and GLMM.}
\label{tab:GOF_BER_case_study}
\centering
\begin{tabular}{@{}lcccc@{}}
\toprule
& \multicolumn{3}{c}{\textbf{SPGLMM}} & \textbf{GLMM} \\
\cline{2-4}
& 
\multicolumn{1}{c}{$\alpha = 0.01$} &
\multicolumn{1}{c}{$\alpha = 0.05$} &
\multicolumn{1}{c}{$\alpha = 0.10$} & \\
\midrule
\textit{AUC}             & 0.9158  & 0.9162 & 0.9162 & 0.9165  \\
\textit{Chosen threshold}& 0.4435  & 0.4397 & 0.4397 & 0.4523  \\
\hline
\textit{Sensitivity}     & 0.8143  & 0.8128 & 0.8127 & 0.8171  \\
\textit{Specificity}     & 0.8851  & 0.8869 & 0.8869 & 0.8807  \\         
\textit{Accuracy}        & 0.8476  & 0.8475 & 0.8475 & 0.8473  \\
\bottomrule
\end{tabular}
\end{table}

\section{Simulation study for the Bernoulli response}
\label{simBern}
\begin{figure}
\centering
\caption{
Barplot for the frequency a certain number of clusters is identified over 500 runs, across different values of $alpha$ 
and $t$, 
for DGPs (i), (ii) and (iii) for Bernoulli response with one fixed slope.}
    \subfloat[\centering DGP (i) - $\alpha$-\textit{criterion}]{
    {\includegraphics[width=7.7cm]{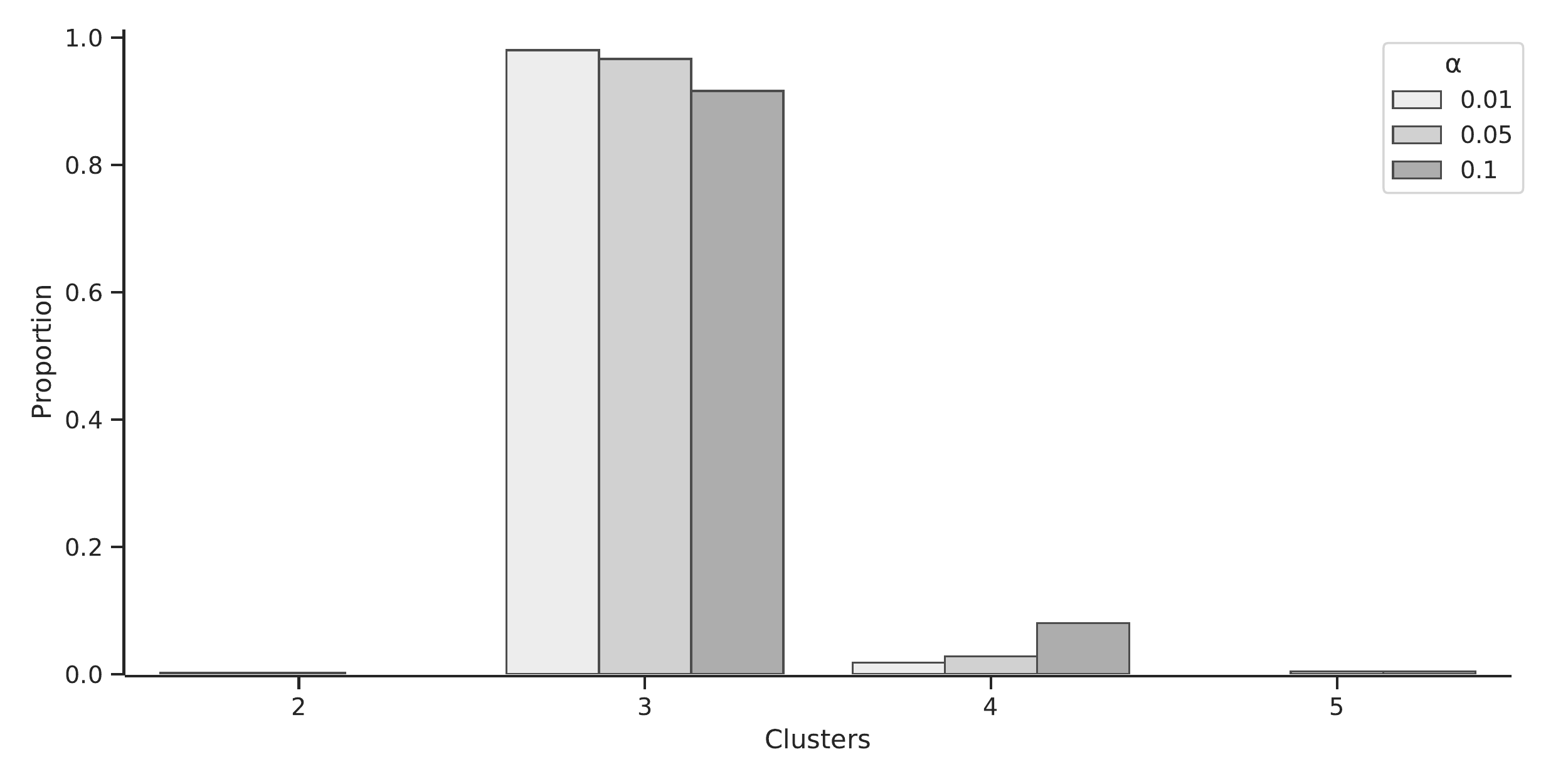} }
    } 
    \subfloat[\centering DGP (i) - \textit{t-criterion}]{
    {\includegraphics[width=7.7cm]{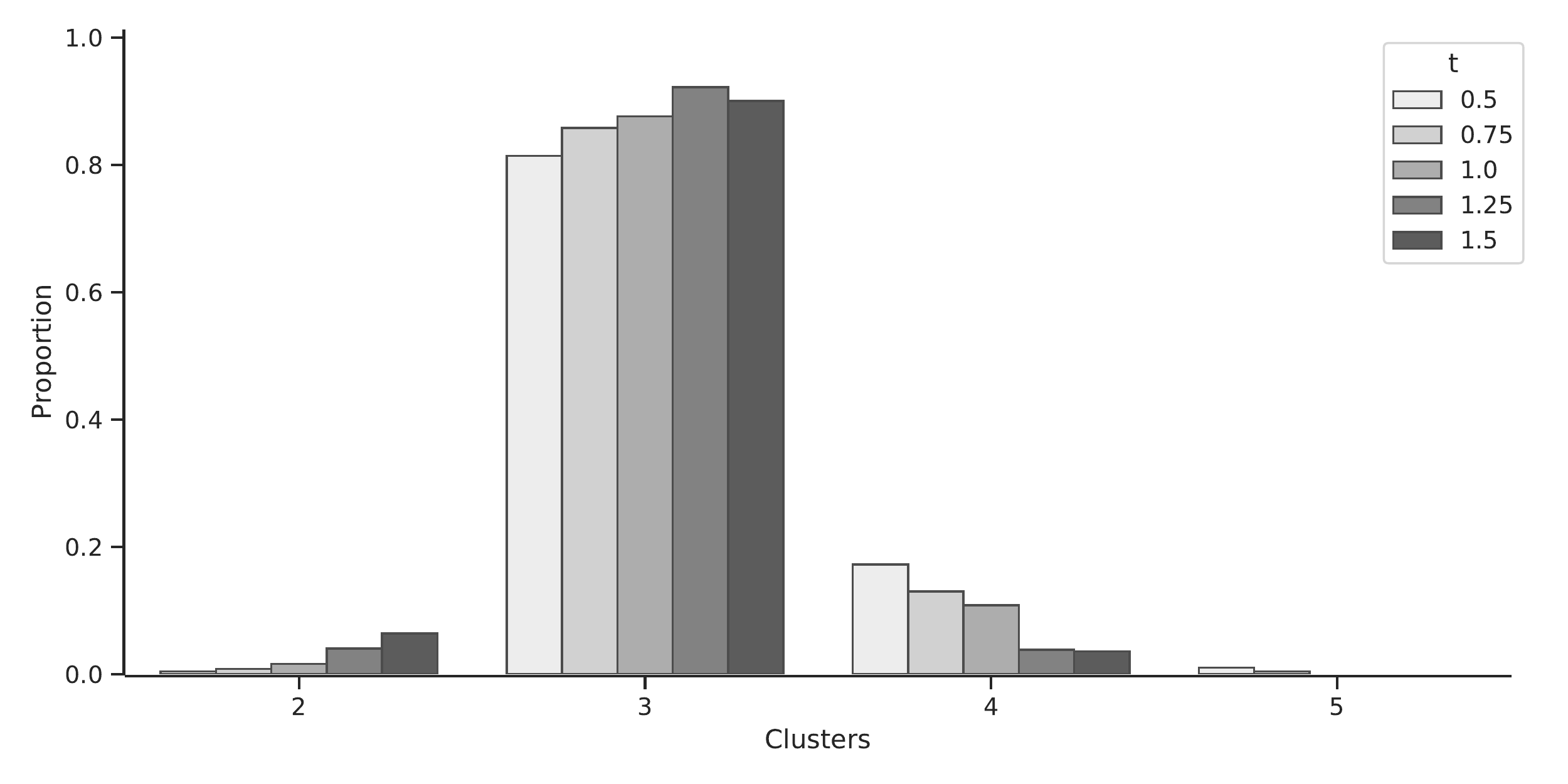} }
    }
    \\
    \subfloat[\centering DGP (i) - $\alpha$-\textit{criterion}]{
    {\includegraphics[width=7.7cm]{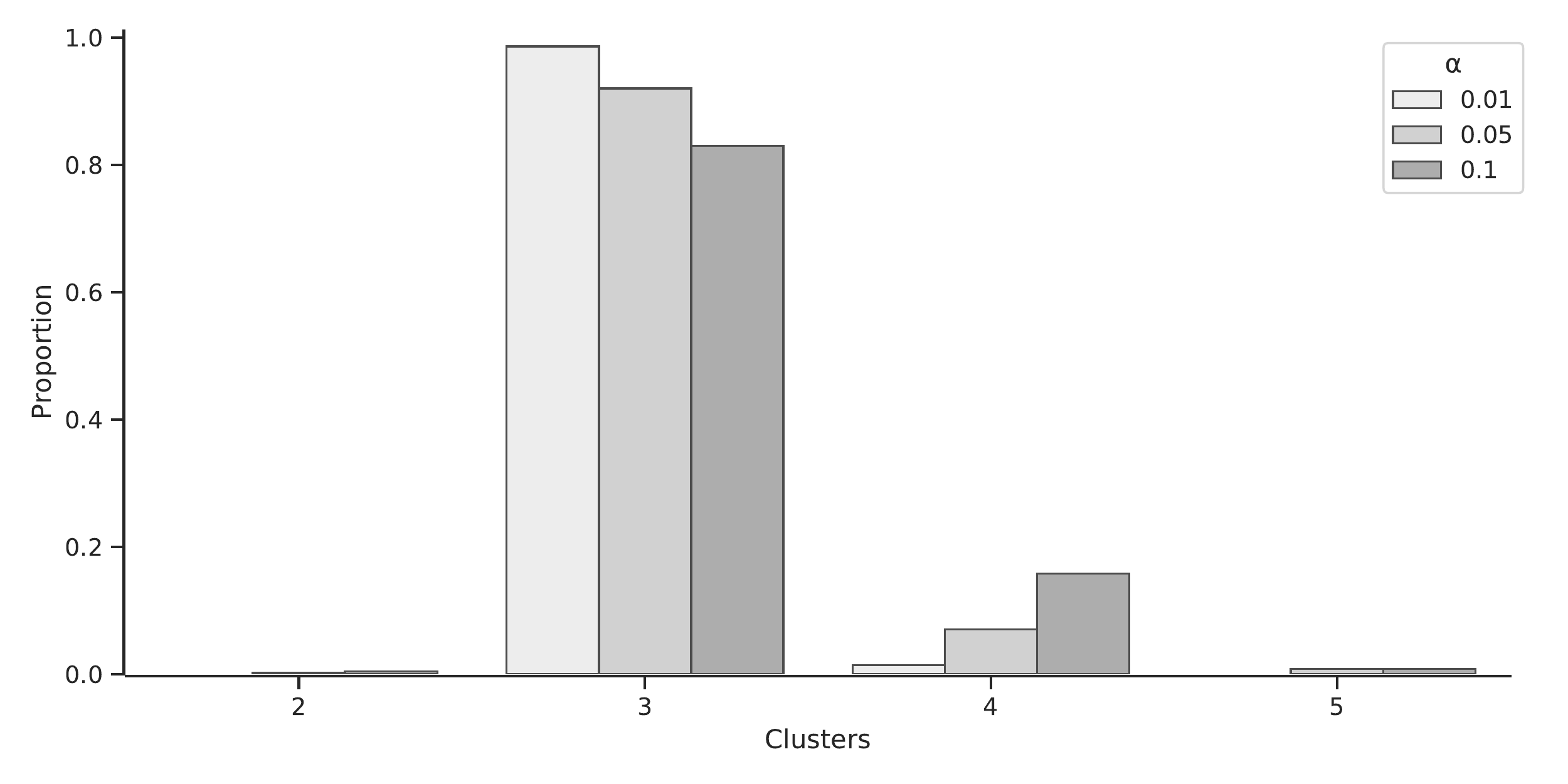} }
    } 
    \subfloat[\centering DGP(ii) - \textit{t-criterion}]{
    {\includegraphics[width=7.7cm]{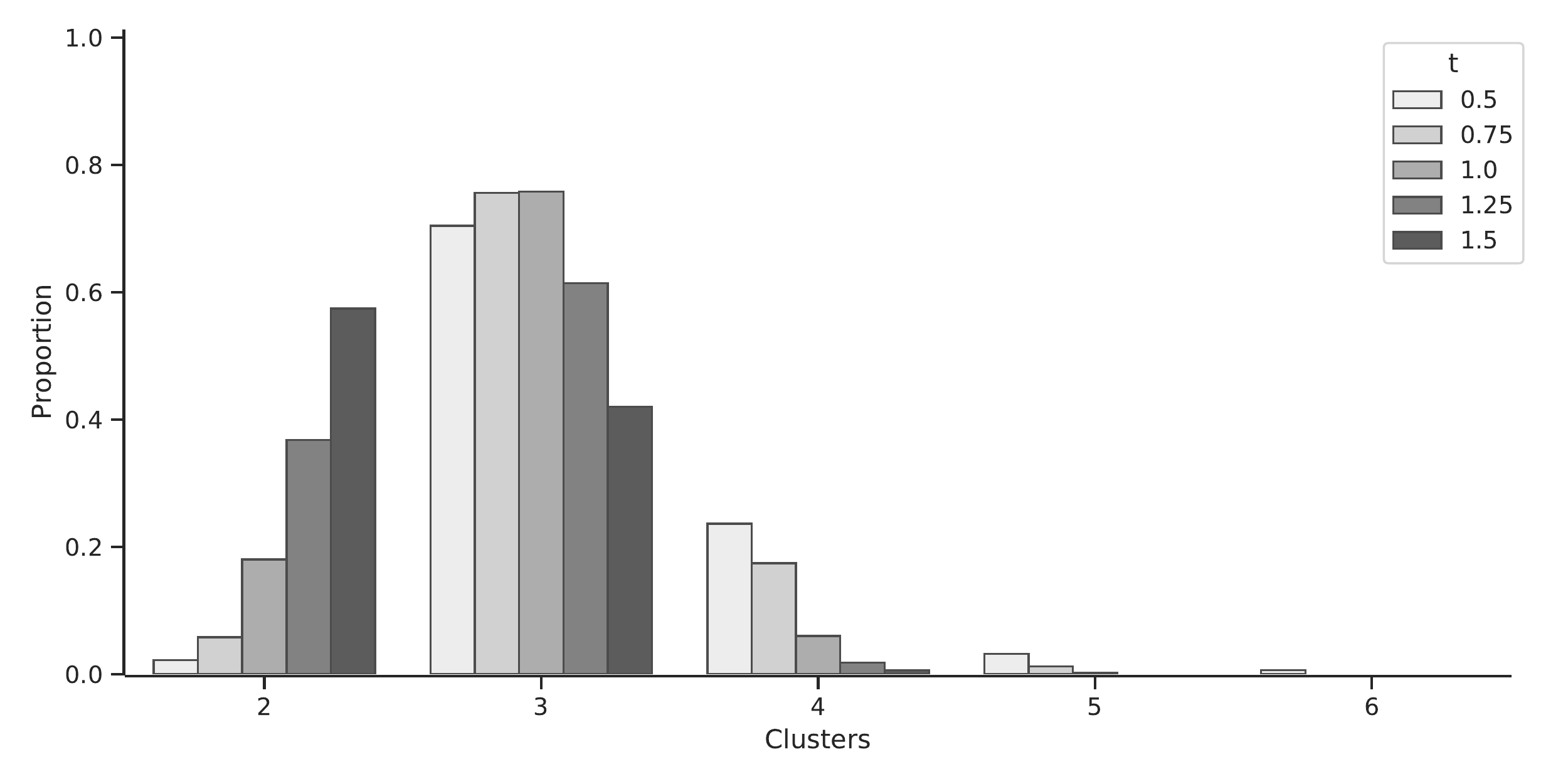} }
    }
    \\
    \subfloat[\centering DGP (iii) - $\alpha$-\textit{criterion}]{
    {\includegraphics[width=7.7cm]{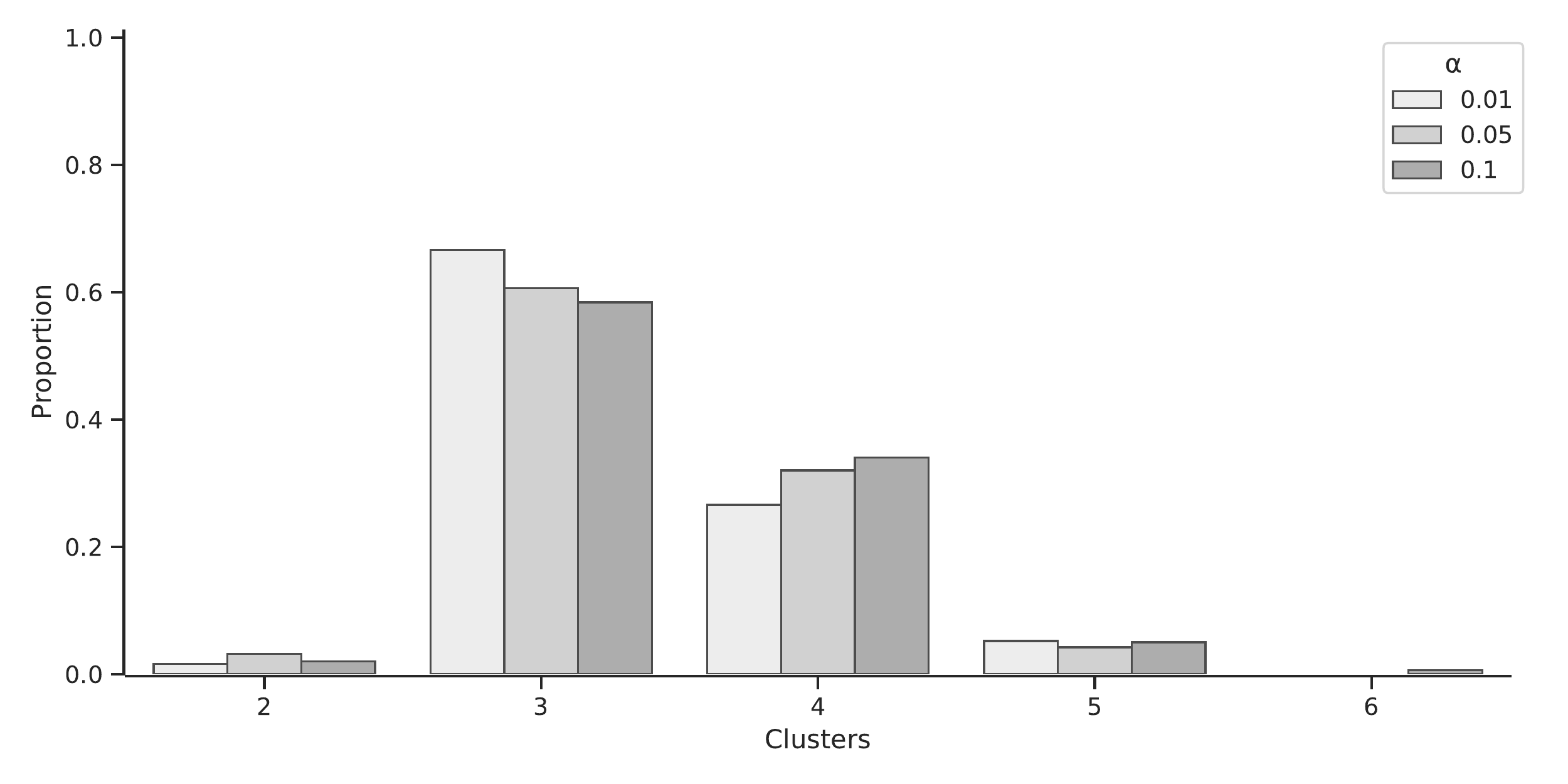} }
    } 
    \subfloat[\centering DGP (iii) - \textit{t-criterion}]{
    {\includegraphics[width=7.7cm]{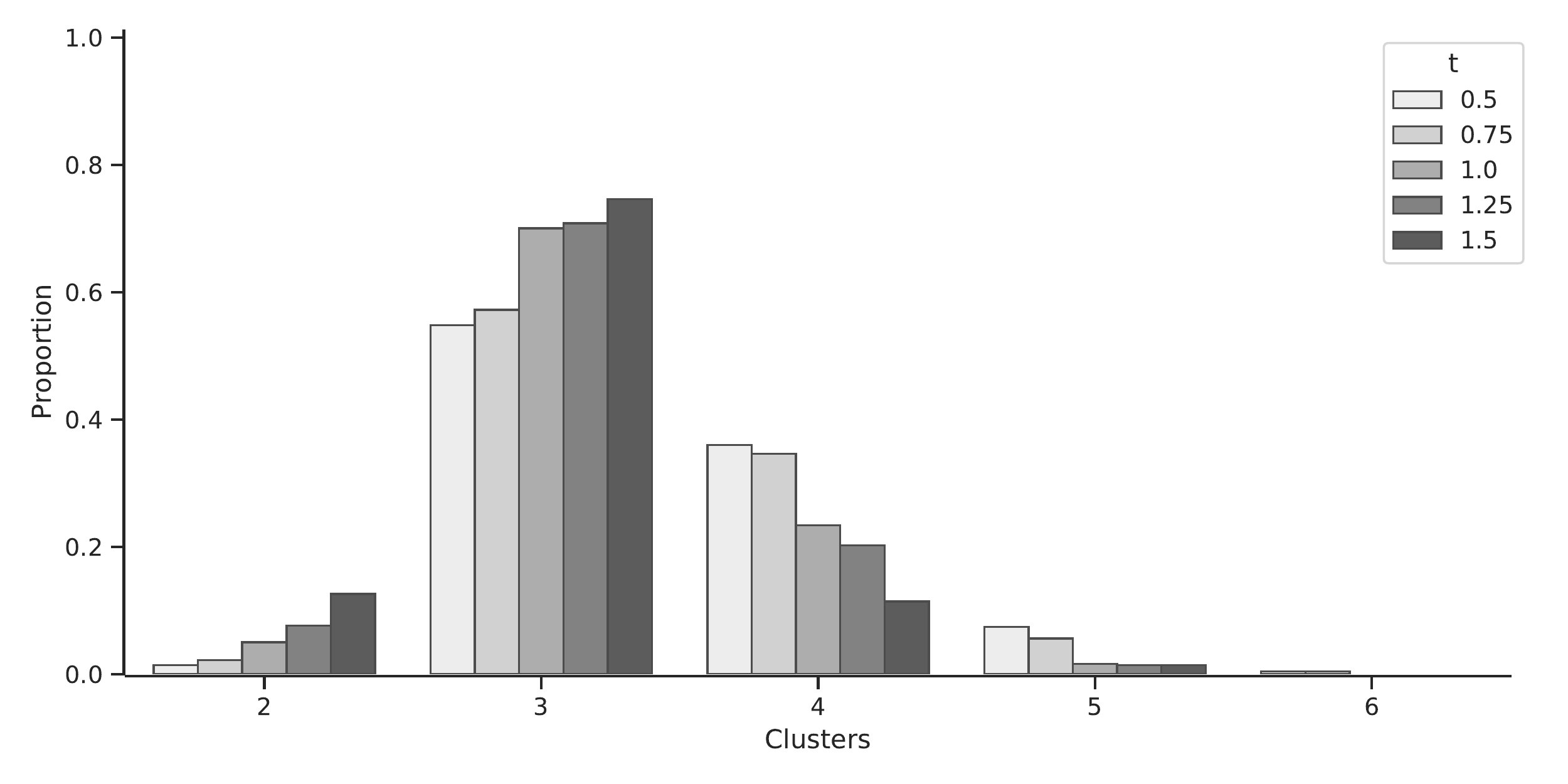} }
    }
    \\
    \footnotesize
Notes: to ease the comparison, the proportions on y-axis are reported on the same scale. The number of identified clusters is reported on the x-axis. In panels (a), (c), and (e), results across $\alpha = 0.01, \, 0.05, \, 0.1$ are represented with different shadows of grey (see legend in the right-up corner). Similarly, in panel (b) results across $t=0.25, \, 0.5, \, 0.75, \, 1, \, 1.25$ are addressed.
\label{fig:proportions_ber}
\end{figure}

\begin{figure}
\centering
\caption{Boxplots for the random intercept $\mathbf{c}_1$ distribution for the DGP (i) for Bernoulli response, with one fixed slope.}
    \subfloat[\centering $\mathbf{c}_1$ (2 groups) - $\alpha$-\textit{criterion}]{
    {\includegraphics[width=7.7cm]{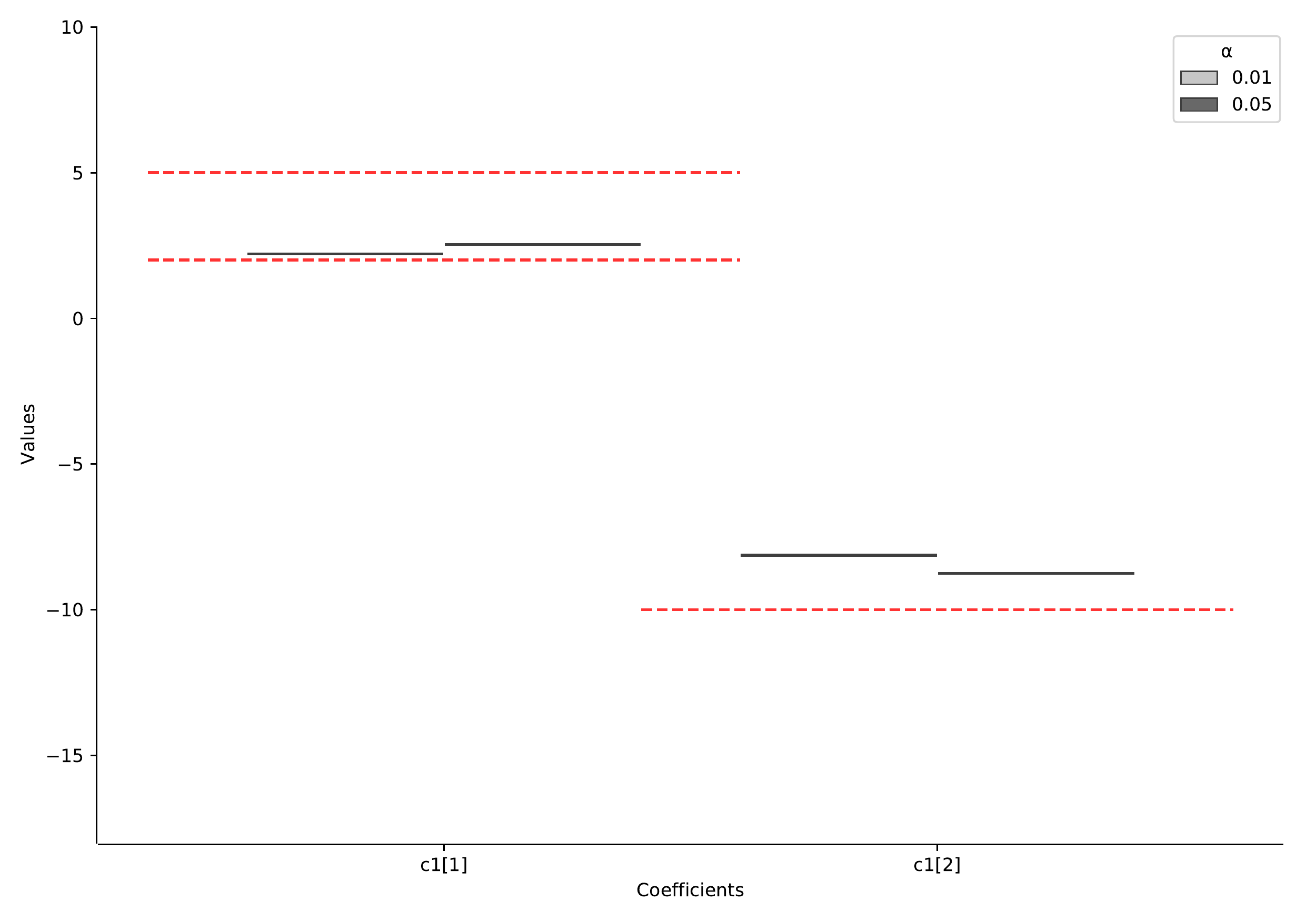} }
    } 
    \subfloat[\centering $\mathbf{c}_1$ (2 groups) - \textit{t}-\textit{criterion}]{
    {\includegraphics[width=7.7cm]{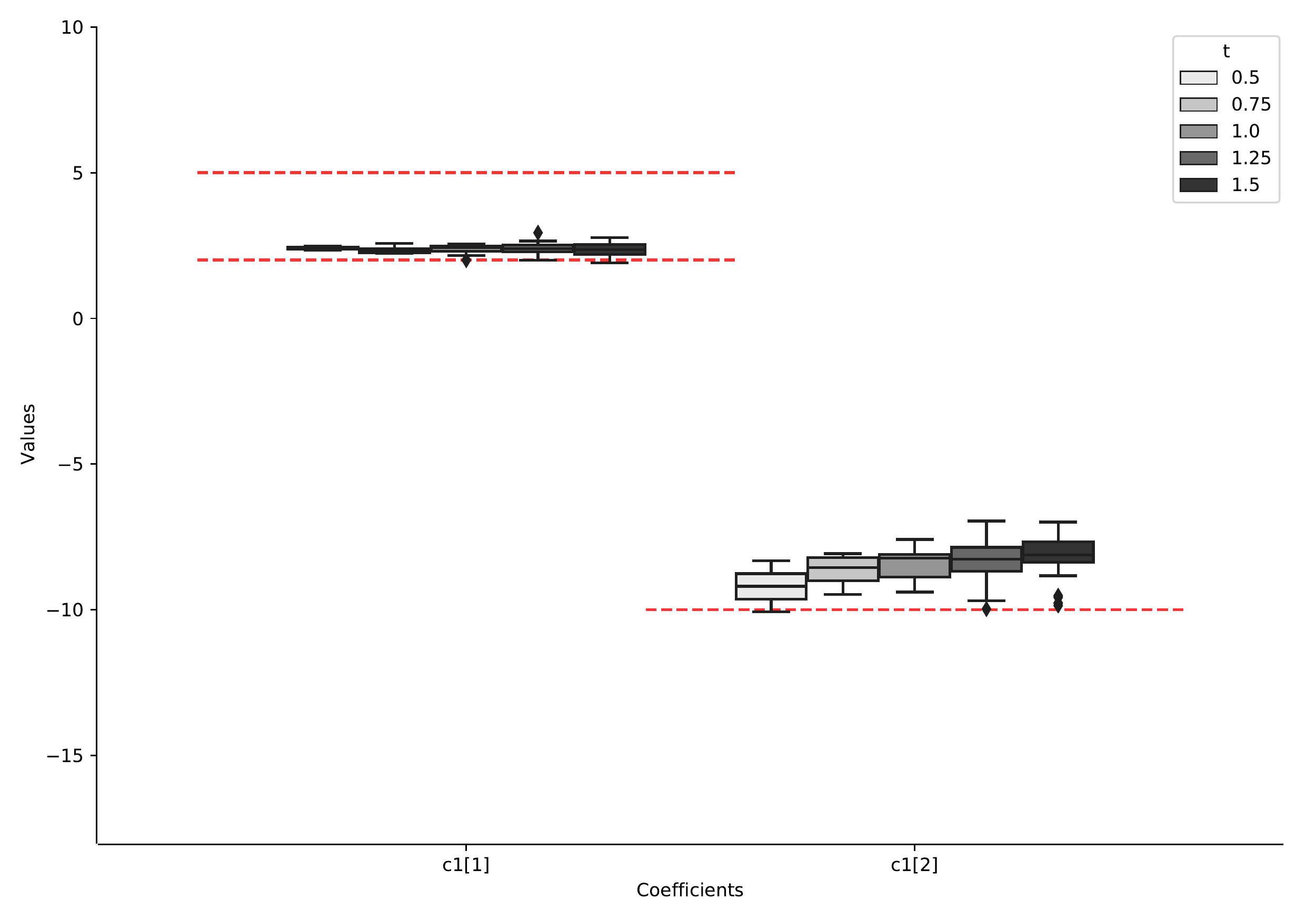} }
    }
    \\
    \subfloat[\centering $\mathbf{c}_1$ (3 groups) - $\alpha$-\textit{criterion}]{
    {\includegraphics[width=7.7cm]{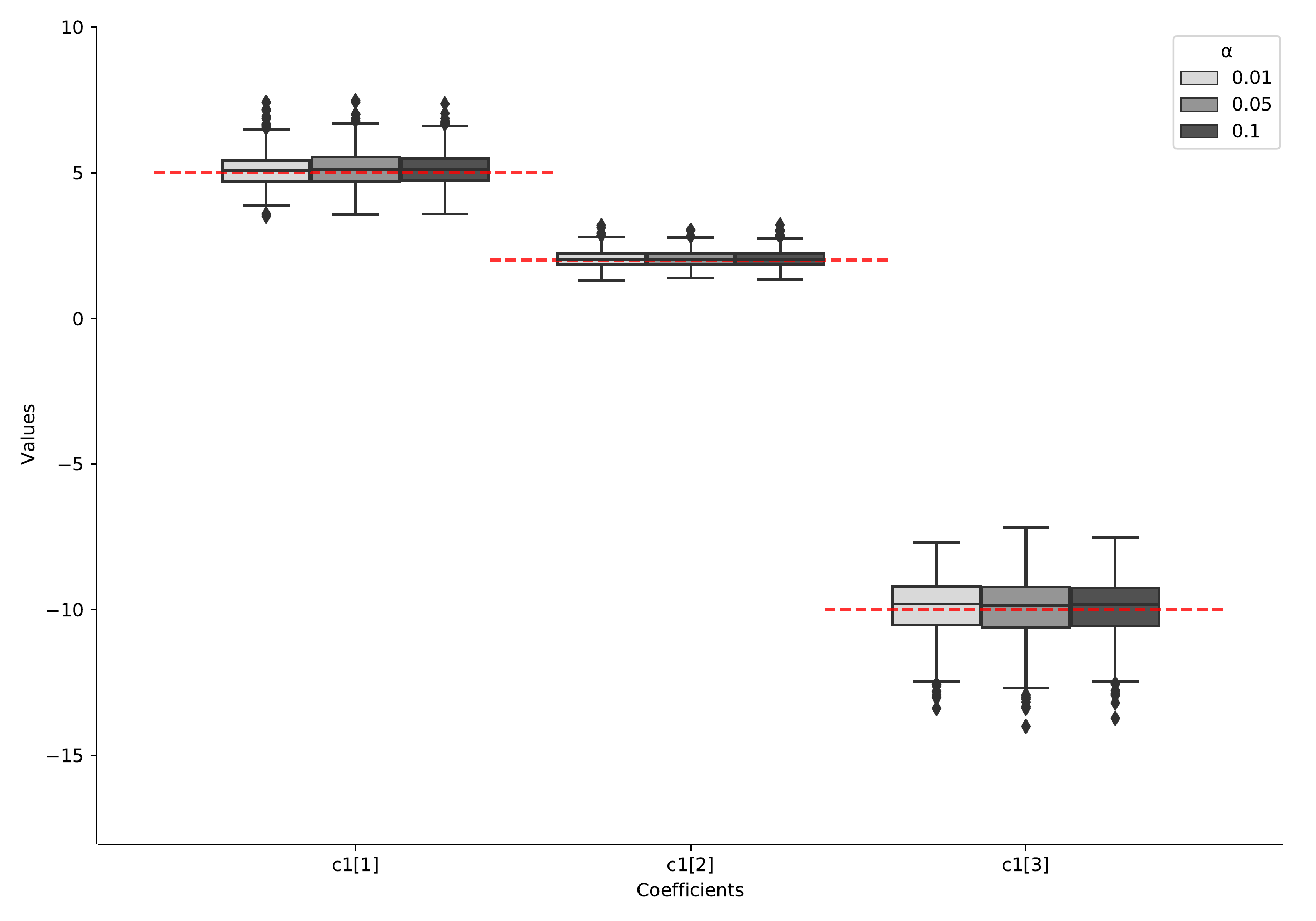} }
    } 
    \subfloat[\centering $\mathbf{c}_1$ (3 groups) - \textit{t}-\textit{criterion}]{
    {\includegraphics[width=7.7cm]{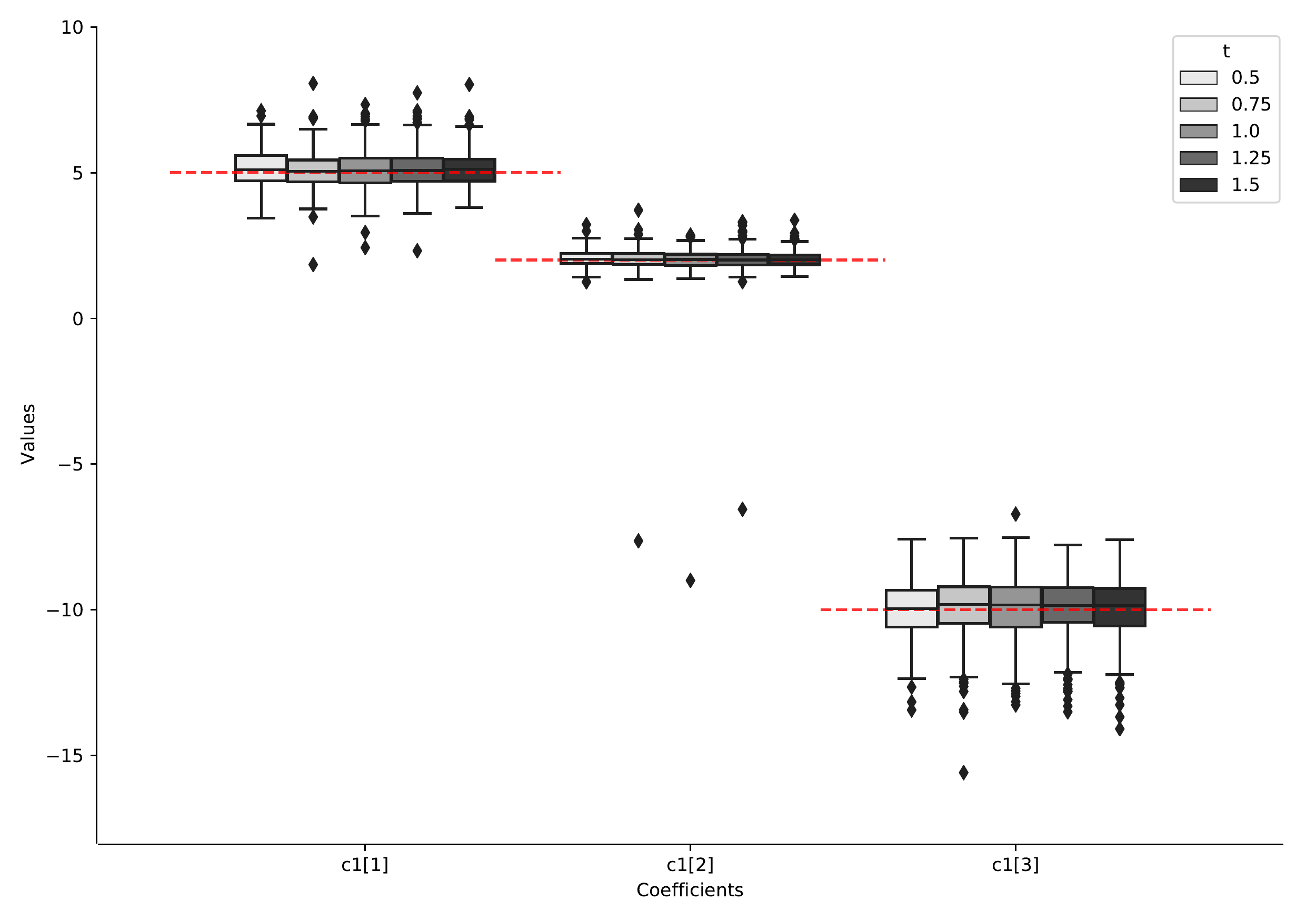} }
    }
    \\
    \subfloat[\centering $\mathbf{c}_1$ (4 groups) - $\alpha$-\textit{criterion}]{
    {\includegraphics[width=7.7cm]{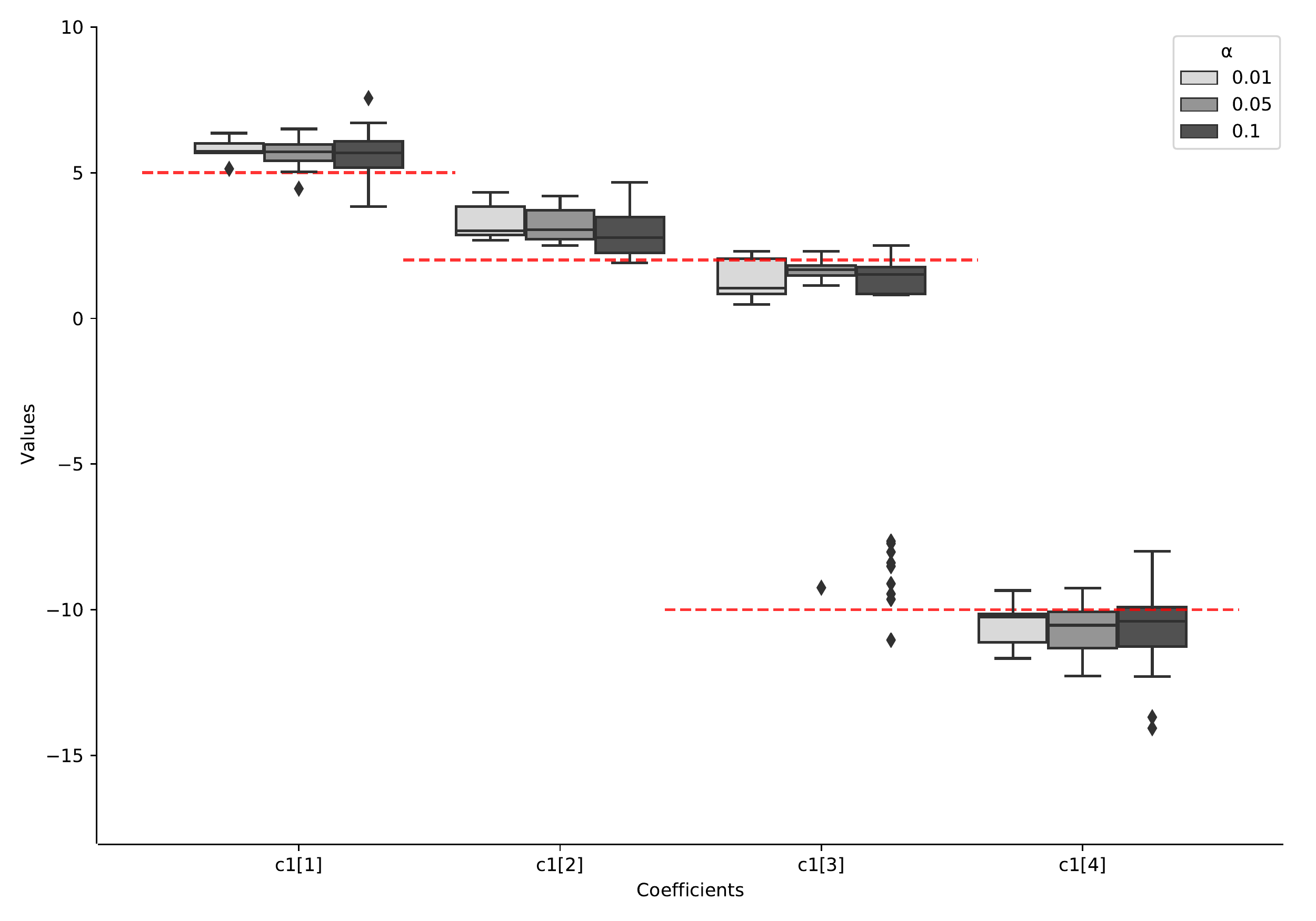} }
    } 
    \subfloat[\centering $\mathbf{c}_1$ (4 groups) - \textit{t}-\textit{criterion}]{
    {\includegraphics[width=7.7cm]{plots_for_tables/False_True_1_c1_4_alpha_B.pdf} }
    }
    \\
    \footnotesize
Notes: For each of the 500 runs with a chosen threshold, we represent boxplots of the value for the components of the random intercept $\mathbf{c}_1$ (y-axis) according to the number of identified clusters (panels (a) and (b) for 2 clusters, panels (c) and (d) for 3 clusters, panels (e) and (f) for 4 clusters).
In the left panels, we report the results of the SPGLMM run via $\alpha$-\textit{criterion}, while in the right panels the results obtained via \textit{t}-\textit{criterion}. The horizontal dotted lines indicate the simulated coefficients.
\label{fig:plots_c1_Bern}
\end{figure}

\begin{figure}
\centering
\caption{Boxplots for the fixed slope $\beta_1$ distribution for the DGP (i) for Bernoulli response, with one fixed slope.}
    \subfloat[\centering $\beta_1$ - $\alpha$-\textit{criterion}]{
    {\includegraphics[width=7.7cm]{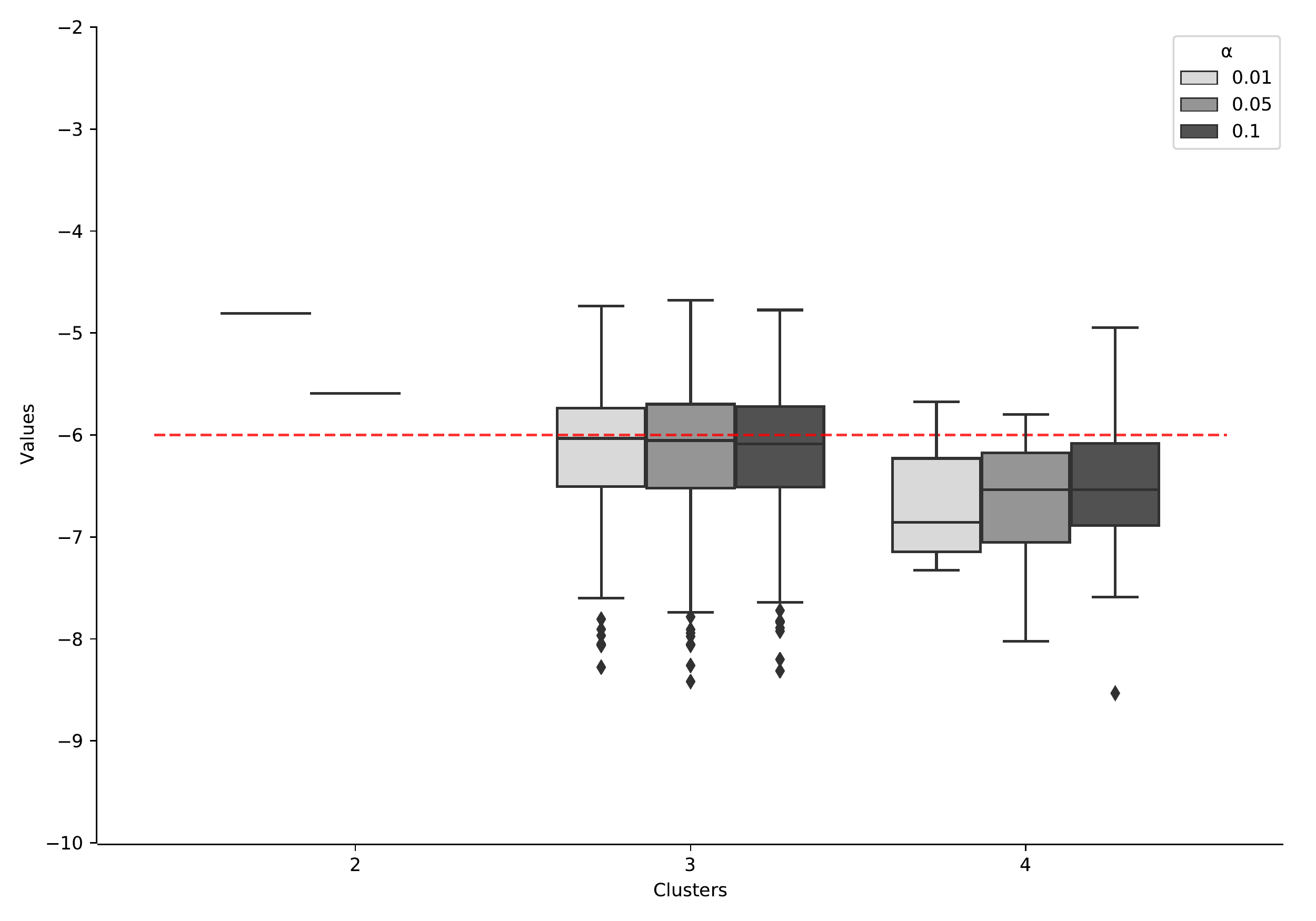} }
    } 
    \subfloat[\centering $\beta_1$ - \textit{t}-\textit{criterion}]{
    {\includegraphics[width=7.7cm]{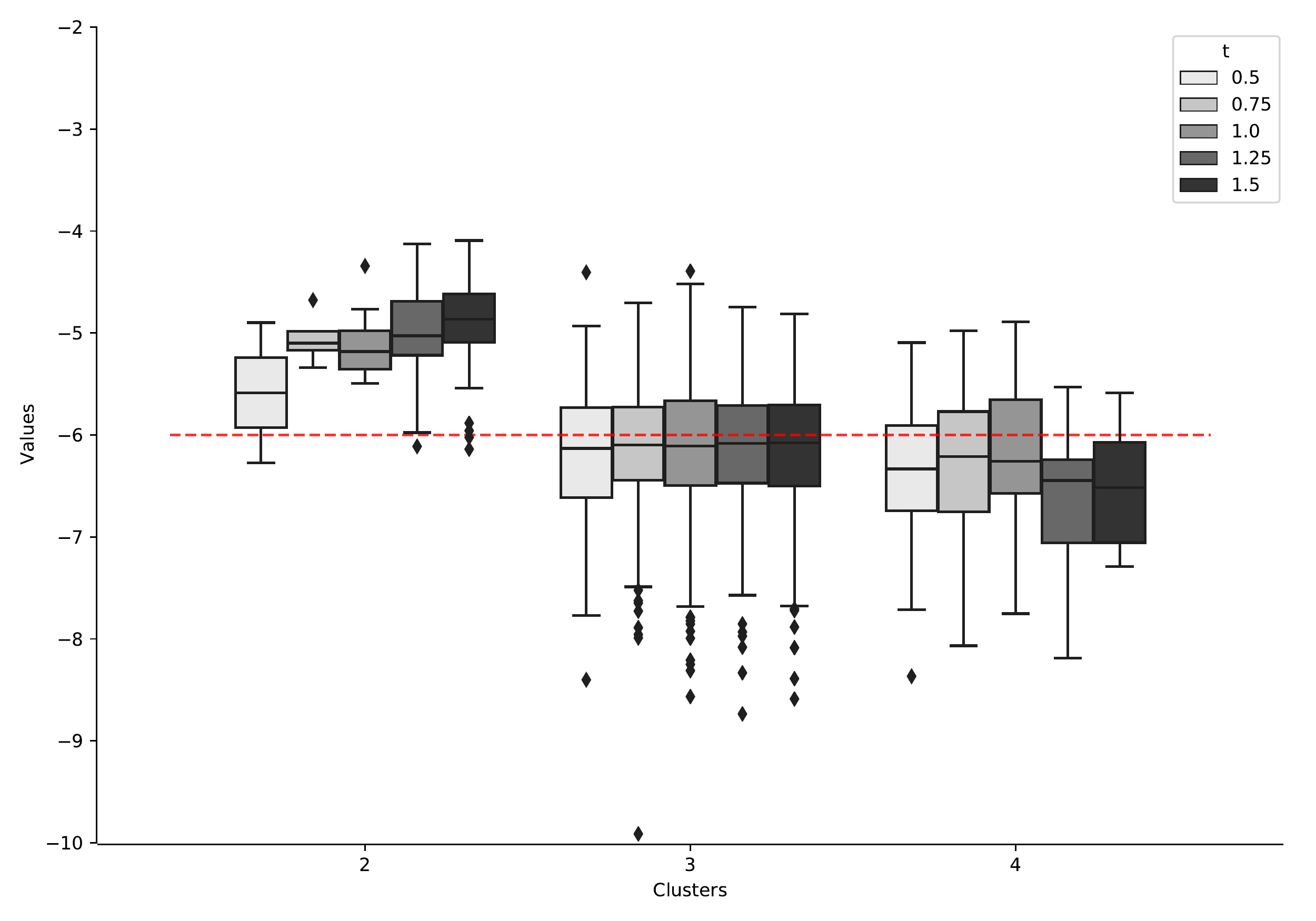} }
    }
    \\
    \footnotesize
Notes: For each of the 500 runs with a chosen threshold, we represent boxplots of the value for the fixed slope $\beta_1$ (y-axis) according to the number of identified clusters (indicated on the x-axis).
In the left panel, we report the results of the SPGLMM run via $\alpha$-\textit{criterion}, while in the right panel the results obtained via \textit{t}-\textit{criterion}. The horizontal dotted lines indicate the simulated coefficients.
\label{fig:plots_beta1_Bern}
\end{figure}

Starting from the simulation study set-up described in Section 4.1, we simulate three different model types, which in turn present:
(i) a random intercept;
(ii) a random slope;
(ii) both a random intercept and a random slope.
For each model, we set up the cases with both one and two fixed slopes (the second fixed slope will be indicated in parenthesis in the following equations), obtaining six different models.
The linear predictor $\boldsymbol{\eta}_i$ is defined by the following Data Generating Processes (DGPs):
\begin{enumerate}[label=(\roman*)]
    \item Random intercept case ($\boldsymbol{\eta}_{i} = \boldsymbol{\beta}_1 \mathbf{x}_{1i} \; + (\boldsymbol{\beta}_2 \mathbf{x}_{2i}) +  c_{1i} \mathds{1}_{n_{i}} $)
    \begin{equation}
    \label{Ber_int_2}
    \boldsymbol{\eta}_i = \begin{cases} 
    -6 \mathbf{x}_{1i} \;+ ( 3 \mathbf{x}_{2i}) + 5\;\mathds{1}_{n_{i}} & \mbox{if } i = 1,2, \\ 
     -6 \mathbf{x}_{1i} \;+ ( 3 \mathbf{x}_{2i}) + 2 \; \mathds{1}_{n_{i}} & \mbox{if } i = 3,4,5,6,7, \\
    -6 \mathbf{x}_{1i} \;+ ( 3 \mathbf{x}_{2i}) -10 \; \mathds{1}_{n_{i}} & \mbox{if } i = 8,9,10
          \end{cases}
\end{equation}

    \item Random slope case ($\boldsymbol{\eta}_{i} = \boldsymbol{\beta}_{1} \mathds{1}_{n_{i}} + \boldsymbol{\beta}_2 \mathbf{x}_{1i} \; + ( \boldsymbol{\beta}_3 \mathbf{x}_{2i})  +  c_{1i} \mathbf{z}_{1i} $)
    \begin{equation}
    \label{Ber_sl_2}
    \boldsymbol{\eta}_i = \begin{cases} 
    +10 \; \mathds{1}_{n_{i}} -6 \mathbf{x}_{1i} \;+ ( 3 \mathbf{x}_{2i}) + 10 \mathbf{z}_{1i} & \mbox{if } i = 1,2, \\ 
     +10 \; \mathds{1}_{n_{i}} -6 \mathbf{x}_{1i} \;+ (3 \mathbf{x}_{2i}) + 5 \mathbf{z}_{1i} & \mbox{if } i = 3,4,5,6,7, \\
    +10 \; \mathds{1}_{n_{i}} -6 \mathbf{x}_{1i} \;+ (3 \mathbf{x}_{2i}) + 0 \mathbf{z}_{1i} & \mbox{if } i = 8,9,10
          \end{cases}
\end{equation}
    
    \item Random intercept and slope case ($\boldsymbol{\eta}_{i} = \boldsymbol{\beta}_1 \mathbf{x}_{1i} \; + (\boldsymbol{\beta}_2 \mathbf{x}_{2i}) +  c_{1i} \mathds{1}_{n_{i}} + c_{2i} \mathbf{z}_{1i} $)
    \begin{equation}
    \label{Ber_intsl_2}
    \boldsymbol{\eta}_i = \begin{cases} 
    -6 \mathbf{x}_{1i} \;+ (3 \mathbf{x}_{2i}) + 5\;\mathds{1}_{n_{i}} +  10 \mathbf{z}_{1i} & \mbox{if } i = 1,2, \\ 
     -6 \mathbf{x}_{1i} \;+ (3 \mathbf{x}_{2i}) + 2 \; \mathds{1}_{n_{i}} + 5 \mathbf{z}_{1i} & \mbox{if } i = 3,4,5,6,7, \\
    -6 \mathbf{x}_{1i} \;+ (3 \mathbf{x}_{2i}) -10 \; \mathds{1}_{n_{i}} + 0 \mathbf{z}_{1i} & \mbox{if } i = 8,9,10
          \end{cases}
\end{equation}
    
\end{enumerate}
Variables $\mathbf{x}_{1i}$, $\mathbf{x}_{2i}$ and $\mathbf{z}_{1i}$ are normally distributed with mean equal to 0 and standard deviation equal to 1.
The choice of the coefficients is driven by the need to simulate situations in which we obtain both balanced and unbalanced proportions of zeros and ones.
In fact, after the computation of $\boldsymbol{\eta}_{i}$ according to each DGP, we retrieve $\mu_{ij}$ by the inverse of the link function $g^{-1}(\eta_{ij})$ 
and we compute $y_{ij}$ for $i=1,...,N$ and $j=1,...,n_i$ as follows:
\begin{equation}
y_{ij} = 
\begin{cases}
0 & \mbox{if } u>\mu_{ij} \\
1 & \mbox{otherwise}
\end{cases}
\end{equation}
where $u$ is randomly extracted from $\mathcal{U}[0,1]$.
Afterwards, we apply SPGLMM with both \textit{t-} and $\alpha$\textit{-criterion}s, performing 500 runs for each of the six settings shown in Eqs. (\ref{Ber_int_2}-\ref{Ber_intsl_2}), for different values of $t$ and $\alpha$.
Results obtained for DGPs (i), (ii) and (iii) with \textbf{one fixed slope} are reported respectively in Tables \ref{tab_Ber_int}, \ref{tab_Ber_sl} and \ref{tab_Ber_intsl} in Section \ref{appC}.
To further prove the generality of our results, we also report the DGP (i) with \textbf{two fixed slopes} in Table \ref{tab_Ber_int2} in Section \ref{appC}.
In the tables, estimates of the proportion of identified clusters, entropy (refer to Section \ref{appEntropy} for the definition), weights $\hat{\omega}$, random and fixed coefficients are reported in terms of mean and standard deviation (sd) across the 500 iterations.
Moreover, we report results concerning DGP (i) with one fixed slope in a more compact way in Figures \ref{fig:proportions_ber}, \ref{fig:plots_c1_Bern} and \ref{fig:plots_beta1_Bern}.
Similar conclusions to the Poisson distributed response can be inferred.
In particular, in Figure \ref{fig:proportions_ber} we notice that the $\alpha$-\textit{criterion} performs much better than the $t$-\textit{criterion} for DGPs (i) and (ii).
For DGP (iii) (last row), the model is more complex since we move to 2-dimensional random effects and the model struggles more in identifying the true number of clusters.
In this case, we are able to identify a $t$ performing better than the $\alpha$-\textit{criterion}. Nevertheless, tuning the most well performing $t$ is computationally expensive, and the $\alpha$-\textit{criterion} gives a well performing alternative.

\begin{figure}[!t]
\centering
\caption{Boxplots representing the distribution of the fixed slope $\beta_1$ and random intercepts for DGP (i) for Bernoulli response, for 100 iterations of the DGP.}
   \subfloat[\centering GLMM (10 groups)]{{\includegraphics[width=15.5cm]{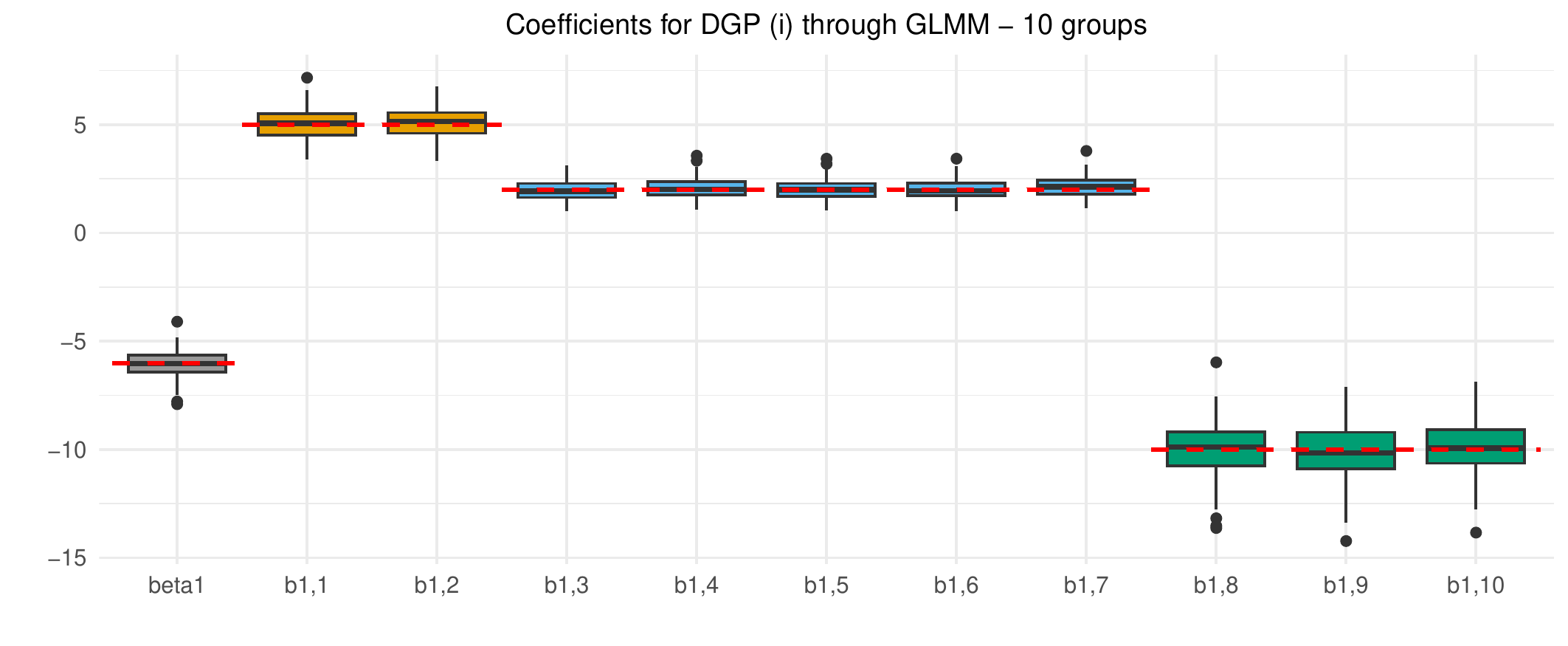} }} \\
    \subfloat[\centering GLMM (3 clusters)]{{\includegraphics[width=7.5cm]{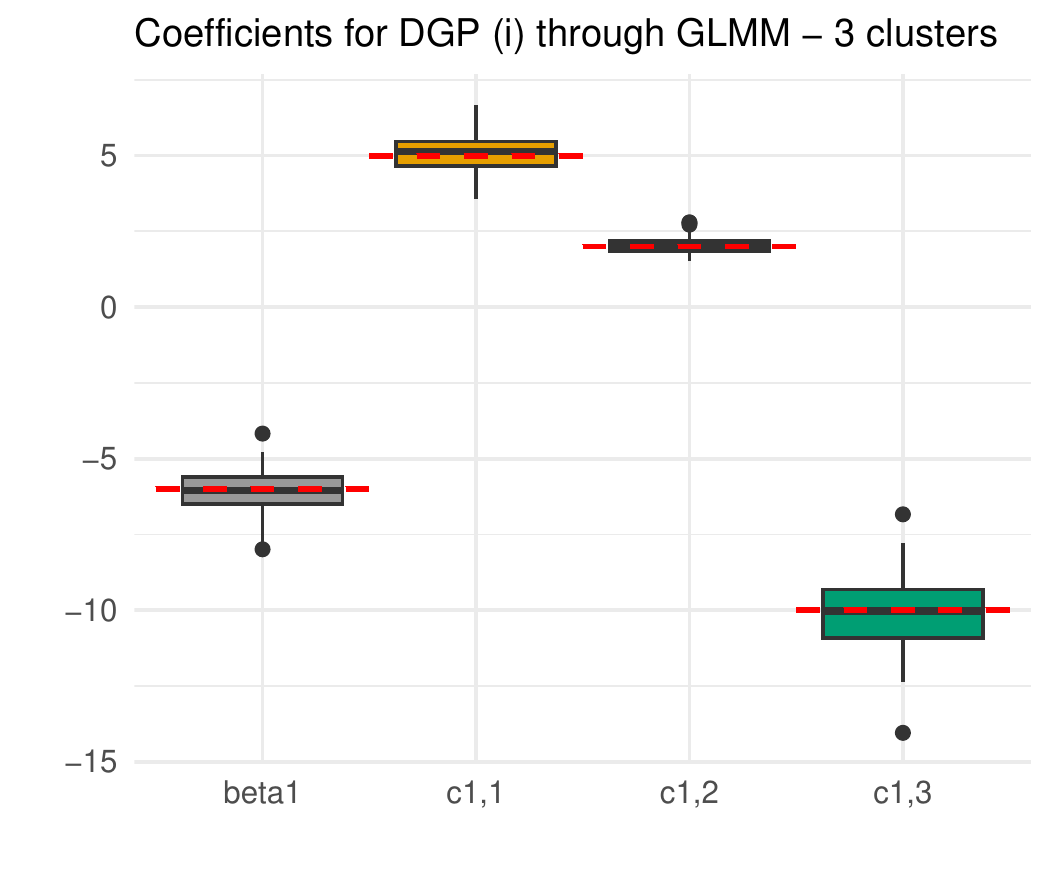} }}
    \;
    \subfloat[\centering SPGLMM (3 clusters)]{{\includegraphics[width=7.5cm]{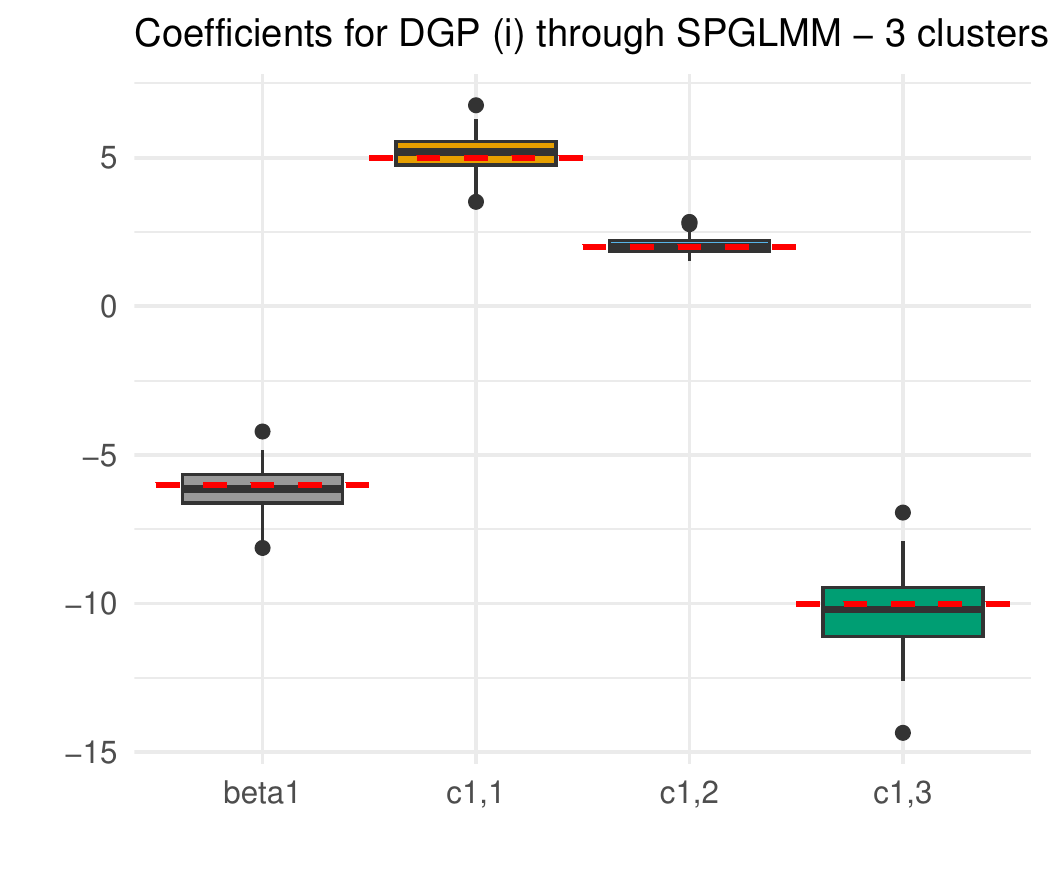} }} \\
    \footnotesize
Notes: Random intercepts are denoted by $b_{1i}$ for $i=1,...,10$ in panel (a) and $c_{1m}$ for $m=1,2,3$ in panels (b) and (c). The horizontal dotted lines indicate the simulated coefficients.
\label{fig:10glmer_3GLMM_3SPGLMM_coef}
\end{figure}

On the other hand, we compare the results of SPGLMM to the ones obtained through a parametric GLMM, implemented in the function \textit{glmer} in R package \textit{lme4} 
(\cite{lme4}, \cite{Rproject}).
Also in this case, we focus on DGP (i) with one fixed slope, to be in line with the case study addressed in Section \ref{bernoulli_case_study}.
We simulate 100 different DGPs as in Eq. (\ref{Ber_int_2}) and we fit, in turn, two different GLMMs, as described in Section 4.1. Moreover, we fit the SPGLMM with $\alpha$-\textit{criterion} with $\alpha=0.05$, knowing that with such a value the algorithm identifies 3 clusters in the $96.6\%$ of the times, as highlighted in Table \ref{tab_Ber_int}. 
For each of the three models (i.e., GLMM with 10 random intercepts, GLMM with 3 random intercepts, SPGLMM), we represent through boxplots the distribution of the obtained coefficients across the 100 DGPs, emphasizing the true values through dotted lines, respectively in Figure \ref{fig:10glmer_3GLMM_3SPGLMM_coef}, panels (a), (b) and (c).
We report in Table \ref{tab:GoF_bernoulli} the summary statistics of the Goodness-of-Fit (GoF) metrics (Sensitivity, Specificity and Accuracy) retrieved by the confusion matrix for each model, which is computed comparing  $y_{ij}$ of the DGP and the estimated $\hat{y}_{ij}$, which assumes value 1 if $\hat{\mu}_{ij} > 0.5$ and 0 otherwise.

Results in Table \ref{tab:GoF_bernoulli} show that the SPGLMM performs almost as well as the GLMM in which 3 clusters are provided to the parametric model. Similar conclusions to the Poisson case can be drawn.
The case in which we run a GLMM with 10 groups performs slightly better, as expected since the models have more flexibility to adapt to the differences in each group. Nevertheless, the difference in the performance can be appreciated only at the third or fourth decimal number in Sensitivity, Specificity and Accuracy.
Concerning the computation of the coefficients, in the boxplots in Figure \ref{fig:10glmer_3GLMM_3SPGLMM_coef} we can appreciate that the actual values are correctly identified in all the cases, with a slightly higher presence of outliers in panels (a) and (c).

\begin{table}
\caption{Summary statistics of the GoF metrics estimates 
for DGP (i), Bernoulli response, with GLMM (10 groups and 3 clusters) and SPGLMM (3 clusters, $\alpha$-\textit{criterion}, $\alpha = 0.05$).}
\label{tab:GoF_bernoulli}
\centering
\begin{tabular}{@{}lcrcrrr@{}}
\hline
&& & &\multicolumn{3}{c}{Quantile} \\
\cline{5-7}
Model &Metric &
\multicolumn{1}{c}{Mean} &
Std. dev.&
\multicolumn{1}{c}{25\%} &
\multicolumn{1}{c}{50\%}&
\multicolumn{1}{c@{}}{75\%} \\
\hline
\multirow{3}{*}{\textbf{GLMM}, 10 groups}  & \textit{Sensitivity} & 0.9421   & 0.0088 & 0.9367 & 0.9414 & 0.9475 \\
                    & \textit{Specificity} & 0.9390   & 0.0109 & 0.9306 & 0.9403 & 0.9466 \\
                    & \textit{Accuracy}    & 0.9406   & 0.0082 & 0.9348 & 0.9418 & 0.9460 \\[6pt]
\multirow{3}{*}{\textbf{GLMM}, 3 clusters} & \textit{Sensitivity} & 0.9403   & 0.0082 & 0.9343 & 0.9410 & 0.9459 \\
                    & \textit{Specificity} & 0.9370   & 0.0111 & 0.9301 & 0.9376 & 0.9450 \\
                    & \textit{Accuracy}    & 0.9387   & 0.0083 & 0.9338 & 0.9390 & 0.9442 \\[6pt]     
\multirow{3}{*}{\textbf{SPGLMM}, 3 clusters}& \textit{Sensitivity} &  0.9400 & 0.0083 &  0.9341  & 0.9410 & 0.9450 \\
                    & \textit{Specificity} &  0.9370 & 0.0111 &  0.9302  & 0.9372 & 0.9444 \\
                    & \textit{Accuracy}    &  0.9385 & 0.0083 &  0.9338  & 0.9386 & 0.9442 \\
\hline
\end{tabular}
\end{table}

\section{The entropy and the elbow method}
\label{appEntropy}

\subsection{Definition of the (average) entropy}
\label{entropy_def}
As mentioned in \cite{masci2022semiparametric}, the uncertainty of classification - with which the algorithm classifies groups into clusters - can be evaluated by measuring the entropy of the rows of the conditional weights matrix $\mathbf{W}$ (see Eq. (5)).
For each group $i$, for $i=1,...,N$, the \textit{entropy} $E_i$ of each array $[W_{i1}, ..., W_{iM}]$ for $i=1,...,N$ is defined as 
    $E_i = - \sum_{m=1}^M W_{im}\; \mathrm{ln}\; W_{im} $.
The \textit{average entropy}\footnote{Simply named \textit{entropy} in the following, without any distinction.} $ E = \frac{1}{N} \sum_{i=1}^N E_i$ assesses the level of uncertainty for which
each group (i.e., the element at the higher hierarchical level) is assigned to a cluster: the closer to $0$, the less uncertain the assignment to the cluster is.
In fact, in the best case, the algorithm assigns each group $i$ to a cluster $m$ with probability $1$ and each row of the matrix $\mathbf{W}$ would be composed of $M-1$ values equal to $0$ and one value equal to $1$ and the entropy $E_i$ would assume value $0$.

\subsection{Discussion on the entropy results in the simulation studies}
\label{entropy_discussion_simstudy}

\begin{figure}
\centering
\caption{Average entropy plot for each of the four simulated cases with one fixed slope.}
    \subfloat[\centering DGP - Poisson - $\alpha$-\textit{criterion}]{
    {\includegraphics[width=7.7cm]{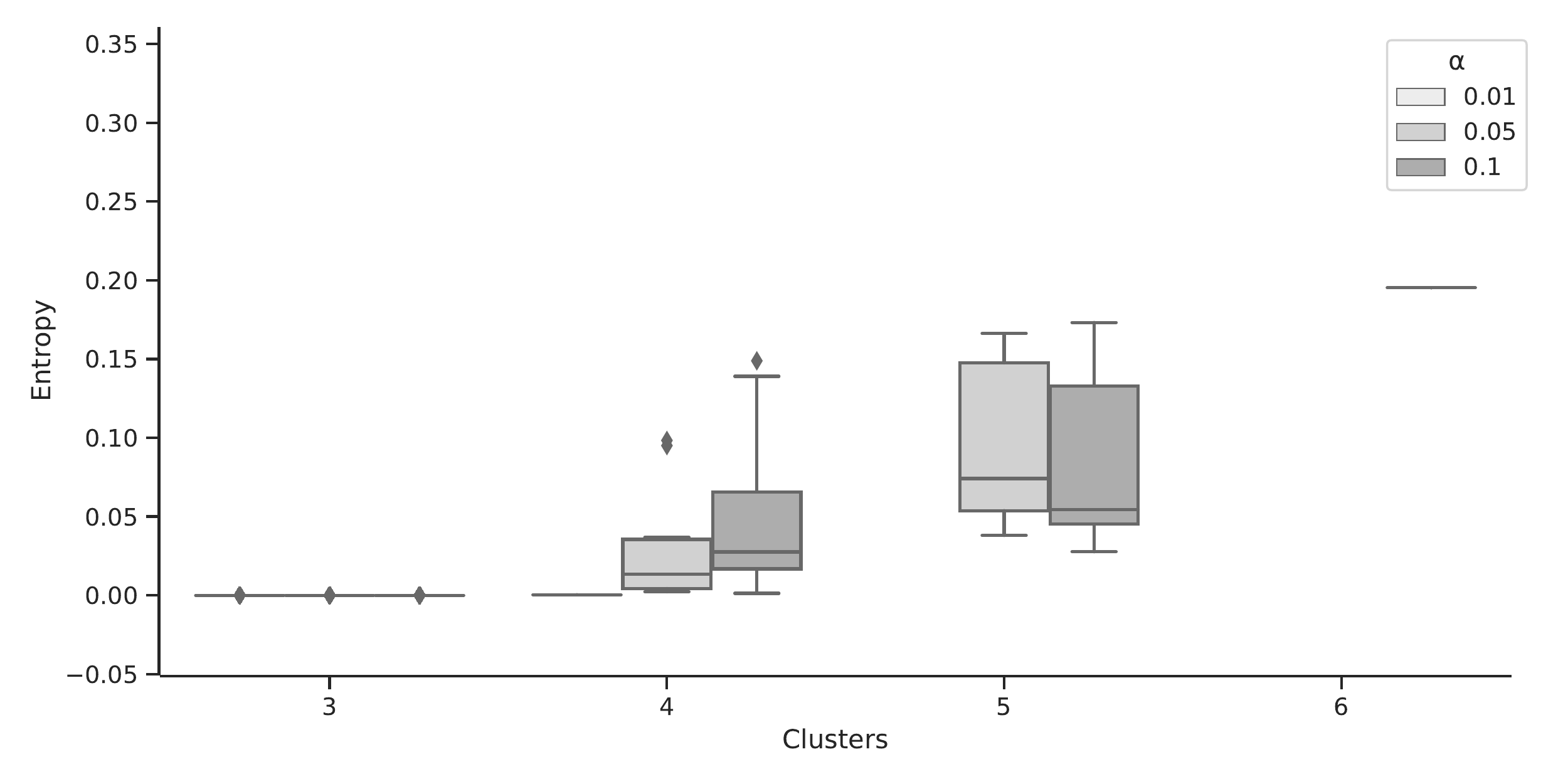} }
    } 
    \subfloat[\centering DGP - Poisson - \textit{t-criterion}]{
    {\includegraphics[width=7.7cm]{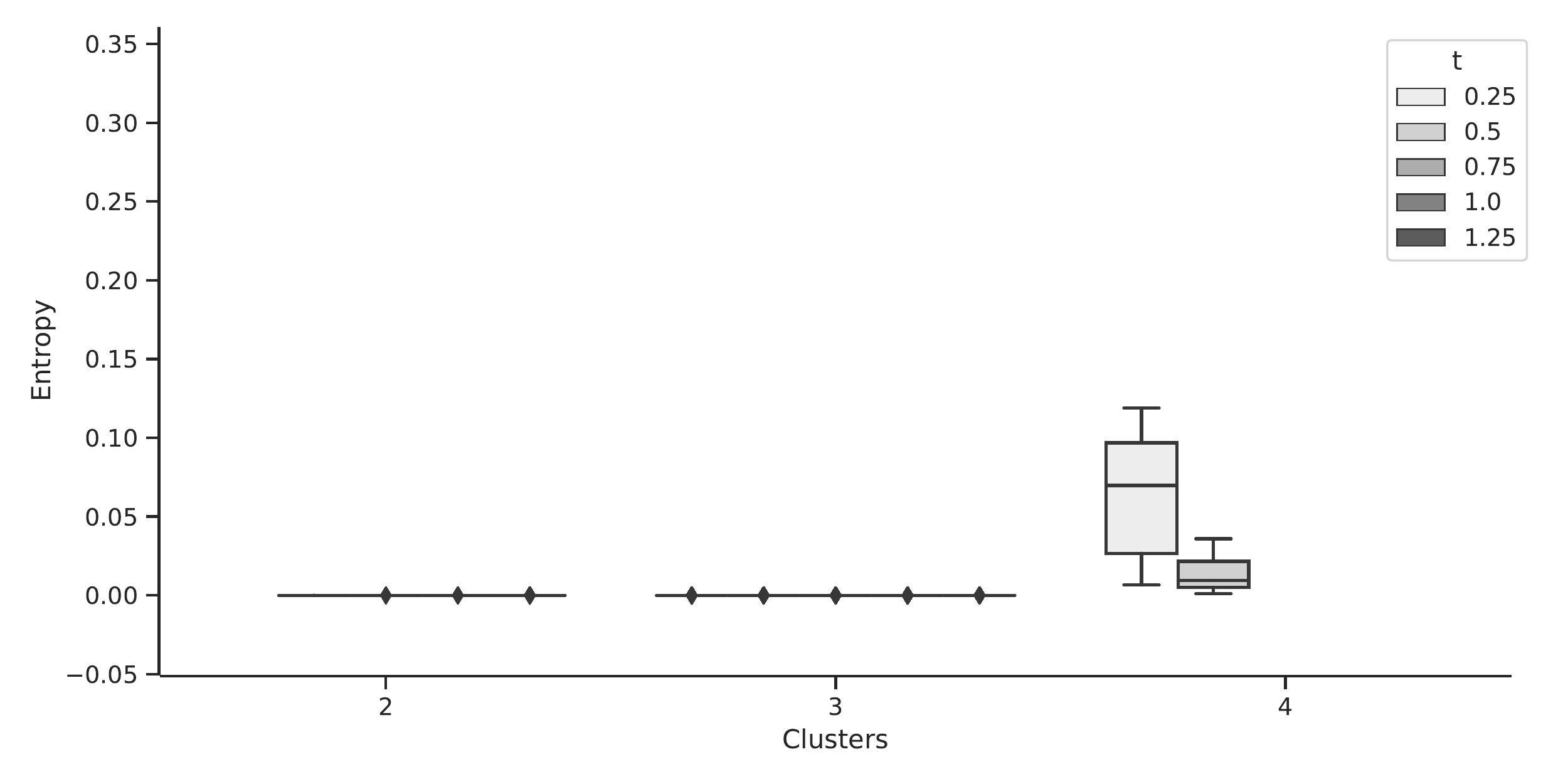} }
    }
    \\
    \subfloat[\centering DGP (i) - Bernoulli - $\alpha$-\textit{criterion}]{
    {\includegraphics[width=7.7cm]{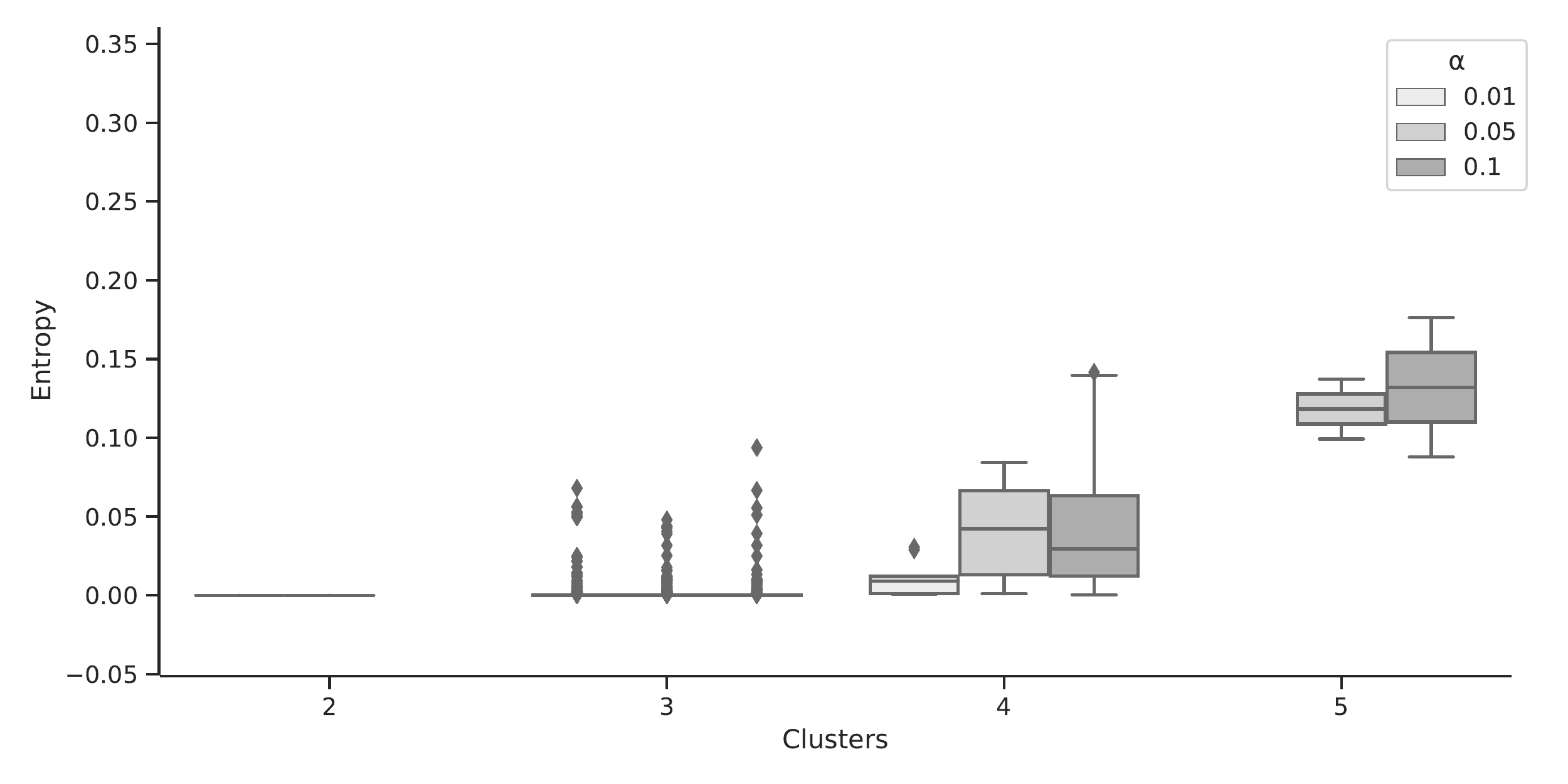} }
    } 
    \subfloat[\centering DGP (i) - Bernoulli - \textit{t-criterion}]{
    {\includegraphics[width=7.7cm]{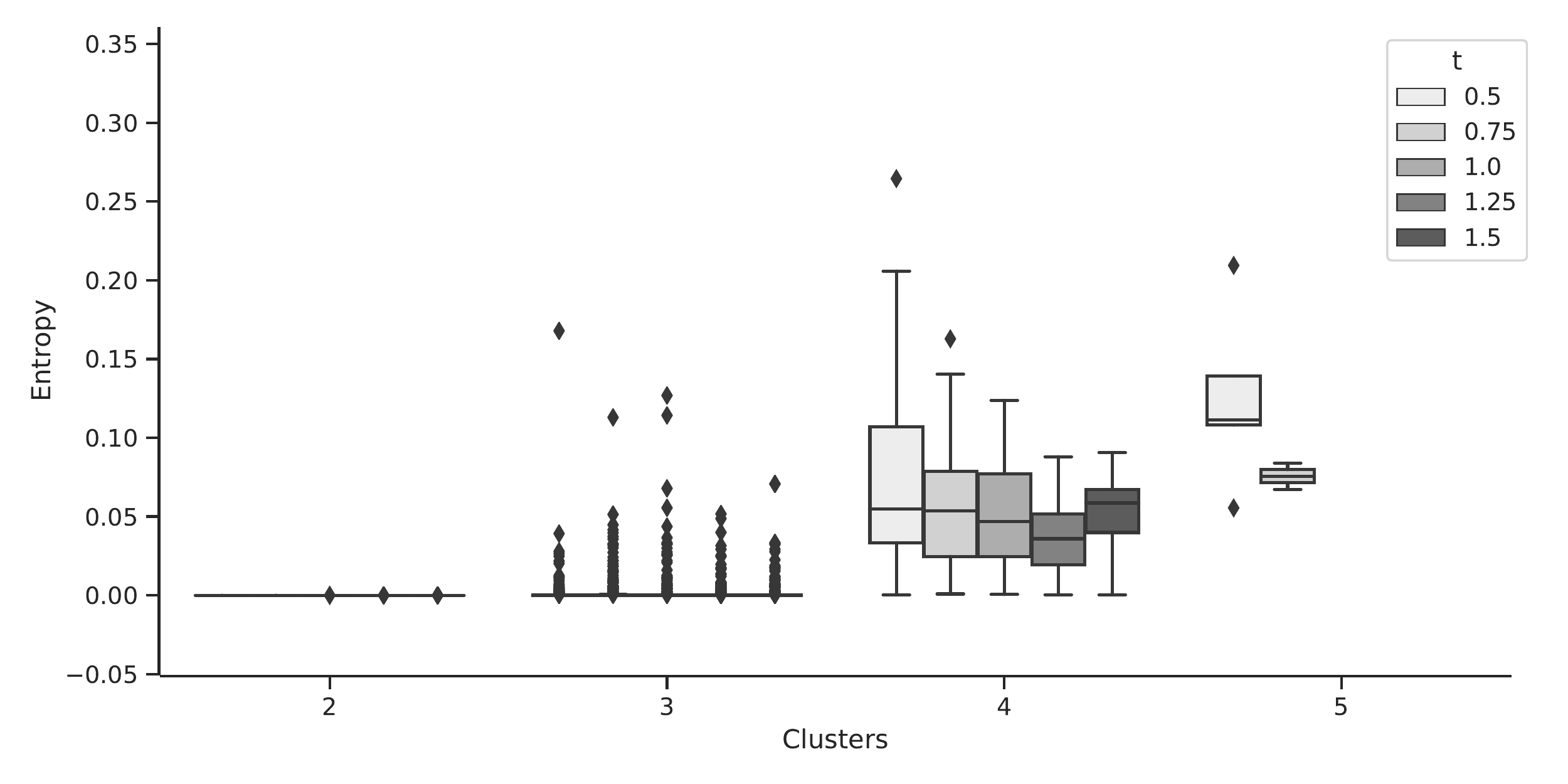} }
    }
    \\
    \subfloat[\centering DGP (ii) - Bernoulli - $\alpha$-\textit{criterion}]{
    {\includegraphics[width=7.7cm]{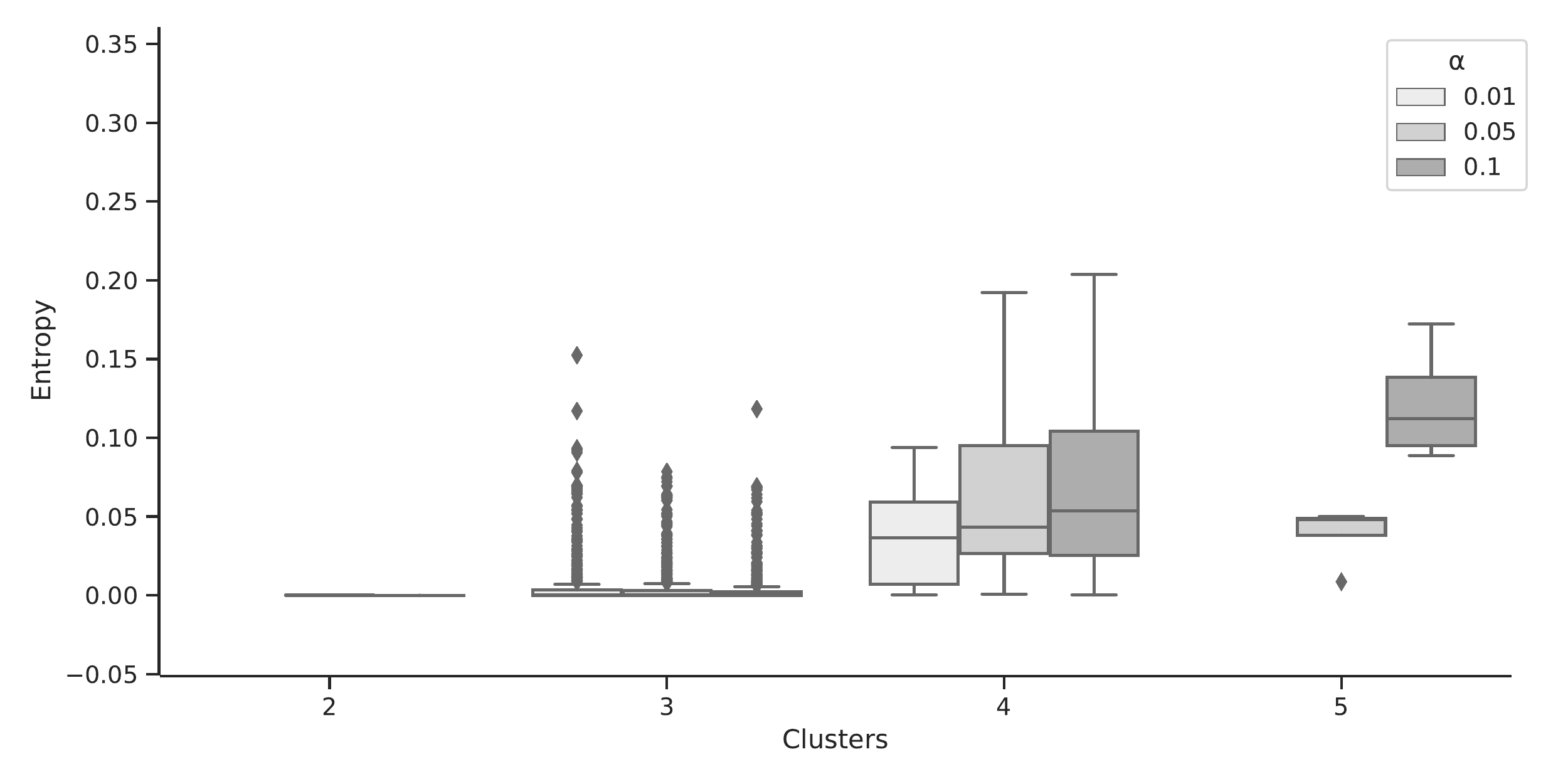} }
    } 
    \subfloat[\centering DGP (ii) - Bernoulli - \textit{t-criterion}]{
    {\includegraphics[width=7.7cm]{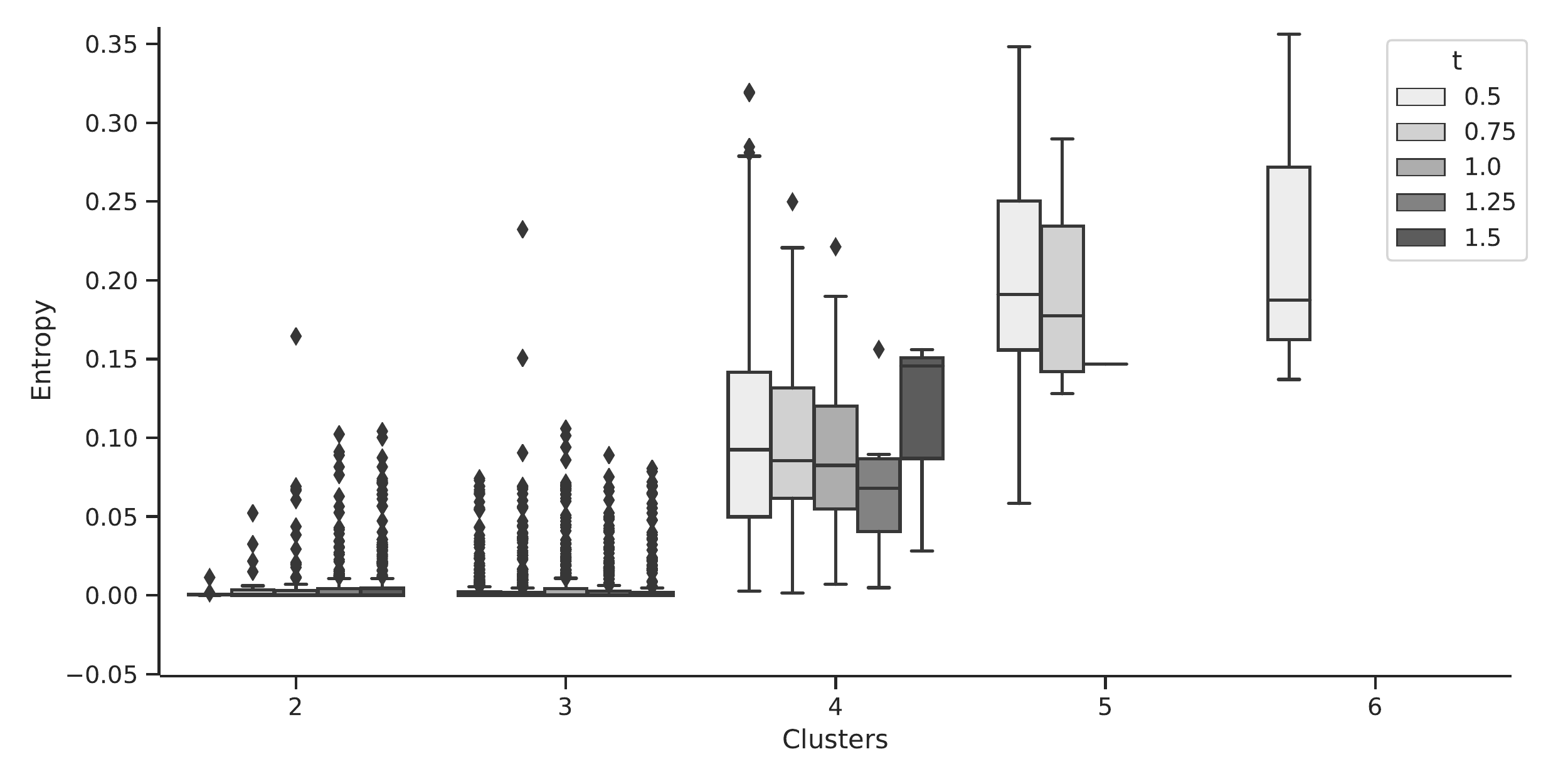} }
    }
    \\
    \subfloat[\centering DGP (iii) - Bernoulli - $\alpha$-\textit{criterion}]{
    {\includegraphics[width=7.7cm]{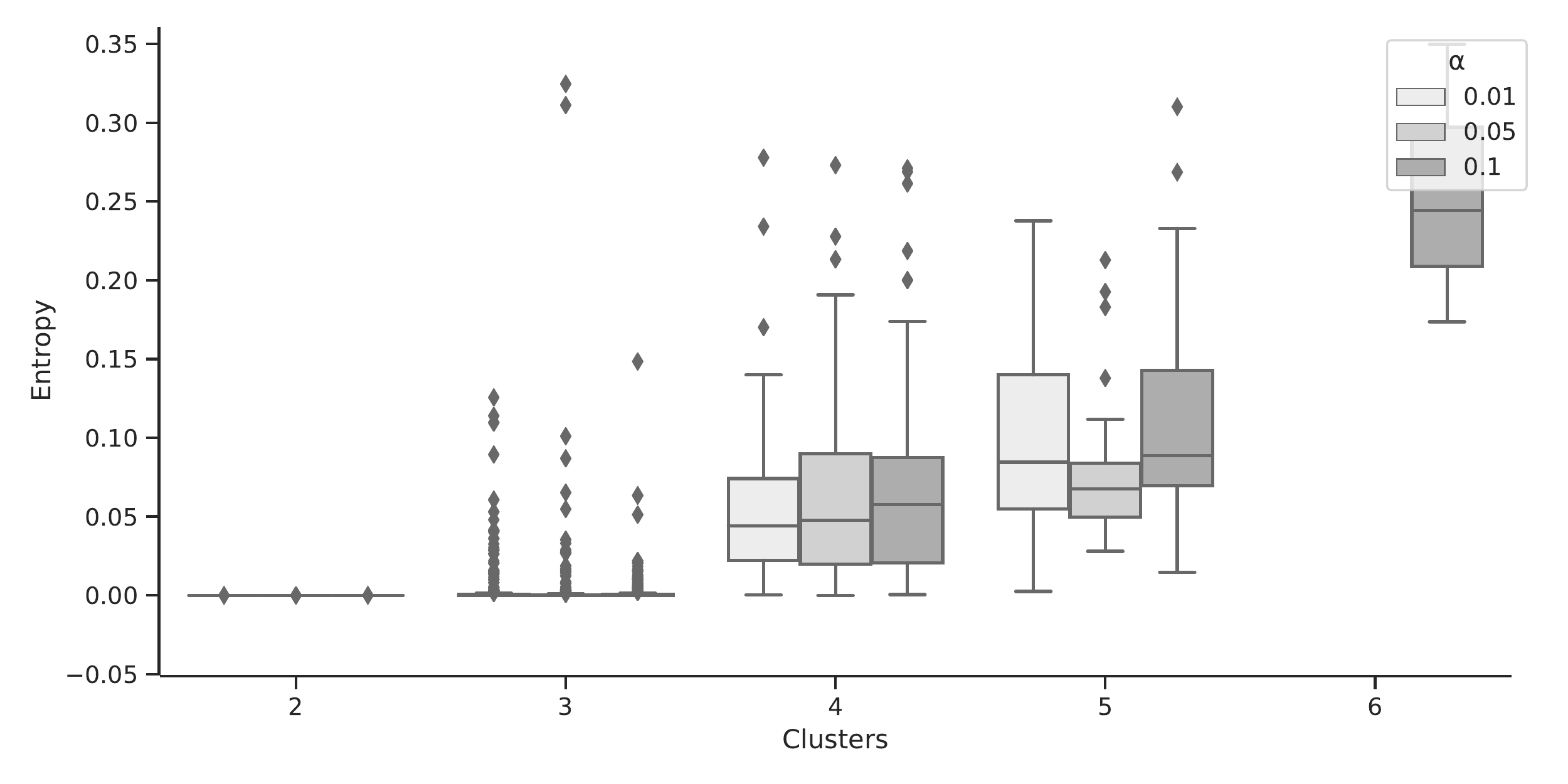} }
    } 
    \subfloat[\centering DGP (iii) - Bernoulli - \textit{t-criterion}]{
    {\includegraphics[width=7.7cm]{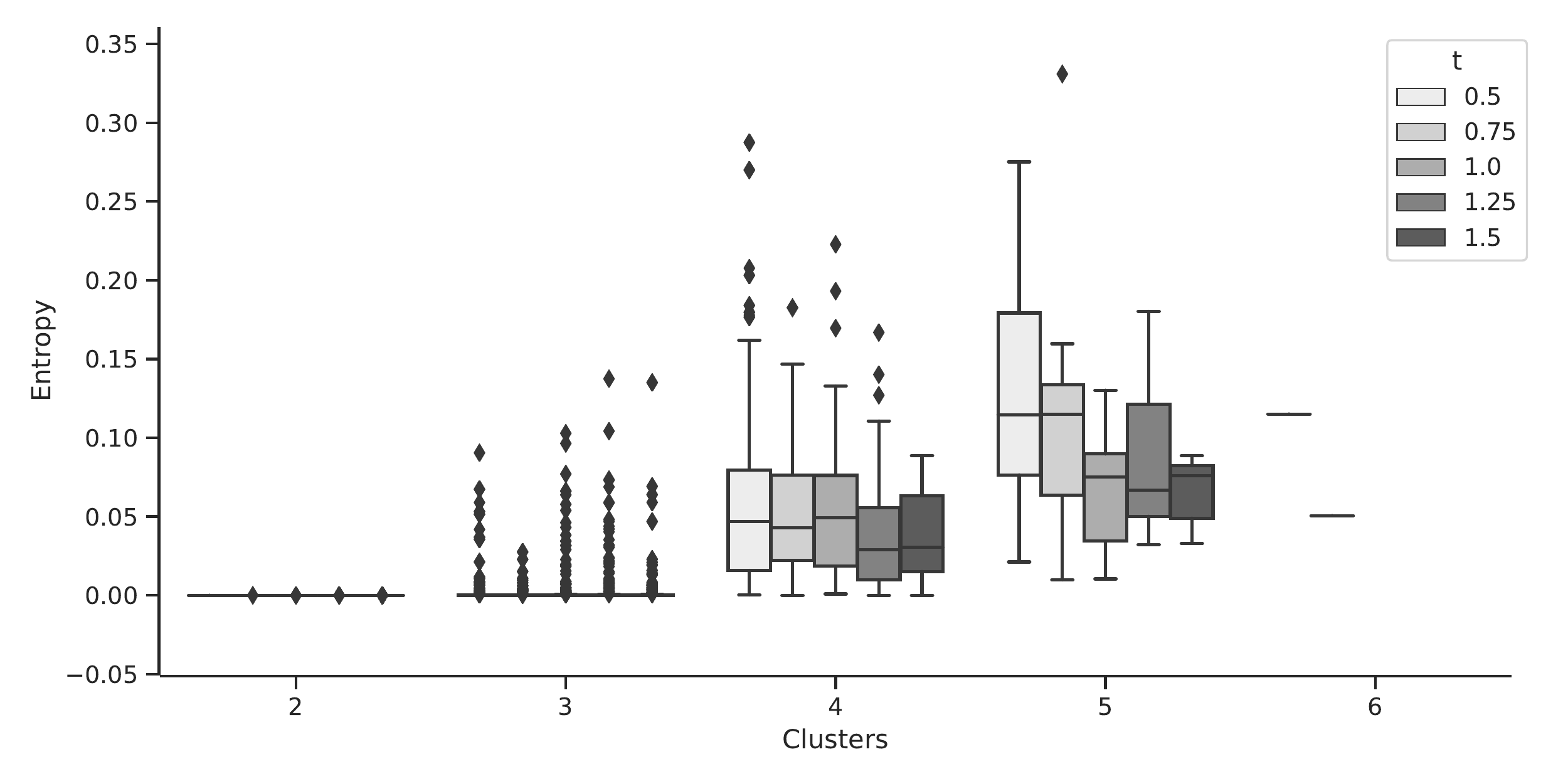} }
    } 
    \\
    \footnotesize
Notes: The first row (panels (a) and (b)) concerns the Poisson response, while the last three rows (panels (c-h)) represent results obtained by fitting an SPGLMM with a Bernoulli response.
For each of the 500 runs with a chosen threshold (see the legend in the right-up corner), we represent boxplots of the average entropy (y-axis) according to the number of clusters $M$ identified by the algorithm (indicated on the x-axis).
In the left panels, we report the results of the SPGLMM run via $\alpha$-\textit{criterion}, while in the right panels the results obtained via \textit{t-criterion}.

\label{fig:entropy}
\end{figure}

For the sake of completeness, we report in Figure \ref{fig:entropy} the boxplots of the average entropy distributions for the simulated cases both for Poisson and Bernoulli responses.
This information can be also found, respectively, in Tables \ref{tab_Poi_int}, \ref{tab_Ber_int}, \ref{tab_Ber_sl} and \ref{tab_Ber_intsl}.
We observe that the entropy is very low for all the cases, for both $\alpha$ and $t$-\textit{criteria}. Moreover, we notice that the entropy assumes higher values in all the cases in which the algorithm identifies 4 clusters (one cluster more than the true number), compared to the cases in which the algorithm identifies 2 and 3 clusters.
These plots should help in understanding the interpretation given in Section \ref{entropy_def}: when the algorithm identifies the true number of clusters (i.e., the latent structure induced by the simulation), the entropy assumes a value of approximately zero. 
This suggests that, as addressed in the next section, the plot of the entropy across different values of $t$ identifies an elbow which could be considered a good indicator for the identification of the optimal number of clusters (equal to 3 in our case).

\subsection{The elbow method}
\label{entropy_elbow method}
As anticipated, the selection of the best performing $t$ within the \textit{t-criterion} is not trivial: the SPGLMM outputs could be very sensitive to it.
The average entropy could serve as a driver for the selection of the best-performing $t$.
In order to identify such a $t$, we could fit distinct SPGLMMs across different values of $t$, and compute, for each of them, the entropy $E(t)$.
By plotting $E(t)$ in function of $t$, inspired by the \textit{elbow}-method in classical clustering algorithms, we could identify the best $t$, namely $t^*$,
in correspondence of an elbow in the piecewise-continuous diagram of $E(t)$. The rationale is to choose a threshold $t^*$ for which the induced number of clusters has low average entropy $E(t^*)$, so that the increase of the threshold to $t^*+\epsilon$  would not make $E(t^*+\epsilon)$ significantly lower than $E(t^*)$ and would not bring particular improvements to the modelling, aside from identifying fewer clusters, which might not be what we are interested in. In other words, we should select the first value of $E$ (in increasing order) after which the decrease in the entropy is negligible.

\begin{figure}
\centering
\caption{Average entropy plot in function of $t$ for each of the four simulated cases with one fixed slope.}
    \subfloat[\centering DGP - Poisson]{
    {\includegraphics[width=7.7cm]{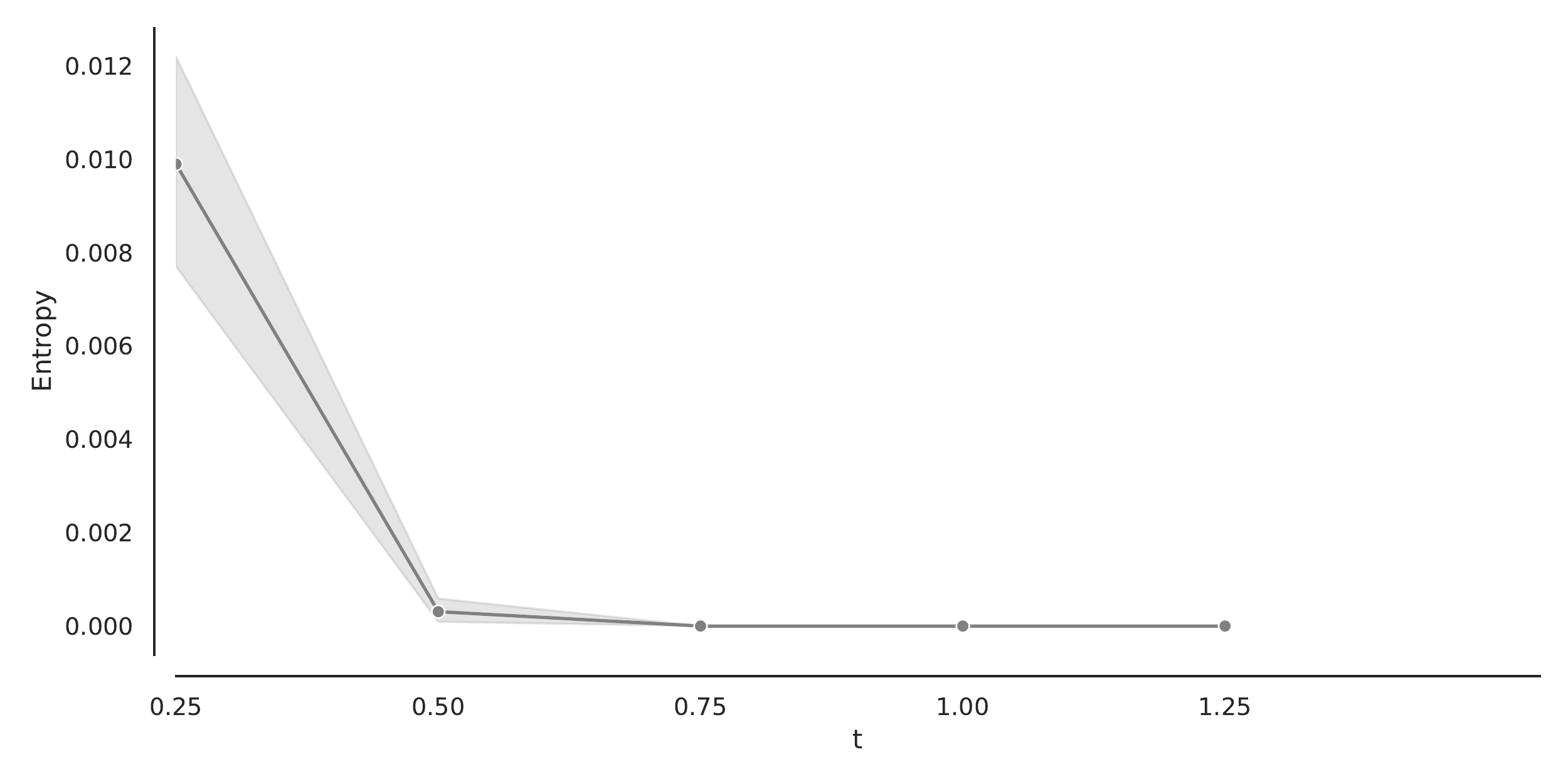} }
    }
    \subfloat[\centering DGP (i) - Bernoulli]{
    {\includegraphics[width=7.7cm]{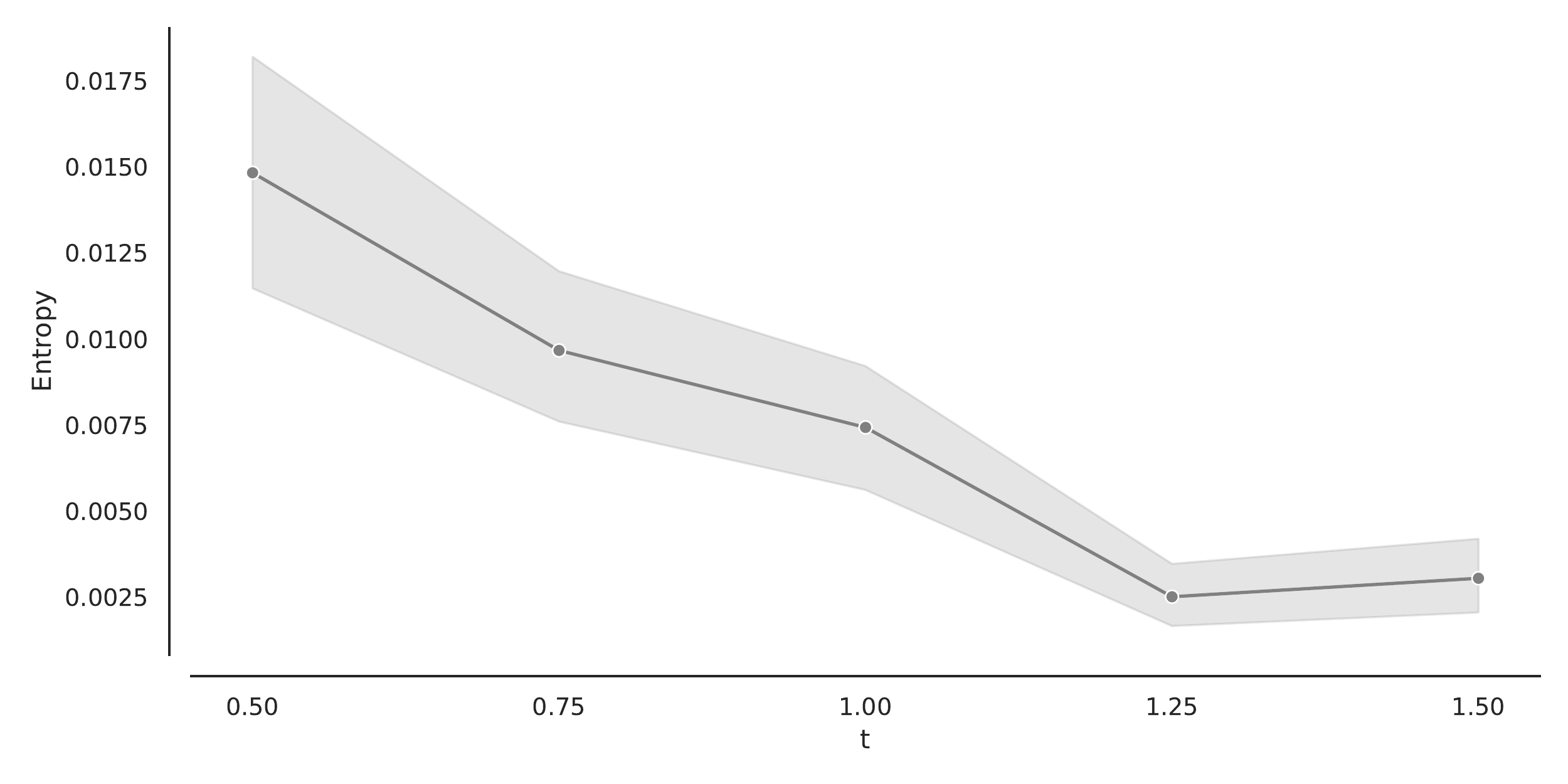} }
    } 
    \\
    \subfloat[\centering DGP (ii) - Bernoulli]{
    {\includegraphics[width=7.7cm]{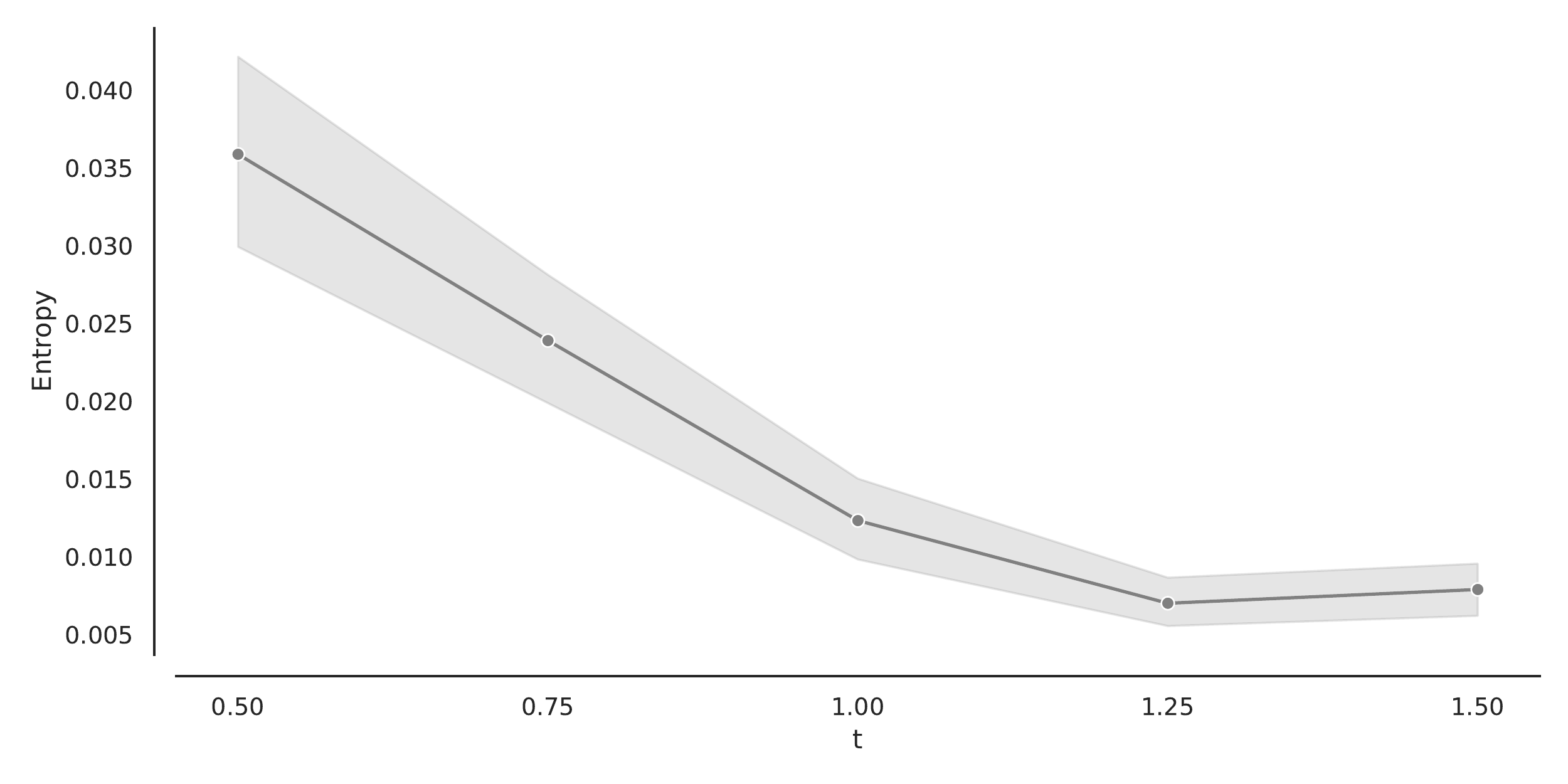} }
    }
    \subfloat[\centering DGP (iii) - Bernoulli]{
    {\includegraphics[width=7.7cm]{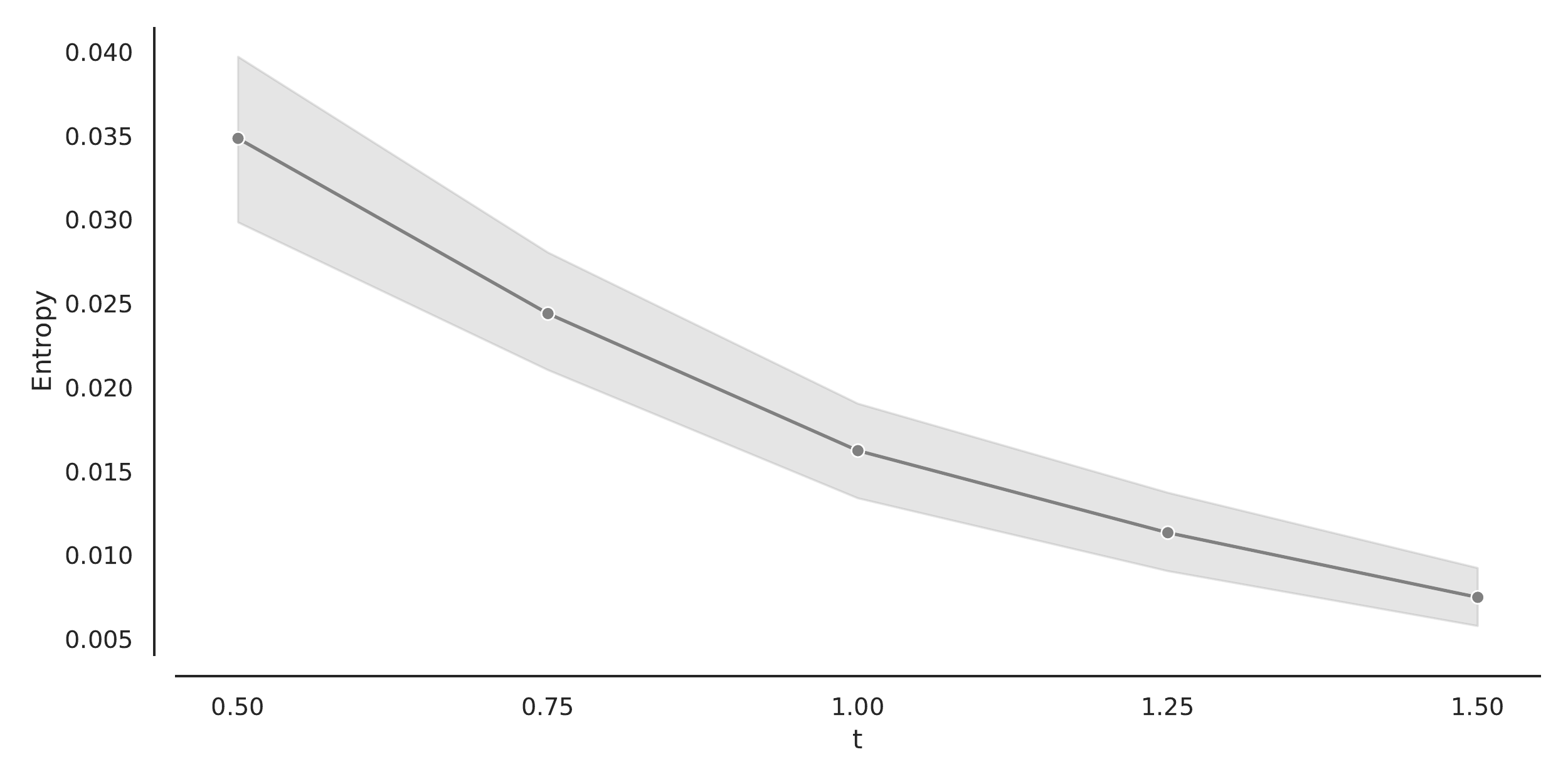} }
    }
    \\
    \footnotesize
Notes: The grey full line represents the average entropy averaged 
across the 500 runs with a certain $t$ (x-axis), while the shadow grey is the $95\%$ confidence interval for the mean.

In panel (a), we clearly identify the elbow at $t=0.50$, which corresponds to the level at which with maximum Proportion in Table \ref{tab_Poi_int} the true number of clusters is identified (95 \%). 

Similarly, in panels (b) and (c), we can identify elbows at $t=1.25$ and $t=1.00$, finding confirmation in Tables \ref{tab_Ber_int} and \ref{tab_Ber_sl} with a maximum Proportion of 92.2\% and 76\%, respectively.

The identification an elbow in the plot in panel (d) is less trivial. This is due to the fact that, as shown in Table \ref{tab_Ber_intsl}, the maximum Proportion (74.6\%) can be identified at $t=1.50$. This means that in order to clearly visualize the elbow we should run the model for $t\geq1.5$. 
\label{fig:elbow_method}
\end{figure}

In Figure \ref{fig:elbow_method} we provide the application of such method to the simulated data (DGP for Poisson response and DGP (i) for Bernoulli response described in Section \ref{simBern}).
The results find a match in the Tables \ref{tab_Poi_int}, \ref{tab_Ber_int}, \ref{tab_Ber_sl} and \ref{tab_Ber_intsl},
proving that this method heuristically seems to work, though it could result in being computationally expensive because requires fitting the model multiple times until when an elbow is clearly identified.

\section{Tables with simulation study outputs}
\label{appC}
We report extended simulation results with:
\begin{itemize}
    \item \textbf{Poisson response} for DGP in Tables \ref{tab_Poi_int} and \ref{tab_Poi_int2} (with one and two fixed slopes, respectively);
    \item  \textbf{Bernoulli response} 
for DGP(i) in Tables \ref{tab_Ber_int} and \ref{tab_Ber_int2}
(with one and two fixed slopes, respectively),
for DGP(ii) in Table \ref{tab_Ber_sl} (with one fixed slope)
and for DGP(iii) in Table \ref{tab_Ber_intsl} (with one fixed slope).
\end{itemize}
In each table, estimates of the proportion of identified clusters, entropy, weights $\hat{\omega}$, random and fixed coefficients are reported in terms of mean and standard deviation (sd) across the 500 iterations.

\begin{landscape}
\begin{table}[!htbp]
 \caption{Results obtained by SPGLMM algorithm for the \textbf{Poisson response} through DGP 
 for $t=0.25, 0.50, 0.75, 1.00, 1.25$, $\alpha=0.01, 0.05, 0.10$, with \textbf{1 fixed slope}.} 
 \begin{center}
\resizebox{\linewidth}{!}{
	  \estauto{tables/Sim_Poi_int.tex}{12}{c} }
\end{center}
\begin{tablenotes}
      \footnotesize
      \item Notes: For each case, 500 runs were performed and results are reported for the cases in which the algorithm identifies 2, 3 and 4 clusters. Estimates of entropy, weights, random and fixed coefficients are reported in terms of mean (sd). True Values (TV) of the coefficients used to simulate data are reported under the relative estimates. Results related to the true number of clusters (i.e. 3) are reported in bold.
 The cases for which the algorithm identifies 1 or more than 4 clusters are not reported in table, but can be identified by complementing with 1 the sum of the three reported Proportions.
 \end{tablenotes}
\end{table}
\end{landscape}

\begin{landscape}
\begin{table}[!htbp]
 \caption{Results obtained by SPGLMM algorithm for the \textbf{Poisson response} through DGP 
 for $t=0.50, 0.75, 1.00, 1.25, 1.50$, $\alpha=0.01, 0.05, 0.10$, with \textbf{2 fixed slopes}.} 
\begin{center}
\resizebox{\linewidth}{!}{
	  \estauto{tables/Sim_Poi_int2.tex}{12}{c} }
\end{center}
\begin{tablenotes}
      \footnotesize
      \item Notes: For each case, 500 runs were performed and results are reported for the cases in which the algorithm identifies 2, 3 and 4 clusters. Estimates of entropy, weights, random and fixed coefficients are reported in terms of mean (sd).
True Values (TV) of the coefficients used to simulate data are reported under the relative estimates. Results related to the true number of clusters (i.e. 3) are reported in bold.
 The cases for which the algorithm identifies 1 or more than 4 clusters are not reported in table, but can be identified by complementing with 1 the sum of the three reported Proportions.
 \end{tablenotes}
\end{table}
\end{landscape}

\begin{landscape}
\begin{table}[!htbp]
 \caption{Results obtained by SPGLMM algorithm for the \textbf{Bernoulli response} through DGP(i) in Eq. (\ref{Ber_int_2}), for $t=0.50, 0.75, 1.00, 1.25, 1.50$, $\alpha=0.01, 0.05, 0.10$, with \textbf{1 fixed slope}.} 
 \begin{center}
\resizebox{\linewidth}{!}{
	  \estauto{tables/Sim_Bin_int.tex}{12}{c} }
\end{center}
\begin{tablenotes}
      \footnotesize
      \item Notes: For each case, 500 runs were performed and results are reported for the cases in which the algorithm identifies 2, 3 and 4 clusters. Estimates of entropy, weights, random and fixed coefficients are reported in terms of mean (sd).
True Values (TV) of the coefficients used to simulate data are reported under the relative estimates. Results related to the true number of clusters (i.e. 3) are reported in bold.
 The cases for which the algorithm identifies 1 or more than 4 clusters are not reported in table, but can be identified by complementing with 1 the sum of the three reported Proportions.
 \end{tablenotes}
\end{table}
\end{landscape}

\begin{landscape}
\begin{table}[!htbp]
 \caption{Results obtained by SPGLMM algorithm for the \textbf{Bernoulli response} through DGP(i) in Eq. (\ref{Ber_int_2}), for $t=0.50, 0.75, 1.00, 1.25, 1.50$, $\alpha=0.01, 0.05, 0.10$, with \textbf{2 fixed slopes}.} 
 \begin{center}
\resizebox{\linewidth}{!}{
	  \estauto{tables/Sim_Bin_int2.tex}{12}{c} }
\end{center}
\begin{tablenotes}
      \footnotesize
      \item Notes: For each case, 500 runs were performed and results are reported for the cases in which the algorithm identifies 2, 3 and 4 clusters. Estimates of entropy, weights, random and fixed coefficients are reported in terms of mean (sd).
True Values (TV) of the coefficients used to simulate data are reported under the relative estimates. Results related to the true number of clusters (i.e. 3) are reported in bold.
 The cases for which the algorithm identifies 1 or more than 4 clusters are not reported in table, but can be identified by complementing with 1 the sum of the three reported Proportions.
 \end{tablenotes}
\end{table}
\end{landscape}

\begin{landscape}
\begin{table}[!htbp]
 \caption{Results obtained by SPGLMM algorithm for the \textbf{Bernoulli response} through DGP(ii) in Eq. (\ref{Ber_sl_2}), for $t=0.50, 0.75, 1.00, 1.25, 1.50$, $\alpha=0.01, 0.05, 0.10$, with \textbf{1 fixed slope}.} 
 \begin{center}
\resizebox{\linewidth}{!}{
	  \estauto{tables/Sim_Bin_slope.tex}{12}{c}}
\end{center}
\begin{tablenotes}
      \footnotesize
      \item Notes: For each case, 500 runs were performed and results are reported for the cases in which the algorithm identifies 2, 3 and 4 clusters. Estimates of entropy, weights, random and fixed coefficients are reported in terms of mean (sd).
True Values (TV) of the coefficients used to simulate data are reported under the relative estimates. Results related to the true number of clusters (i.e. 3) are reported in bold.
 The cases for which the algorithm identifies 1 or more than 4 clusters are not reported in table, but can be identified by complementing with 1 the sum of the three reported Proportions.
 \end{tablenotes}
\end{table}
\end{landscape}

\begin{landscape}
\begin{table}[!htbp]
 \caption{Results obtained by SPGLMM algorithm for the \textbf{Bernoulli response} through DGP(iii) in Eq. (\ref{Ber_intsl_2}), for $t= 0.50, 0.75, 1.00, 1.25, 1.50$, $\alpha=0.01, 0.05, 0.10$, with \textbf{1 fixed slope}.} 
 \begin{center}
\resizebox{\linewidth}{!}{
	  \estauto{tables/Sim_Bin_intslope.tex}{12}{c}}
\end{center}
\begin{tablenotes}
      \footnotesize
      \item Notes: For each case, 500 runs were performed and results are reported for the cases in which the algorithm identifies 2, 3 and 4 clusters. Estimates of entropy, weights, random and fixed coefficients are reported in terms of mean (sd).
True Values (TV) of the coefficients used to simulate data are reported under the relative estimates. Results related to the true number of clusters (i.e. 3) are reported in bold.
 The cases for which the algorithm identifies 1 or more than 4 clusters are not reported in table, but can be identified by complementing with 1 the sum of the three reported Proportions.
 \end{tablenotes}
\end{table}
\end{landscape}

\bibliographystyle{chicago}
\bibliography{bibliography}

\end{document}